\documentclass[final,3p,times]{elsarticle}
\usepackage{setspace}
\usepackage{listings}
\usepackage{xcolor}
\usepackage{graphicx}
\usepackage{float}
\usepackage{pifont}
\usepackage{caption}
\usepackage{subcaption}
\usepackage{tikz,siunitx}
\usepackage{amsmath}
\usepackage{amssymb}
\usepackage{amsthm}
\usepackage{bm}
\usepackage{soul}
\usepackage[ruled,vlined]{algorithm2e}
\usepackage{appendix}
\usepackage{wrapfig, blindtext}
\usepackage[utf8x]{inputenc}
\usepackage[noend]{algpseudocode}
\usepackage{comment}
\usepackage{lineno}
\usepackage{multirow}
\usepackage{adjustbox}
\usepackage{parskip} 
\usepackage[version=4]{mhchem}
\journal{Applied Soft Computing Journal}
\theoremstyle{definition}

\title{Sparse deep neural networks for modeling aluminum electrolysis dynamics}
\definecolor{darkOrange}{rgb}{1.0,0.55,0}

\definecolor{blue}{rgb}{0.,0.,1.0}

\definecolor{defaultblue}{RGB}{31,119.,180}

\definecolor{defaultorange}{RGB}{255,168,91}

\usepackage{xcolor}
\definecolor{Tblue}{HTML}{1f77b4}
\definecolor{Torange}{HTML}{ff7f0e}
\definecolor{Tgreen}{HTML}{2ca02c}
\definecolor{Tred}{HTML}{d62728}
\definecolor{Tpurple}{HTML}{9467bd}
\definecolor{Tbrown}{HTML}{8c564b}

\usepackage{tikz}
\usepackage{pgfplots}
\pgfplotsset{compat=1.17}

\makeatletter
\newenvironment{customlegend}[1][]{%
    \begingroup
    \pgfplots@init@cleared@structures
    \pgfplotsset{#1}%
}{%
    \pgfplots@createlegend
    \endgroup
}%

\def\addlegendimage{\pgfplots@addlegendimage}
\newcommand{\todo}[1]{}

\begin{document} 
\begin{frontmatter}
\author{Erlend Torje Berg Lundby\corref{cor1}\fnref{label1}}
\ead{erlend.t.b.lundby@ntnu.no}
\cortext[cor1]{Corresponding author}
\author[label1,label2]{Adil Rasheed}
\ead{adil.rasheed@ntnu.no}
\author[label1]{Jan Tommy Gravdahl}
\ead{jan.tommy.gravdahl@ntnu.no}
\author[label2]{Ivar Johan Halvorsen}
\ead{Ivar.J.Halvorsen@sintef.no}
\address[label1]{Norwegian University of Science and Technology, Department of Engineering Cybernetics, Trondheim, Norway}
\address[label2]{SINTEF Digital, Department of Mathematics and Cybernetics,  Trondheim, Norway}

\begin{abstract}
    Deep neural networks have become very popular in modeling complex nonlinear processes due to their extraordinary ability to fit arbitrary nonlinear functions from data with minimal expert intervention. However, they are almost always overparameterized and challenging to interpret due to their internal complexity. Furthermore, the optimization process to find the learned model parameters can be unstable due to the process getting stuck in local minima. In this work, we demonstrate the value of sparse regularization techniques to significantly reduce the model complexity. We demonstrate this for the case of an aluminium extraction process, which is highly nonlinear system with many interrelated subprocesses. We trained a densely connected deep neural network to model the process and then compared the effects of sparsity promoting $\ell_{1}$ regularization on generalizability, interpretability, and training stability. We found that the regularization significantly reduces model complexity compared to a corresponding dense neural network. We argue that this makes the model more interpretable, and show that training an ensemble of sparse neural networks with different parameter initializations often converges to similar model structures with similar learned input features. Furthermore, the empirical study shows that the resulting sparse models generalize better from small training sets than their dense counterparts.
\end{abstract}

\end{frontmatter}

\section{Introduction}
    First principle physics-based models (PBMs) have been the workhorse in modeling complex dynamical systems. However, owing to an incomplete understanding of the phenomena, computationally expensive nature, and uncertainty associated with the factors influencing the dynamics, the PBMs are slowly yielding their place to their data-driven modeling counterpart. The increased availability of copious amounts of data, cheap computational resources, and significant algorithmic advancements have further fuelled interest in data-driven models (DDMs). The DDM has the potential to accurately model even poorly understood complex phenomena directly from the data. As a result, a broad array of scientific communities have explored their applicability in many engineering applications.
    Some examples of DDM approaches to modeling dynamical systems are the use of tensor-product based model \cite{hedrea2021tensor}, auto-regressive models \cite{QuocNguyenARMAX2022} and Sparse Identification of Nonlinear Dynamics (SINDy) from a dictionary of candidate functions \cite{brunton2016discovering}. Lately, there has been an increased interest in using neural networks to model nonlinear dynamical systems due to their highly expressive power. Examples are using neural networks to model simulated dynamics of a pressurized water nuclear reactor \cite{Naimi2020100}, identification of the dynamics of the production and purification process of bio-ethanol \cite{9723038}, prediction of chemical reactions \cite{C7ME00107J}, and determine chlorinated compounds in fish \cite{PAPADOPOULOS20115155}. Authors in \cite{blakseth2022dnn} convincingly demonstrated how DDM outperforms PBM in the absence of the full understanding of physics. However, despite their advantages, they suffer from certain shortcomings \cite{san2021hybrid}; they are challenging to interpret, difficult to generalize to solve previously unseen problems, and unstable to train. These are critical shortcomings to overcome before the models can be used in high-stake applications. We discuss each of these briefly.
    
    \textit{Interpretability:} This can be defined as the ability of a model to express itself in human interpretable form \cite{zhang2021survey}. A simple model like linear regression having very few trainable parameters can be a good example of an interpretable model. However, a Deep Neural Network (DNN), constituting millions of trainable parameters, can be extremely difficult or almost impossible to interpret. We can attempt to remedy this by training DNNs with a sparsity prior, thereby reducing the number of parameters and revealing a more parsimonious structure.
    
    \textit{Generalizability:} This refers to the model's ability to predict outcome values for unseen data from the same distribution as the training data. Highly complex and overparameterized models are prone to overfitting, meaning they do not generalize to unseen data not adequately represented during the training. The overfitting can be mitigated by increasing the amount and variety of data. However, in many complex physical systems, the cost and challenges of data acquisition limit the amount of training data. As a result, the trained DNN can fail to generalize. However, if the overparameterization issue is addressed, then there is tentative evidence that the DNN will be better at generalization \cite{liu2019improving}. 
    
    \textit{Stability:} The training of a DNN requires solving an optimization problem in a multidimensional space. Depending upon the complexity of the problem, the dimension can easily be in thousands or even millions, and multiple local minima might be encountered. Even the same DNN with just a slightly different initialization of the parameters can end up in very different minima, and the risk of bad ending up in a bad minima is high. By stability in the context of the current work, we mean that the optimization leads to a reasonable minima, which, even if not global, yields a similar loss value. On the contrary, an unstable DNN will be the one where the optimization process yields inconsistent results, gets stuck in bad local minima, or fails to converge to an acceptable parameter configuration.   

    Combining PBM and DDM in a hybrid modeling approach is an emerging strategy to address the individual shortcomings of PBM and DDM \cite{san2021hybrid}. In \cite{blakseth2022dnn, blaksethcpb2022, robinson2022anc}, the equations known from PBM are augmented by a corrective source term generated by a DNN. In \cite{LUNDBY202162}, the prediction of a PBM is subtracted from a coarse simulation to obtain residual measurements. A compressed sensing algorithm uses these residual measurements to model the unknown disturbance signal of the dynamics. In \cite{RAISSI2019686}, a novel hybrid modeling approach is proposed called Physics Informed Neural Network (PINN). PINN utilizes known first principle knowledge expressed in partial differential equations (PDE) and their corresponding boundary conditions to regularize a neural network that is trained to approximate the solution of the PDE. Authors of \cite{pawar2021physics} introduce a method called Physics Guided Machine Learning (PGML). The training of the DNN is augmented with simplified theories relevant to the system's dynamics. Instead of passing these features as inputs to the network, they are concatenated with the hidden layers of the network, avoiding information loss in earlier layers. As most examples above illustrate, neural networks are often vital in hybrid models. Thus addressing the three issues with DNN mentioned above will benefit not only DDMs but also the hybrid models.  

     In this work we address the challenges of generalizability, interpretability, and stability by training sparse neural networks. Authors in \cite{liu2019improving} show empirically that sparse neural networks can generalize better than dense neural networks on classification tasks. This line of reasoning is intuitively related to Occam's Razor, and even early research such as \cite{NIPS1988_07e1cd7d} has investigated trimming small neural networks to increase interpretability. However, most recent research in deep learning focuses on high-dimensional data and use architectures with millions to billions of parameters. Fully interpreting these models is unfortunately intractable and is not given much attention in this work. On the other hand, dynamical systems can often be expressed in a relatively low dimensional state space despite their rich and complex behaviour. This makes them a good candidate for further investigation. Unfortunately, the research on sparse neural networks for modeling dynamical systems is limited. The authors in \cite{ZHOU2022110489} propose a sparse Bayesian deep learning algorithm for system identification. The method was tested on a simulator of a cascade tank \cite{schoukens2016cascaded} with two states and one input, and on a simulator of coupled electric drives \cite{wigren2017coupled} with three states and one input. Besides this, little research has been done, and the critical shortcomings of DNN mentioned above remain mostly unaddressed. In this work we attempt to address the following research questions:
    \begin{itemize}
    \item What effect can sparsity promoting regularization have on the complexity of neural networks?
    \item Can one generate insight from the interpretation of sparse neural networks, or are they as difficult to interpret as their dense counterparts?
    \item Can sparsity promoting regularization improve the data efficiency of neural networks?
    \item Can the model uncertainty of neural networks be reduced, and their accuracy improved so that they are better suited for modeling the complex dynamics over both short and long-term horizons? 
    \end{itemize}
     
    To conduct the study, we have chosen the dynamics involved in Hall-H\'eroult process for aluminium electrolysis, which is a reasonably complicated system with many states and inputs. In this process, alumina $(Al_2O_3)$ is dissolved in cryolite $(Na_3AlF_6)$ and then reduced to aluminum. The reaction is driven by a line current sent through the electrolytic bath \cite{grjotheim1993introduction}. Accurate dynamic models are crucial for optimizing product quality and energy consumption. The dynamics are highly nonlinear, and interrelated sub-processes make modeling even more challenging. Furthermore, the harsh environment in the electrolytic cells requires extra effort to ensure safe operation. Most recently developed models for this system are PBMs. For example, authors in \cite{gusberti2012modeling} developed a mass and energy balance model based on the first law of thermodynamics. The resulting model includes, among other things, a complete control volume analysis, an extensive material balance, a 3D finite element model (FEM) for modeling resistance in the cell lining and shell, and a Computational Fluid Dynamics (CFD) simulation for computing gas velocity streamlines. In \cite{EINARSRUD20173}, a multi-scale, multi-physics modeling framework including magneto-hydrodynamics, bubble flow, thermal convection, melting and solidification based on a set of chemical reactions was developed. Although highly interpretable, these kinds of PBMs are based on numerous assumptions, an incomplete understanding of the physics, discretization errors, and uncertainties in the input parameters. For example, the magneto-hydrodynamic phenomena or the reactivity and species concentration distribution are challenging to model \cite{mandin2009industrial}. Thus these first-principle model predictions might deviate significantly from measurements of the true system. For the aluminum electrolysis process, DNNs have been applied to predict essential variables that are difficult to measure continuously. In \cite{Chermont2016}, a dense, single-layer neural network with more than 200 neurons was used to simulate bath chemistry variables in the aluminum electrolysis. The paper mainly addresses the training speed of neural networks using an extreme learning machine. In \cite{souza2019soft} dense neural networks were used to predict variables in the electrolysis cell. The study accounted for the changing properties of the electrolysis cells by collecting data over the course of their life-cycle. In \cite{bhattacharyay2017artificial}, dense neural networks with two hidden layers were used to model the properties of the carbon anode. While the literature mentioned above addresses important challenges in modeling this particular process, none consider the issues of interpretability, generalizability, and training stability.
    
    The article is structured as follows. Section~\ref{sec:Theory} presents the relevant theory for the work in the case study. Section~\ref{sec:Method_experiment} presents the method applied in the paper and the experimental setup of the simulator for data generation. In section~\ref{sec:resultsanddiscussion}, the results are presented and discussed. Finally, in section~\ref{sec:conclusions}, conclusions are given, and potential future work is presented.   

\section{Theory}
\label{sec:Theory}
    \subsection{Deep neural networks (DNN)}
    \label{subsec:dnn}
        A DNN is a supervised machine learning algorithm \cite{goodfellow2016deep} that can be denoted by
        \begin{equation}
            \mathbf{y} = \mathbf{\hat{f}}(\mathbf{x}; \boldsymbol{\theta}),
        \end{equation}
        where $\mathbf{y} \in \mathbb{R}^s$ is the output vector of the network model and $s$ its length. $\mathbf{x} \in \mathbb{R}^d$ is the input vector to the network model, and $d$ is the input dimension. Here, $\boldsymbol{\theta} \in \mathbb{R}^p$ denotes all trainable parameters in the network model where $p$ is the number of the parameters. Each layer \textit{j+1} operates on the output vector from the previous layer $\mathbf{Z}^{j} \in R^{L_{j}}$ and outputs a vector $\mathbf{Z}^{j+1} \in \mathbb{R}^{L_{j+1}}$:
        \begin{equation}
            \mathbf{Z}^{j+1} = \sigma (\mathbf{W}^{j+1} \mathbf{Z}^{j} + \mathbf{b}^{j+1}).
        \end{equation}
        $\mathbf{W}^{j+1} \in \mathbb{R}^{L_{j+1} \times L_{j}}$ is called the weight matrix, and $\mathbf{b}^{j+1} \in \mathbb{R}^{L_{j+1}}$ is the called the bias vector of layer $j+1$. $\boldsymbol{\theta} = \{\boldsymbol{\theta^1}, \; ..., \; \boldsymbol{\theta}^{j+1},\; ...\}$, and $\boldsymbol{\theta}^{j+1} = \{\mathbf{W}^{j+1}, \;\mathbf{b}^{j+1}\}$. $\sigma$ is a non-linear activation function. That is,
        \begin{align}
            \sigma(\mathbf{W}^{j+1}\mathbf{Z}^{j} + \mathbf{b}^{j+1}) = (\sigma(W^{j+1}_1\mathbf{Z}^{j} + b^{j+1}_1),\; ...,\nonumber\\ \;
            \sigma(W^{j+1}_i\mathbf{Z}^{j} +  b^{j+1}_i),\; ...,\; \sigma(W^{j+1}_{L_{j+1}}\mathbf{Z}^{j} + b^{j+1}_{L_{j+1}}))^T.
        \end{align}
        $W^{j+1}_i$ are the row vectors of the weight matrix $\mathbf{W}_{j+1}$ and $b^{j+1}_i, \; i=1,\;...,\; L_{j+1}$ are the entries of the bias vector $\mathbf{b}^{j+1}$. Thus, the activation function calculates the output of each neuron in layer $j+1$ as a nonlinear function of the weighted sum of outputs from the neurons in the previous layer plus a bias. Each neuron outputs one value, and the weight in the consecutive layer determines the importance of the output of each neuron in the current layer. The nonlinear activation function $\sigma$ can, for example, be the sigmoid function, hyperbolic tangent function or the binary step function to mention a few. For the last decade or so, the popularity of the piece-wise linear (PWL) activation function Rectified Linear Unit (ReLU) has grown exponentially. This is in part due to its computational simplicity, representational sparsity and non-vanishing gradients. The ReLU activation function is given by:
        \begin{equation}
            \sigma(z) = max\{0,\;z\}. 
        \end{equation}
        ReLU is the only activation function used in the current work.
    
    \subsection{Sparse neural networks and regularization}
    \label{subsec:Sparse_NN}
    Dense neural networks are often overparameterized models, meaning that they have more parameters than can be estimated from the data and thus often suffer from overfitting. In \cite{frankle2018lottery}, it is shown empirically that randomly initialized dense neural networks contain subnetworks that can improve generalization compared to the dense networks. These subnetworks, characterized by significantly fewer non-zero trainable parameters than their dense counterpart, are called sparse neural networks. Their utility can further be seen in terms of increased computational performance for inference and training, and increased storage and energy efficiency. Typically large-scale models that require millions to billions of parameters and arithmetic operations can highly benefit from such sparsification. To conclude sparsification of complex models will lead to simpler models which are relatively easier to interpret, generalize, and train.  
    
    There are many existing schemes and methods for training sparse neural networks. Coarsely speaking, model sparsity can be divided into structured sparsity, which includes pruning for example entire neurons, and unstructured sparsity, which deals with pruning individual weights. Furthermore, methods can be classified into one out of three: \textit{data-free} (such as magnitude pruning \cite{https://doi.org/10.48550/arxiv.1710.01878}), \textit{data-driven} (such as selection methods based on the input or output sensitivity of neurons \cite{zeng2006hidden} )  and \textit{training-aware} methods (like weight regularization) based on the methods way of selecting candidates for removal. A comprehensive review is given in \cite{hoefler2021sparsity}. Among the methods that can be used to sparsify a complex network, regularization techniques are the most popular ones, with a solid mathematical foundation. In regularization, penalty terms $R(\mathbf{w})$ defined on the weights are added to the cost function $C$:
    \begin{equation}
        C(\mathbf{x}_i, \mathbf{y}_i, \boldsymbol{\theta}) =  L(\mathbf{y}_i, \mathbf{\hat{f}}(\mathbf{x}; \boldsymbol{\theta})) + \lambda R(\mathbf{w}).
    \end{equation}
    The vector $\boldsymbol{\theta} = \{\mathbf{w}, \; \mathbf{b}\}$ denotes the adaptable parameters, namely the weights $\mathbf{w}$ and biases $\mathbf{b}$ in the network model. Furthermore, $\{(\mathbf{x}_i, \mathbf{y}_i)\}_{i=1}^N$ is the training data, $L(\cdot, \cdot)$ is the loss function to minimize and the positive scalar coefficient $\lambda$ is a hyperparameter to weight the terms $L(\cdot, \cdot)$ and $R(\cdot)$. The standard choice of loss function $L(\cdot, \cdot)$ for regression tasks is the Mean Squared Error (MSE). In the training process, the cost function is minimized to find optimal values of the parameters:
    \begin{equation}
        \begin{aligned}
        \boldsymbol{\theta}^* = \underset{\boldsymbol{\theta}}{\mathrm{argmin}}\left\{\frac{1}{N}\sum_{i=1}^N C(\mathbf{x}_i, \mathbf{y}_i, \boldsymbol{\theta})   \right\}.
        \end{aligned}
        \label{eq:general_opt_NN}
    \end{equation}
    
    The most intuitive sparsity promoting regularizer is the $\ell_0$ norm, often referred to as the sparsity norm:
    \begin{equation}
        R_{\ell_0}(\mathbf{w}) = ||\mathbf{w}||_0 = \sum_i \begin{cases}
         1 & w_i \neq 0,\\
         0 & w_i = 0.
        \end{cases}
    \end{equation}
    The $\ell_0$ norm counts the number of nonzero weights. Unfortunately, the $\ell_0$ norm has several drawbacks that make it less suitable for optimization. The $\ell_0$ norm is nondifferentiable and sensitive to measurement noise. Furthermore, in terms of computational complexity, the problem of $\ell_0$ norm is shown to be NP-hard\cite{natarajan1995sparse}. The $\ell_1$ norm is a convex relaxation of the $\ell_0$ norm, and is given by:
    \begin{equation}
        R_{\ell_1}(\mathbf{w}) = ||\mathbf{w}||_1 = \sum_i|w_i|.
    \end{equation}
    Due to its geometrical characteristics, $\ell_1$ minimization is sparsity promoting. However, the $\ell_1$ norm usually does not reduce the weights to zero but rather to very small magnitudes. Thus, magnitude pruning can be applied after $\ell_1$ minimization to achieve true sparse models. 
    \begin{figure}[H]
        \centering
        \includegraphics[width=0.5\linewidth]{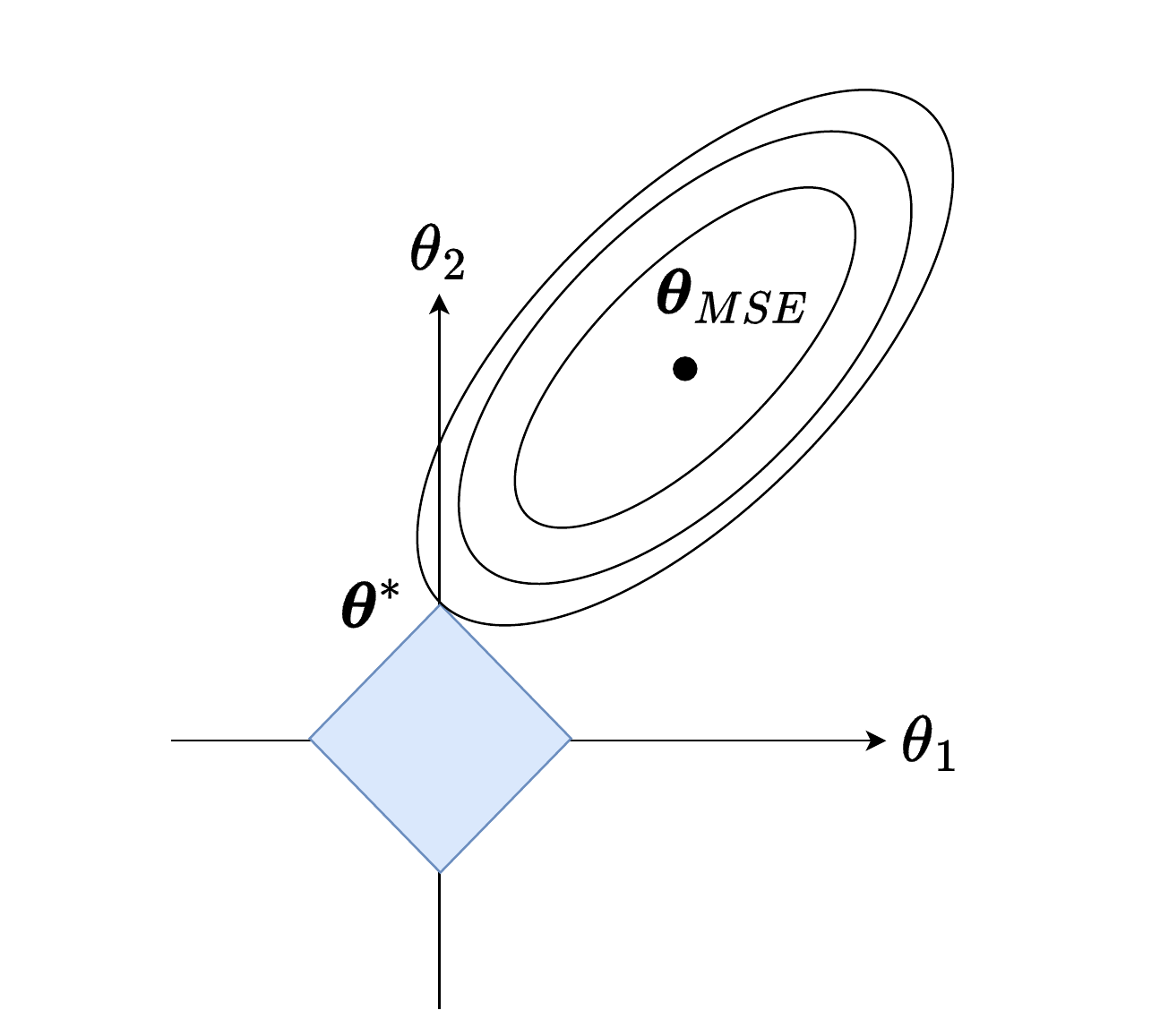}
        \caption{Illustration of $\ell_1$ regularization. $\theta_1$ and $\theta_2$ are the model parameters. $\boldsymbol{\theta}_{MSE}$ is the MSE estimate. The ellipses show the contours of the error from the MSE estimate. The blue diamond illustrates the $\ell_1$ constraints. $\boldsymbol{\theta}^*$ is the parameter estimate when $\ell_1$ regularization is added to the optimization.}
        \label{fig:l1_reg_illustration}
    \end{figure}
    Fig.~\ref{fig:l1_reg_illustration} illustrates how $\ell_1$ regularization can lead to a sparse solution. The contours represented by the ellipse correspond to the mean squared error part ($L(\mathbf{y}_i,\mathbf{\hat{f}}(\mathbf{x};\boldsymbol{\theta}))$) of the cost function $C(\mathbf{x}_i,\mathbf{y}_i,\boldsymbol{\theta})$ while the contours represented by the rhombus correspond to the regularization term. The stronger the regularization parameter $\lambda$ the more the weights $\theta$ will be pushed towards the origin. It can be clearly seen that the two contours corresponding to the two parts of the cost function has a good chance interesting on the axis which will results in many $\theta$ values being zero.
    
    \subsection{Region bounds for PWA neural networks}
    \label{subsec:regionsPWA}
        The complexity of neural networks with Piecewise Affine (PWA) activation functions such as ReLU can be analyzed by looking at how the network partitions the models' input space to an exponential number of linear response regions \cite{montufar2014LinReg, pascanu2014numberResponsReg}. For each region in the input space, the PWA neural network has a linear response for the output. 
        Authors in \cite{SerraBoundingLinear2018} present asymptotic upper and lower bounds for maximum number of regions for a DNN with ReLU activation:
        \begin{align}
            \begin{aligned}
            \textrm{Lower} &: \Omega\left(\left(\frac{n}{d} \right)^{(L-1)d}n^d \right),\\
            \textrm{Upper} &:\mathcal{O}\left(n^{dL} \right). \\
            \end{aligned}
            \label{eq:upper_lower_bound_regions}
        \end{align}
        $d$ is the input dimension, $L$ is the number of hidden layers, and $n$ is the number of activation functions or neurons in each layer. The bounds in Eq.~(\ref{eq:upper_lower_bound_regions}) are valid for networks with the same number of neurons in each layer. The bonds for networks with an uneven number of neurons show similar exponential results and are thus not included for convenience. Eq.~(\ref{eq:upper_lower_bound_regions}) illustrates the exponential relation between the input dimension, number of neurons, and the depth of the network. For realistic amounts of data sampled from a physical system, the number of linear regions that a relatively small dense neural network partition the input space into exceeds the sampled data by several orders of magnitude. Thus, in order to generalize to larger areas of the models' input space, the number of regions needs to be reduced drastically. This motivates sparsifying the model. 
    
    \subsection{Simulation model}
    \label{subsec:simulaitonmodel}
        The simulation model used in this article is based on the mass and energy balance of an aluminum electrolysis cell. The derivation of the model is found in the \ref{appendix:a}. In this section, the simulation model is put in a state space form, where constants, system states, and control inputs are denoted by the symbols $k_i$, $x_i$, or $u_i$ respectively. The simulation model can be expressed as a nonlinear system of ODE's with $8$ states $\mathbf{x}\in \mathbb{R}^8$ and $5$ inputs $\mathbf{u} \in \mathbb{R}^5$ on the form:
        \begin{equation}
            \Dot{\mathbf{x}} = f(\mathbf{x}, \mathbf{u}),
            \label{eq:nonlin_state_space}
        \end{equation}
        where $\Dot{\mathbf{x}} \in \mathbb{R}^8$ is the time derivative of the states $\mathbf{x}$, and $f(\mathbf{x}, \mathbf{u})$ is a nonlinear function. Table \ref{table:states_inputs} gives the physical meaning and units of all the state and input variables. Fig. \ref{fig:schematic} gives a schematic diagram of the process.
        \begin{figure}[ht]
        \centering
        \includegraphics[width=\linewidth]{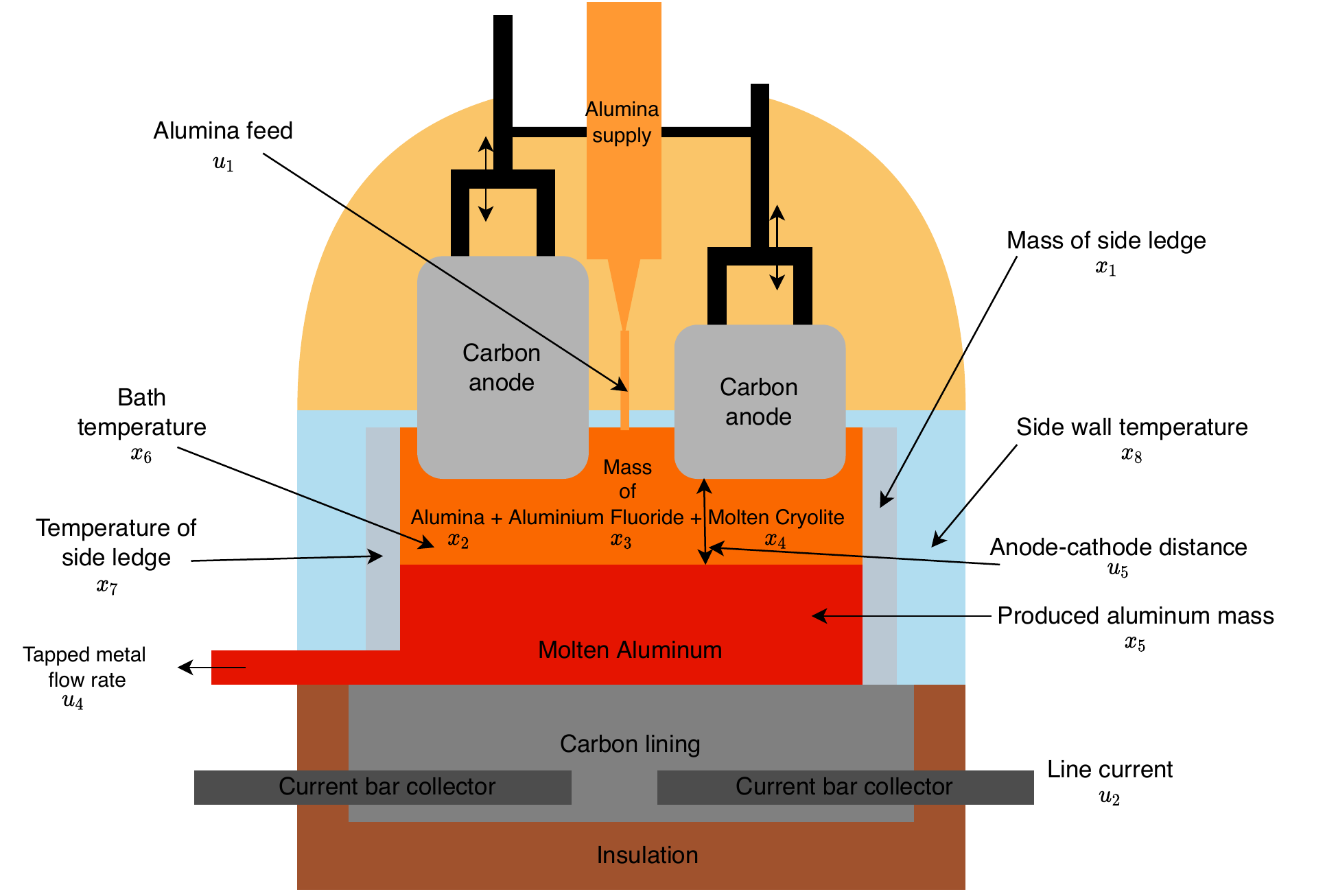}
        \caption{Schematic of the setup. The figure show the states and inputs in Table~\ref{table:states_inputs}.}
        \label{fig:schematicofthesetup}
        \end{figure}

        \begin{table}[H]
        \begin{center}
        \caption{Table of states and inputs}
        \begin{tabular}{l|l|l}
        \hline
         Variable&Physical meaning & Unit  \\ \hline
        $x_1$ & mass side ledge & $kg$ \\
        $x_2$ & mass $Al_2O3$ & $kg$  \\
        $x_3$& mass $ALF_3$ & $kg$ \\
        $x_4$ & mass $Na_3 AlF_6$ & $kg$  \\
        $x_5$ & mass metal & $kg$ \\
        $x_6$ & temperature bath & $^\circ C$ \\
        $x_7$ & temperature side ledge & $^\circ C$ \\
        $x_8$ & temperature wall & $^\circ C$\\
        \hline
        $u_1$ &  $Al_2O_3$ feed &$kg$ \\
        $u_2$ & Line current  & $kA$\\
        $u_3$ & $AlF_3$ feed & $kg$\\
        $u_4$ & Metal tapping & $kg$\\
        $u_5$ & Anode-cathode distance & $cm$ \\
        \hline
        \end{tabular}
        \label{table:states_inputs}
        \end{center}
        \end{table}
        
        The nonlinear functions in Eqs. (\ref{eq:g_1}) - (\ref{eq:g_5}) that partly describes the dynamics of the system states are defined in advance of presenting the system dynamics in order to simplify the expressions in Eq. \ref{eq:nonlin_state_space}:
        
        \begin{align}
            g_1 &= \,  991.2 + 112c_{x_3} + 61c_{x_3}^{1.5} - 3265.5c_{x_3}^{2.2} \nonumber \\
            &\, - \frac{793 c_{x_2}}{- 23 c_{x_2} c_{x_3} - 17 c_{x_3}^{2} + 9.36 c_{x_3} + 1}\label{eq:g_1}
        \end{align}
        \begin{align}
            g_2 &=\,  \text{exp}\,\left(2.496 - \frac{2068.4}{273+x_6} - 2.07c_{x_2}\right) \label{eq:g_2}
        \end{align}
        \begin{align}
            g_3 &=\, 0.531 + 3.06 \cdot 10^{-18} u_{1}^{3} - 2.51 \cdot 10^{-12} u_{1}^{2} \nonumber \\ &+ 6.96 \cdot 10^{-7} u_{1} 
        - \frac{14.37 (c_{x_2} - c_{x2,crit}) - 0.431}{735.3 (c_{x_2} - c_{x2,crit}) + 1} \label{eq:g_3}
        \end{align}
        \begin{align}
            g_4 &=\, \frac{0.5517 + 3.8168 \cdot 10^{-6}u_2}{1 + 8.271 \cdot 10^{-6}u_2}\label{eq:g_4}
        \end{align}
        \begin{align}
            g_5 &=\, \frac{3.8168\cdot 10^{-6}\cdot g_3 \cdot g_4 \cdot u_2}{g_2(1 -g_3)}\label{eq:g_5}  
        \end{align}
        
        $g_1$ is the liquidus temperature $T_{liq}$ defined in Eq.~(\ref{eq:T_liq}), $g_2$ is the electrical conductivity $\kappa$ defined in Eq. (\ref{eq:el_cond}), $g_3$ is the bubble coverage $\phi$ defined in Eq. (\ref{eq:bub_cov}), $g_4$ is the bubble thickness $d_{bub}$ defined in Eq. ($\ref{eq:d_bub}$), and $g_5$ is the bubble voltage drop $U_{bub}$ defined in Eq.~(\ref{eq:U_bub}).
        The expressions and coefficients of the physical properties in Eqs.~(\ref{eq:g_1})-(\ref{eq:g_5}) are taken from different academic works that have estimated these quantities. These academic works are cited in the Appendix.
        Notice that the mass ratios in the electrolyte of $x_2$ (\ce{Al2O3}) and $x_3$ (\ce{AlF_3}) can be expressed as: 
        \begin{align}
            c_{x_2} &= x_2/(x_2 + x_3 + x_4) 
            \label{eq:alu_ratios_x2}
        \end{align}
        \begin{align}
           c_{x_3} &= x_3/(x_2 + x_3 + x_4)
           \label{eq:alu_ratios_x3}
        \end{align}
        With the nonlinear functions $g_1,\; ... \; g_5$ described, the state space equations in Eq. \ref{eq:nonlin_state_space} is described by the following equations:
        \begin{align}
        \Dot{x}_1 &=\,\frac{k_1(g_1 - x_7)}{x_1 k_0} - k_2 (x_6 - g_1) \label{eq:xdot1}
        \end{align}
        \begin{align}
        \Dot{x}_2 &=\, u_1 - k_3 u_2 \label{eq:xdot2}
        \end{align}
        \begin{align}
        \Dot{x}_3 &=\, u_3 - k_4 u_1 \label{eq:xdot3}
        \end{align}
        \begin{align}
        \Dot{x}_4 &=\, -\left(\frac{k_1 (g_1 - x_7)}{x_1 k_0} - k_2 (x_6 - g_1) \right) + k_5 u_1 \label{eq:xdot4}
        \end{align}
        \begin{align}
        \Dot{x}_5 &=\, k_6 u_2 - u_4 \label{eq:xdot5}
        \end{align}
        \begin{align}
        \Dot{x}_6 &=\, \frac{\alpha}{x_2+x_3+x_4} \biggl[ u_2\left( g_5 + \frac{u_2 u_5}{2620 g_2}\right) \nonumber\\
        &\, \; - k_9 \frac{x_6 - x_7}{k_{10} + k_{11} k_0  x_1}  \nonumber\\
        & \, \; - \left (k_7 (x_6 - g_1)^2 - k_8 \frac{(x_6 - g_1)(g_1 - x_7)}{k_0x_1}\right)\biggr] \label{eq:xdot6} 
        \end{align}
        \begin{align}
        \Dot{x}_7 &=\, \frac{\beta}{x_1} \biggl[-\biggl(\quad k_{12}(x_6 - g_1)(g_1 - x_7) \nonumber \\
        & \, \; - k_{13} \frac{(g_1 - x_7)^2}{k_0x_1}  \biggr) \nonumber\\
        &\, + \frac{k_{9} (g_1 - x_7)}{k_{15}k_0 x_1} - \frac{k_9(x_7 - x_8)}{k_{14} + k_{15}k_0  x_1} \biggr] \label{eq:xdot7}
        \end{align}
        \begin{align}
        \Dot{x}_8 &=\, k_{17} k_9 \left(\frac{x_7 - x_8}{k_{14} + k_{15} k_0 \cdot x_1} - \frac{x_8 - k_{16}}{k_{14} + k_{18}}\right) \label{eq:xdot8}
        \end{align}
Eq.~(\ref{eq:xdot1}) - (\ref{eq:xdot8}) are derived in ~\ref{appendix:a}, where Eq.~(\ref{eq:xdot1}) - (\ref{eq:xdot8}) corresponds to Eq.~\ref{eq:m_dot_sl}, Eq.~\ref{eq:m_dot_al2o3}, Eq.~\ref{eq:m_dot_alf3}, Eq.~\ref{eq:m_dot_cry}, Eq.~\ref{eq:m_dot_al}, Eq.~\ref{eq:T_dot_bath_sim}, Eq.~\ref{eq:T_dot_sl}, and Eq.~\ref{eq:T_dot_wall} respectively.
\section{Method and experimental setup}
\label{sec:Method_experiment}
    \subsection{Training with sparsity promoting regularization}
    \label{subsec:sparsity}
\begin{figure}[H]
            \centering
            \includegraphics[width=0.95\linewidth]{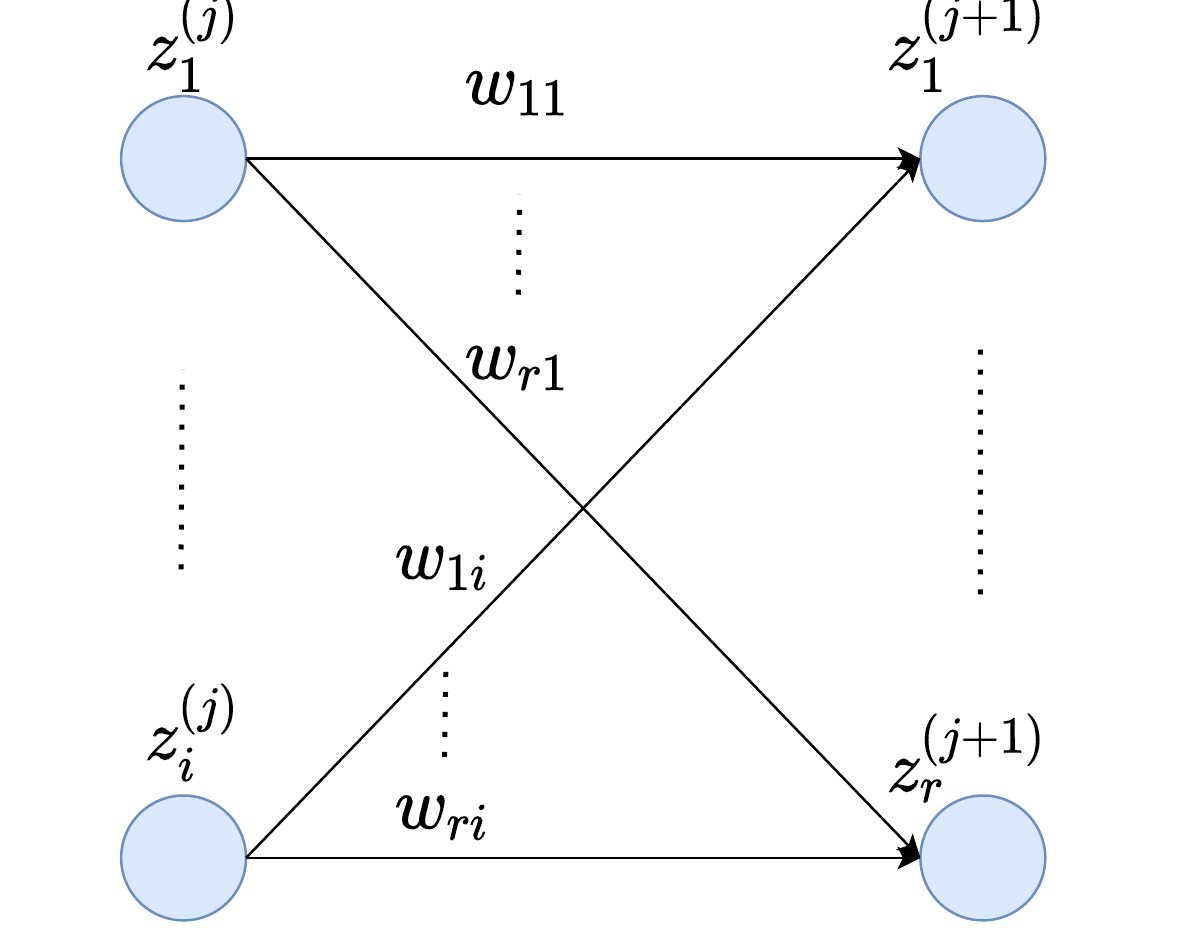}
            \caption{Enumerated weights according to their input and output nodes between layer $j$ and $(j+1)$}
            \label{fig:enum_weights_neurons}
        \end{figure}

        In this article, sparse DNN models are utilized to predict state variables in the aluminum electrolysis simulator. All the weight matrices in a DNN model are $\ell_1$ regularized to impose sparsity. Fig.~\ref{fig:enum_weights_neurons} illustrates how weights are enumerated according to their input and output nodes. Layer $j$ has $i$ nodes and layer $(j+1)$ has $r$ nodes. Layer $j=0$ corresponds to the input layer, and consist of measured or estimated states $\mathbf{x}(t)$ and control inputs $\mathbf{u}(t)$ at time step $t$. The output layer consists of the estimated time derivatives of the states $\mathbf{\Dot{x}}(t)$ at time step $t$. 
        The weight matrix $\mathbf{W}_{j+1}$ corresponds to the weights that connect layer $j$ to $j+1$. $\mathbf{W}_{j+1}$ is arranged as follows:
        \begin{equation}
        \mathbf{W}_{j+1} =
        \begin{bmatrix}
         w_{11} & w_{12} &\; ...\; & w_{1i}\\
         w_{21} & w_{22} &\; ...\; & w_{2i}\\
         \vdots & \ddots & \\
         w_{r1} & w_{r2} & \hdots & w_{ri}
        \end{bmatrix} .
        \label{eq:weight_matrix}
        \end{equation}
        Regularization terms $R_{\ell_1, j+1}$ are defined for each weight matrix $\mathbf{W}_{j+1}$ :
        \begin{equation}
            R_{\ell_1, j+1} = \sum_{i,k \in W_{j+1}} |w_{i,k}|.
        \end{equation}
        where $w_{i,k}$  are the model weights from layer $j$ to $j+1$, or equivalently, the elements of $\mathbf{W}_{j+1}$. The regularization terms for each layer are added to the cost function:
        \begin{equation}
                \begin{aligned}
            \mathbf{w}^* &=\ \underset{\mathbf{w}}{\mathrm{argmin}}\biggl\{\frac{1}{N}\sum_{i=1}^N (\mathbf{y_i}-\mathbf{f(x_i}))^2 + \lambda_1 R_{\ell_1, 1} ...\\
            &\ \quad + \lambda_{j+1}R_{\ell_1, j+1} + ... + \lambda_L R_{\ell_1, L}  \biggr\},
            \end{aligned}
            \label{eq:cost_fn_sparse_nn}
        \end{equation}
        where $\frac{1}{N}\sum_{i=1}^N (\mathbf{y_i}-\mathbf{f(x_i}))^2$ is the MSE, and $\lambda_{j+1}, j=0,\; ..., 3$ is a layer-specific hyperparameter that determines how the weights in $\mathbf{W}_{j+1}$ are penalized.
    \subsection{Experimental setup and data generation}
    \label{sec:setup}
        \begin{table}[h]
        \begin{center}
        \caption{Constants in the simulator}
        \begin{tabular}{l|l|l}
        \hline
        Constant &Physical meaning &Nummeric value  \\ \hline
        $k_0$ & $1/(\rho_{sl}A_{sl})$ &$2\cdot 10^{-5}$ \\
        $k_1$ &$2k_{sl}A_{sl}/\Delta_{fus} H_{cry}$ &$7.5\cdot 10^{-4}$ \\
        $k_2$ &$h_{bath-sl}A_{sl}/\Delta_{fus} H_{cry}$ &$0.18$  \\
        $k_3$& $0.002\frac{M_{Al_2O_3}\cdot CE}{z\cdot F}$ &$1.7\cdot 10^{-7}$ \\
        $k_4$ & $C_{Na_2O}\frac{4M_{AlF_3}}{3M_{Na_2O}}$ &$0.036$  \\
        $k_5$ & $C_{Na_2O}\frac{2M_{cry}}{3M_{Na_2O}}$ &$0.03$  \\
        $k_6$ & $0.002\frac{M_{Al}\cdot CE}{z\cdot F}$&$4.43\cdot 10^{-8}$ \\
        $k_7$ & $k_2 \cdot c_{p_{cry, \; liq}}$ &$338$ \\
        $k_8$ & $k_1 \cdot c_{p_{cry, \; liq}}$ & $1.41$ \\
        $k_9$ & $A_{sl}$ &$17.92$ \\
        $k_{10}$ & $1/h_{bath-sl}$ &$0.00083$ \\
        $k_{11}$ & $1/(2k_{sl})$ &$0.2$ \\
        $k_{12}$ & $k_2\cdot c_{p_{cry,\; s}}$ &$237.5$ \\
        $k_{13}$ & $k_1\cdot c_{p_{cry,\; s}}$&$0.99$ \\
        $k_{14}$ & $x_{wall}/(2k_{wall})$ &$0.0077$ \\
        $k_{15}$ & $1/(2k_{sl})$ &$0.2$ \\
        $k_{16}$ & $T_{0}$ &$35$ \\
        $k_{17}$ & $1/(m_{wall}c_{p, \; wall})$ &$5.8 \cdot 10^{-7}$ \\
        $k_{18}$ & $1/h_{wall-0}$ &$0.04$ \\
        $\alpha$ & $1/c_{p_{bath, \; liq}}$ &$5.66\cdot 10^{-4}$\\
        $\beta$ & $1/c_{p_{cry, \; sol}}$ &$7.58\cdot 10^{-4}$\\
        \hline
        \end{tabular}
        \label{table:const_simulator}
        \end{center}
        \end{table}
        Data for the aluminium electrolysis process is generated by integrating the non-linear ODEs given by Eqs.~(\ref{eq:xdot1}) - (\ref{eq:xdot8}) with a set of chosen initial values for the state variables $\mathbf{x}(t_0)$, and fourth-order Runge-Kutta (RK4) algorithm. The initial conditions of each variable $x_i$ for each simulation are randomly chosen from a given interval of possible initial conditions given in Table~\ref{table:init_conditions}. For $x_2$ and $x_3$, concentrations $c_{x_2}$ and $c_{x_3}$ are given.
        \begin{table}[h]
        \begin{center}\caption{Initial conditions for system variables}
        \begin{tabular}[width=\textwidth]{l|l}
        \hline
        Variable &  Initial condition interval \\ \hline
        $x_1$ & $[3260,\;3260]$\\
        $c_{x_2}$ & $[0.02,\; 0.03]$ \\
        $c_{x_3}$ & $[0.10,\; 0.12]$  \\
        $x_4$& $[13500,\; 14000]$  \\
        $x_5$ & $[9950,\; 10000]$ \\
        $x_6$ & $[975,\; 975]$ \\
        $x_7$ & $[816,\; 816]$ \\
        $x_8$ & $[580,\; 580]$ \\
        \hline
        \end{tabular}
        \label{table:init_conditions}
        \end{center}
        \end{table}
        Data-driven models depend on a high degree of variation in the training data to be reliable and valid in a large area of the input space. The input signal determines how the system is excited and thus what data is available for modeling and parameter estimation. Operational data from a controlled, stable process is generally characterized by a low degree of variation. Even large amounts of data sampled over a long period from a controlled process can not guarantee that the variation in the training data i large enough to ensure that the trained model generalizes to unseen data. Therefore, random excitation are added to the input signals to increase the variation in the sampled data. The intuition is that the random excitation will push the dynamics out of the standard operating condition so that variation in the training data increases. In general, each control input $i$ is given by:
        \begin{equation}
            u_i = \textrm{Deterministic term + Random term}.
        \end{equation}
        The control inputs $u_1, \; u_3$ and $u_4$ are impulses. The random term is zero for these control inputs when the deterministic term is zero. The deterministic term is a proportional controller. The control inputs $u_2$ and $u_5$ are always nonzero. These control inputs have constant deterministic terms and a random term that changes periodically. The random term stays constant for a certain period $\Delta T_{rand}$ before changing to a new randomly determined constant. Choosing the period $\Delta T_{rand}$ is a matter of balancing different objectives. On one hand, it is desirable to choose a large period $\Delta T_{rand}$ so that the system can stabilize and evolve under the given conditions to reveal the system dynamics under the given conditions. On the other hand, it is desirable to test the systems under many different operational conditions. By empirically testing different periods $\Delta T_{rand}$, and seeing how the dynamics evolve in simulation, it turns out that setting $\Delta T_{rand} =30\Delta T$ is a fair compromise between the two.
        \begin{table*}[h]
        \begin{center}
        \caption{Control functions}
        \label{table:control_inputs}
        \begin{tabular}[width=0.7\textwidth]{l|l|l|l}
        \hline
        Input & Deterministic term & Random term interval &  $\Delta T_{rand}$   \\ \hline
        $u_1$ & $3e4(0.023 - c_{x_2})$ &$[-2.0,2.0]$ &$\Delta T$ \\
        $u_2$ & $14e3$ &$[-7e3,7e3]$ & $30\cdot\Delta T$\\
        $u_3$ & $13e3(0.105-c_{x_3})$ &$[-0.5, \;0.5]$ & $\Delta T$  \\
        $u_4$& $2(x_5 - 10e3)$ & $[-2.0, \;2.0]$&$\Delta T$  \\
        $u_5$ & $0.05$ &$[-0.015, \;0.015]$& $30\cdot\Delta T$  \\
        \end{tabular}
        \end{center}
        \end{table*}
        Table~\ref{table:control_inputs} gives the numerical values of the deterministic term of the control input, the interval of values for the random terms, and the duration $\Delta T_{rand}$ of how long the random term is constant before either becoming zero ($u_1, u_3, u_4$) or changing to a new randomly chosen value ($u_2, u_5$). One simulation $i$ with a given set of initial conditions is simulated for $1000$ time steps, and each time step $\Delta T = 30s$. The simulation generate the data matrix as in Eq. (\ref{eq:X_mat_single}):
        
        \begin{equation}
        \mathbf{X} =
        \resizebox{.37\textwidth}{!}{
        $
        \begin{bmatrix}
         x_1(0) & x_2(0) &\; ...\; &x_8(0) & u_1(0)& \;...\;&u_5(0) \\
         \\
         x_1(1) & x_2(1) &\; ...\; &x_8(1) & u_1(1)& \;...\;&u_5(1)\\
         \vdots & \vdots & \ddots & \vdots & \vdots & \ddots &\vdots \\
         x_1(k)&\hdots  & x_i(j) & \hdots& u_1(j) &\hdots  &u_5(j)  \\
         \vdots & \vdots & \ddots & \vdots & \vdots & \ddots &\vdots \\
         x_1(999) & x_2(999) &\; ...\; &x_8(999) & u_1(999)& \;...\;&u_5(999)\\
        \end{bmatrix} .$
        }
        \label{eq:X_mat_single}
        \end{equation}
        
        The number $j$ within the parenthesis of variable $i$ indicates the time step for when $x_i(j)$ is sampled. The target values are then calculated as
        \begin{equation}
        \resizebox{.41\textwidth}{!}{
        $
        \mathbf{Y} =
        \begin{bmatrix}
         \frac{x_1(1) - x_1(0)}{\Delta T} &\; ...\; &\frac{x_8(1) - x_8(0)}{\Delta T} \\
                    \\
         \frac{x_1(2) - x_1(1)}{\Delta T} &\; ...\; &\frac{x_8(2) - x_8(1)}{\Delta T} \\
         \vdots &  \ddots&\vdots \\
         \frac{x_1(k) - x_1(j-1)}{\Delta T} &\; ...\; &\frac{x_8(j) - x_8(k-1)}{\Delta T} \\
         \vdots &  \ddots&\vdots \\
          \frac{x_1(1000) - x_1(999)}{\Delta T} &\; ...\; &\frac{x_8(1000) - x_8(999)}{\Delta T} \\
        \end{bmatrix} .$
        }
        \label{eq:Y_mat_single}
        \end{equation}
        
        Each training set $\mathcal{S}_k$ from simulation $k$ are put in input and outputs are put in pairs:
        
        \begin{equation}
            \mathcal{S}_k = [\mathbf{X}, \mathbf{Y}] = 
            \begin{bmatrix}
              [\mathbf{x}^T(0), \mathbf{u}^T(0)]^T, & \mathbf{y}(0) \\
             \vdots & \vdots \\
             [\mathbf{x}^T(j), \mathbf{u}^T(j)]^T, & \mathbf{y}(j) \\
             \vdots & \vdots \\
             [\mathbf{x}^T(999), \mathbf{u}^T(999)]^T, & \mathbf{y}(999) \\
            \end{bmatrix}.
            \label{eq:train_set_sim}
        \end{equation}
        The training sets from each simulation are normalized before they are stacked
        \begin{equation}
            \mathcal{S}_{stack} = \begin{bmatrix} \mathcal{S}_1^T,\; \mathcal{S}_2^T,\; ...,\; \mathcal{S}_k^T,\; ....,\; \mathcal{S}_n^T\end{bmatrix}^T.
            \label{eq:stack_train_set}
        \end{equation}
        Here, $n$ is the number of simulated time-series $\mathbf{X}$. The number of time series simulations $n$ varies in the experiments to evaluate the model performance as a function of training size. Then, all input-output pairs in the stacked training set are shuffled:
        \begin{equation}
            \mathcal{S}_{train} = \text{shuffle}(\mathcal{S}_{stack}).
        \end{equation}
         The shuffled training set is put in mini-batches $[\mathbf{X}_{batch, \;i}, \;\mathbf{Y}_{batch, \;i}]$, and the models are trained on these mini-batches. The test set also consists of several time series simulations generated in the same way described above. The test set is given by:
        \begin{equation}
            \mathcal{S}_{test} = \{ \{X_1\},\; \{X_2\},\; ...,\;\{X_p\}\}.
            \label{eq:test_set}
        \end{equation}
        $\mathcal{S}_{test}(i) = \{X_i\}=\{([\mathbf{x}_k,\; \mathbf{u}_k]\}_{k=1}^{1000}$ is one simulated time series, and $\{\mathbf{x}_k\}_{k=1}^{1000}$ is being forcasted by the models. In all experiments, 20 models of each dense and sparse networks with different initialization are trained on the training set and then evaluated on the test set. 
        
    \subsection{Performance metrics}
    \label{subsec:performancemetrics}
        In aluminum electrolysis, data is generally sampled at rare instants during the operation. Thus, a model needs to accurately forecast several time steps without feedback from measurements to ensure safe and optimal operation. Therefore, the models' capability to estimate the states $\mathbf{x}$ over a given time horizon without measurement feedback becomes an important measure of performance. The initial conditions $\mathbf{x}(t_0)$ are given to the models. Then the consecutive $n$ time steps of the states are estimated $\{\mathbf{\hat{x}}(t_1),\;...,\;\mathbf{\hat{x}}(t_n)\}$. This is called a \textbf{rolling forecast}. The model estimates the time derivatives of the states $d\mathbf{\hat{x}_i}/dt$ based on the current control inputs $\mathbf{u}(t_i)$ and initial conditions $\mathbf{x}_0(t)$ if $t=t_0$, or the estimate of the current state variables $\mathbf{\hat{x}}(t_{i})$ if $t>t_0$:
         \begin{equation}
            \frac{d\mathbf{\hat{x}}(t_i)}{dt} = 
            \begin{cases}
            \hat{f}(\mathbf{\hat{x}}(t_i), \; \mathbf{u}(t_i)),& \text{if } t_i> t_0\\
            \hat{f}(\mathbf{x}_0(t_i), \; \mathbf{u}(t_i)),& \text{if } t_i = t_0
            \end{cases}
         \end{equation}
         Then, the next state estimate $\mathbf{x}(t_{i+1})$ is calculated as
         \begin{equation}
             \mathbf{\hat{x}}(t_{i+1}) = \mathbf{\hat{x}}(t_i) + \frac{d\mathbf{\hat{x}}(t_i)}{dt}\cdot \Delta T.
         \end{equation}
        
        The rolling forecast can be computed for each of the states $x_i$ for one set of test trajectories $\mathcal{S}_{test}$. However, presenting the rolling forecast of multiple test sets would render the interpretation difficult. By introducing a measure called Average Normalized Rolling Forecast Mean Squared Error (AN-RFMSE) that compresses the information about model performance, the models can easily be evaluated on a large number of test sets. The AN-RFMSE is a scalar defined as:
         \begin{equation}
             \textrm{AN-RFMSE} = \frac{1}{p}\sum_{i=1}^p\frac{1}{n}\sum_{j=1}^n\left(\frac{\hat{x}_i(t_j) - x_i(t_j)}{std(x_i)}\right)^2,
             \label{eq:AN-RFMSE}
         \end{equation}
        where $\hat{x}_i(t_j)$ is the model estimate of the simulated state variable $x_i$ at time step $t_j$, $std(x_i)$ is the standard deviation of variable $x_i$ in the training set $\mathcal{S}_{train}$, $p=8$ is the number of state variables and $n$ is the number of time steps the normalized rolling forecast MSE is averaged over. Hence, for every model $\hat{f}_j$ and every test set time series $\mathcal{S}_{test}(i)$, there is a corresponding AN-RFMSE. This generates a matrix, where each row represents individual model instances, and every column represents one test set simulation $\mathcal{S}_{test}(i)$. Each entry in the matrix is the AN-RFMSE for a given model instance and a given $\mathcal{S}_{test}(i)$. The matrix is given by:
        \begin{equation}
        \resizebox{.40\textwidth}{!}{
        $
            \textrm{AN-RFMSE}_{mat} = 
            \begin{bmatrix}
             \textrm{AN-RFMSE}_{11} &  \hdots & \textrm{AN-RFMSE}_{1n}\\
             \vdots & \vdots
             & \vdots  \\
             \textrm{AN-RFMSE}_{k1} &  \hdots& \textrm{AN-RFMSE}_{kn}
            \end{bmatrix}
        $}.
        \end{equation}
        $k$ is the number of models, and $n$ is the number of time series simulations in the test set $\mathcal{S}_{test}$. There are two $\textrm{AN-RFMSE}_{mat}$ matrices, one for sparse models and one for dense models. Averaging over all columns at each row, that is, averaging over all test set time series for each model instance, generates a  vector  
        \begin{equation}
        \resizebox{.40\textwidth}{!}{
        $\overline{\textrm{AN-RFMSE}}_{vec} = 
            \begin{bmatrix}
             \frac{1}{n}\sum_{i_i}^n\textrm{AN-RFMSE}_{1i}\\
             \vdots \\
             \frac{1}{n}\sum_{i_i}^n\textrm{AN-RFMSE}_{ki}
            \end{bmatrix}
            = 
            \begin{bmatrix}
             \overline{\textrm{AN-RFMSE}_{1}}\\
             \vdots \\
             \overline{\textrm{AN-RFMSE}_{k}}
            \end{bmatrix}.
        $}
        \end{equation}
        The elements of $\overline{\textrm{AN-RFMSE}}_{vec}$ is the average $\textrm{AN-RFMSE}$ over all test set time series for every model instance.

 \begin{figure}[H]
     \centering
     \includegraphics[width=\linewidth]{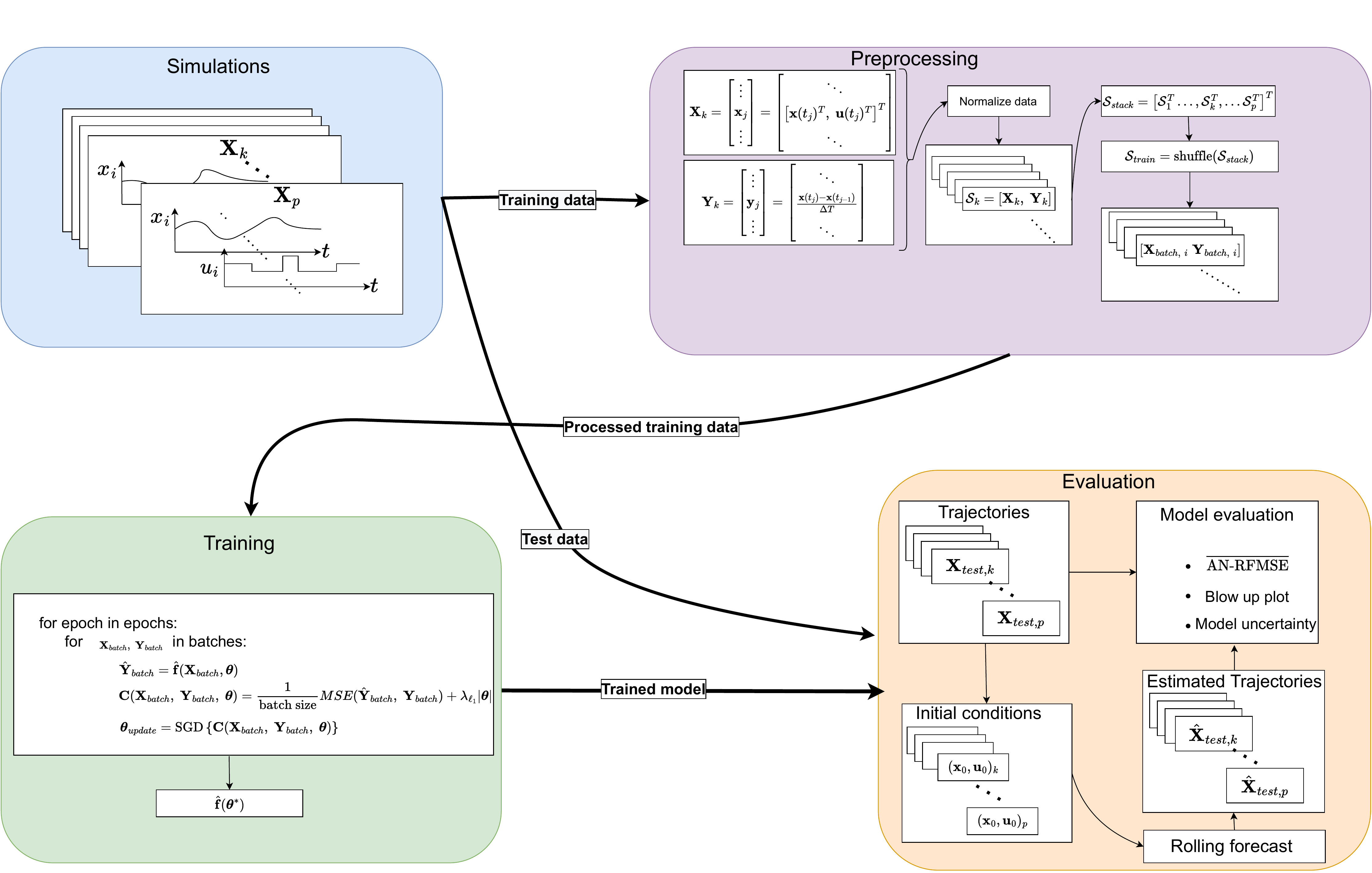}
     \caption{Schematic presentation of the experimental setup. This includes the data simulation, preprocessing of training data, model training and model evaluation. At the first step (blue box), training and test data is simulated. Each frame in the blue box corresponds to a simulation of the dynamics from some random initial conditions. The training and test data are separated, where the test data is given directly to the evaluation part of the case study, while the training data is sent to preprocessing. In the preprocessing stage, input features are arranged from a given simulation is put in a input matric, output features are calculated (see Eq.~(\ref{eq:Y_mat_single})) before they are put in a corresponding output matrix. Then, both input and output features are normalized and put in pairs. After all simulations are arranged in input output pairs, all pairs are shuffled before put in mini-batches for training. In the training procedure, the model parameters are optimized on the mini-batches. The trained models are sent to evaluation. In this stage, models are given the initial conditions from the test set. In addition, the models are given the control inputs in the test set at every time step. The model the forecasts the test set trajectories in a rolling forecast. The estimated trajectories are compared to the test set trajectories and evaluated according to accuracy measures, uncertainty in terms of dissagreement between models with different initial parameters trained on the same data with same hyperparameters, and according to the number of blow ups of a given model type (models trained with same hyperparameters), meaning the number of times that model type estimat diverges from the test set they are estimating.}
     \label{fig:schematic}
 \end{figure}
 
 Fig.~\ref{fig:schematic} summarizes the workflow in the case study. Each step is briefly explained in the figure text, and more thoroughly throughout Sec.~\ref{sec:Method_experiment}.
 
\section{Results and discussion}
\label{sec:resultsanddiscussion}
    In this section, we present and discuss the main findings of the work. In doing so, we will analyze the results from the perspective of interpretability, generalizability and training stability.
    
    \subsection{Interpretability perspective}
\begin{figure}[H]
        \centering
        \includegraphics[width=0.695\textwidth]{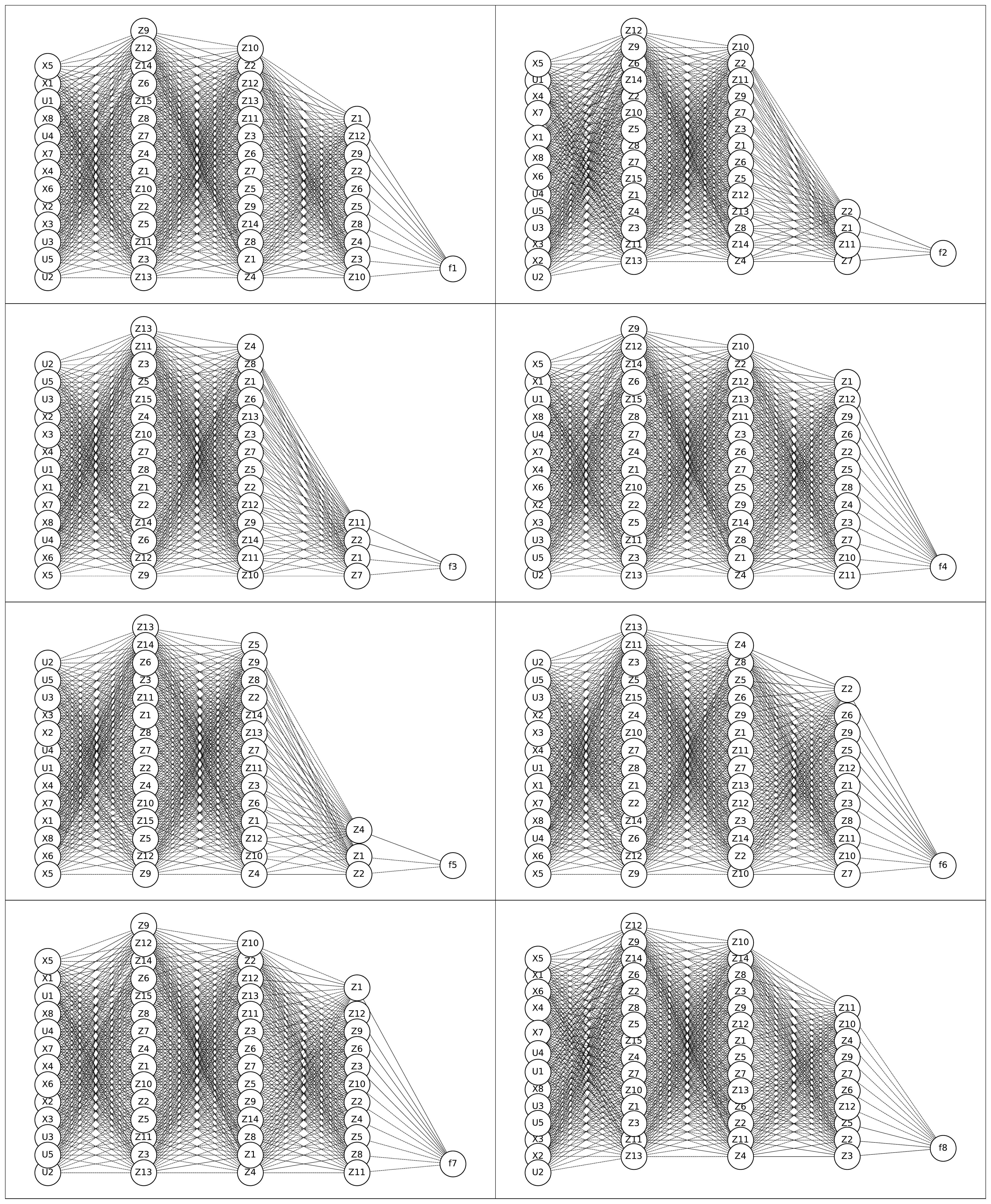}
        \caption{Model structure of each output function $\{f_1, .., f_8\}$ for one of the trained dense neural network models. Blue circles represent neurons, inputs and outputs in the model. $X_i$ represent system variable $i$ in the input layer, $U_i$ represents control input $i$ in the input layer, and $Z_i$ represents latent variable $i$ in the layer it is visualized. The directed edges indicate weights in the model.}
        \label{fig:dense_nn_structure}
    \end{figure}

    As discussed earlier, the interpretability of a model is the key to its acceptability in high-stake applications like the aluminum extraction process considered here. Unfortunately, highly complex dense neural networks having thousands to millions of parameters which were the starting point for the modeling here are almost impossible to interpret. Fig.~\ref{fig:dense_nn_structure} shows the model structure of a dense DNN model learned for the generated data. The figure illustrates how densely trained neural networks yield unintepretable model structures.
    
\begin{figure}[H]
            \centering
            \includegraphics[width=0.8\linewidth]{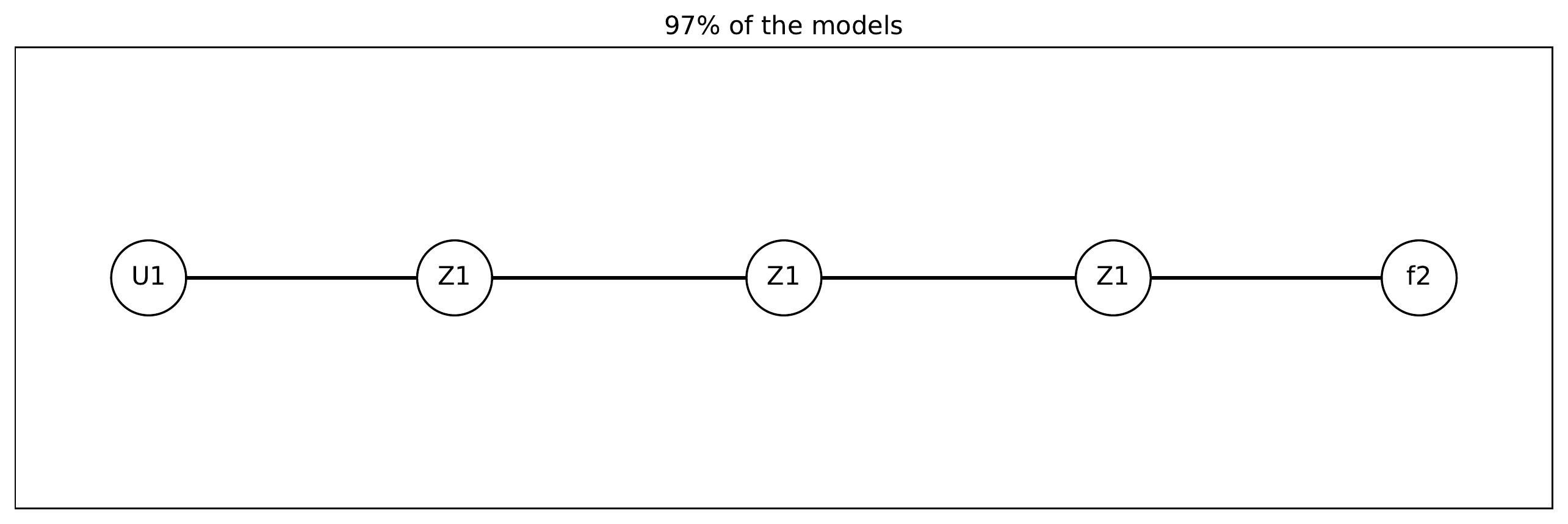}
            \caption{Most common learned structure for $\hat{f}_2(\mathbf{x}, \mathbf{u})$. $97 \%$ of $f_2$ ends up with this structure.}
            \label{fig:sparse_f2_most_learned_struct}
        \end{figure}

    Fortunately, through the regularization it was possible to significantly reduce the model complexity resulting in a drastically reduced number of trainable parameters (see the Figs.~\ref{fig:sparse_f1_most_learned_struct}-\ref{fig:sparse_f8_most_learned_struct}). It can be argued that the reduced model complexity of sparse neural networks increases the model interpretability. With domain knowledge about the aluminum electrolysis process, the sparse models can be evaluated as we will do in the remainder of this section. 
    
    The results related to the interpretability aspect is presented in the form of model structure plots which can be used to explain the input-output mapping of the models. If the model structures are very sensitive to the initialization, then there interpretation will not make sense therefore, 100 DNNs with different initialization are trained independently and their common trends are emphasized in the discussions. We now present each of the model outputs $\{\hat{f}_1(\mathbf{x}, \mathbf{u}),\; ...,\;\hat{f}_8(\mathbf{x}, \mathbf{u})\}$. It is worth mentioning that these outputs are estimates of each of the time derivatives of the states $\{\Dot{x}_1, \; ..., \; \Dot{x}_8\}$ respectively. 
    
    \begin{table*}[h]
    \centering
    \caption{Frequency of learned features for each output $\{\hat{f}_1, \; ...,\; \hat{f}_8\}$. Each column $i$ correspond to an output $\hat{f}_i$ of the neural network. Each row element $j$ correspond to one of the features $\{x_1,\; x_2, \;...,\; x_8,\; u_1,\; u_2,\;...,\; u_5\}$. The value of the table element $(i,j)$ is the percent of how many out of one hundred models of the output $f_i$ that feature $j$ occurs. }
    \label{table:freq_features}
        \begin{tabular}{|c|c|c|c|c|c|c|c|c|}
        \hline
        \multirow{2}{*}{Feature} & \multicolumn{8}{c|}{Output functions}\\
        \cline{2-9}
        &$\hat{f}_1$ &$\hat{f}_2$&$\hat{f}_3$&$\hat{f}_4$&$\hat{f}_5$&$\hat{f}_6$&$\hat{f}_7$&$\hat{f}_8$  \\ 
        \cline{1-9}
        $x_1$ & $86$ &$1$ &$1$&$86$&$2$&$100$&$99$&$100$ \\\hline
        $x_2$ &$100$ &$2$& $2$ & $100$ & $1$ & $100$ & $100$ & $100$\\\hline
        $x_3$ &$100$ & $2$ & $2$ & $100$ & $0$ & $93$ & $100$ & $87$\\\hline
        $x_4$ & $100$& $0$ & $0$ & $100$ & $0$ & $87$ & $100$ & $87$\\\hline
        $x_5$ &$2$& $0$ & $0$ & $2$ & $0$ & $2$ & $2$ & $2$\\\hline
        $x_6$ & $100$ & $2$ & $2$ & $100$ & $1$ & $100$ & $100$ & $87$ \\\hline
        $x_7$ &$22$ & $1$ & $1$ & $21$ & $1$ & $7$ & $89$ & $89$\\\hline
        $x_8$ & $100$& $1$ & $1$ & $100$ & $2$ & $87$ & $100$ & $100$ \\\hline
        $u_1$ &$100$& $100$ & $100$ & $100$ & $0$ & $100$ & $100$ & $87$\\\hline
        $u_2$ & $4$ & $0$ & $0$ & $4$ & $1$ & $100$ & $18$ & $18$\\\hline
        $u_3$ & $2$ & $0$ & $100$ & $2$ & $0$ & $2$ & $4$ & $4$\\\hline
        $u_4$ & $2$ & $0$ & $0$ & $2$ & $100$ & $3$ & $3$ & $3$\\\hline
        $u_5$ & $3$ & $0$ & $0$ &$3$ & $1$ & $100$ & $18$ & $18$ \\\hline
        \end{tabular}
    \end{table*}
    
    \subsubsection{Model output $\hat{f}_1$}
        The simulation model for the first output $f_1$ defined in Eq.~(\ref{eq:xdot1}) is a function of the features $\{x_1, x_2, x_3, x_4, x_6, x_7\}$. $f_1$ can further be divided into three sums $f_1 = h_1(x_1, x_2, x_3, x_4, x_7) + h_2(x_6) + h_3(x_2, x_3, x_4)$, where $h_1 = k_1\frac{g_1(x_2,x_3, x_4) - x_7}{k_0 x_1}$, $h_2 = -k_2x_6$ and $h_3 = g_1(x_2, x_3, x_4)$. $g_1$ is  a nonlinear function defined in Eq.~(\ref{eq:g_1}).
        \begin{figure}[H]
            \centering
            \includegraphics[width=0.8\linewidth]{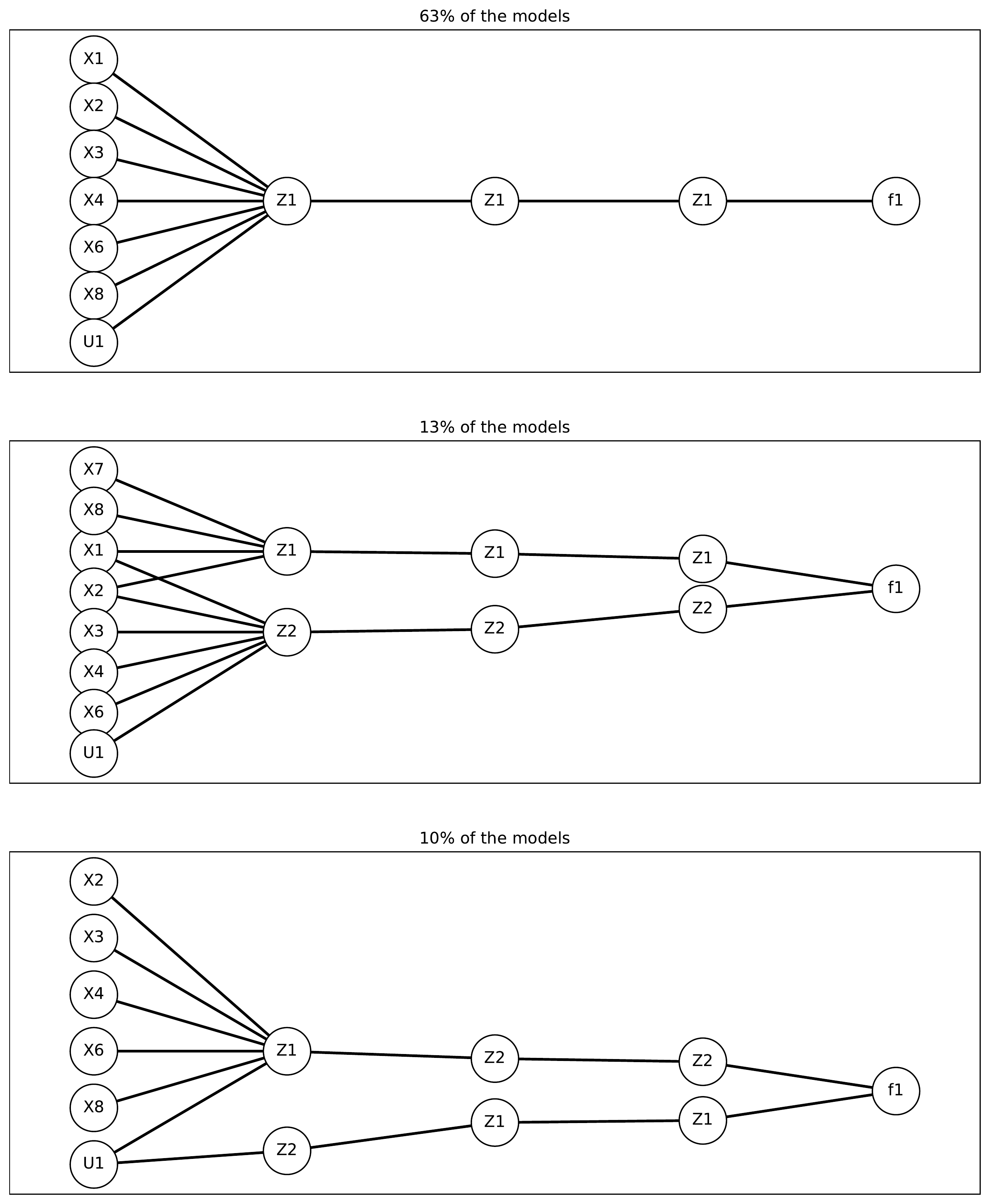}
            \caption{Most common learned structures for $\hat{f}_1(\mathbf{x}, \mathbf{u})$.}
            \label{fig:sparse_f1_most_learned_struct}
        \end{figure}
        Fig.~\ref{fig:sparse_f1_most_learned_struct} shows the three most common learned structures of the first output of the neural network $\hat{f}_1(\mathbf{x}, \mathbf{u})$. In total, these structures account for $86\%$ of all learned structures of $\hat{f}_1$. The top structure forms the resulting structure for $63\%$ of the models. It is a function of seven inputs $\hat{f}_1 = f(x_1, x_2, x_3, x_4, x_6,  x_8, u_1)$. All input features are connected to the same neuron in the first layer. Moreover, it is only one neuron in each hidden layer.  The upper bound in Eq.~(\ref{eq:upper_lower_bound_regions}) states that this model structure only has $n^{dL} = 1^{7\cdot3}= 1$ region. This is equivalent to stating that the model collapses to one linear model. The middle and the bottom structures of Fig.~\ref{fig:sparse_f1_most_learned_struct} has more than one neuron in the hidden layers. However, the structures can be divided into two disconnected subnetworks since the hidden neurons are not connected before they are added in the output layer. Hence, also these models collapse to linear models with a single region. This means that the neural networks do not capture the nonlinear dynamics in the simulation model. All structures of Fig.~\ref{fig:sparse_f1_most_learned_struct} are erroneously including  $x_8$ and $u_1$ in their feature basis. Besides, $x_7$ which is present in the simulation model $f_1$ is not found by the top and bottom structures in Fig.~\ref{fig:sparse_f1_most_learned_struct}. In fact, $x_7$ is found only in $22\%$ of the models $\hat{f}_1$ according to Table~\ref{table:freq_features}. The exact cause of this erroneous feature selection is not trivial. However, $x_8$ which is the wall temperature correlates highly with the side ledge temperature $x_7$. Thus, $x_8$ can possible have been learned as a feature of $f_1$ instead of $x_7$. Moreover, the alumina feed $u_1$ on the other hand affects the time derivative of the mass of alumina $\Dot{x}_2$, the time derivative of mass of aluminum fluoride $\Dot{x}_3$ and the time derivative of cryolite $\Dot{x}_4$ directly. All these variables affect $\Dot{x}_1$ through the liquidus temperature $g_1(x_2, x_3, x_4)$. To understand how this might cause the learning algorithm to find $u_1$ as a feature of $\hat{f}_1$, consider the following: let $u_1$ be zero until time $t$. Then, $\{x_2, x_3, x_4\}$ will be updated due to $u_1$ at the next sampled time step $t+1$. However, the fourth-order Runge Kutta solver splits the sampling interval into 4 smaller intervals $\{t + 0.25,\; t+0.5, \; t+0.75, \; t+1\}$ and solve the ODE eqautions at all these time steps. Thus, the state variables $\{x_2, x_3, x_4\}$ are updated already at time $t+0.25$. Since $\Dot{x}_1$ is depending on these variables, $x_1$ will be updated at $t+0.5$. Therefore, at time $t+1$, when data is sampled, $x_1$ would also be changed. Hence, the learning algorithm finds $u_1$ to affect the time derivative $\Dot{x}_1$. This could might have been solved by shortening the sampling interval. Furthermore, $x_1$ not included as a feature in $14\%$ of the models. This might be a combination of parameter initialization and that $x_1$ is multiplied by the small constant $k_0 = 2\cdot 10^{-5}$. 
        
    \subsubsection{Model output $\hat{f}_2$}
        Fig.~\ref{fig:sparse_f2_most_learned_struct} shows the most common learned structure among the models $\hat{f}_2$ that models the time derivative of alumina $f_2 = \Dot{x}_2$. $97\%$ of the structures end up as the structure in Fig.~\ref{fig:sparse_f2_most_learned_struct}. The simulation model $f_2$  in Eq.~(\ref{eq:xdot2}) is a linear model dependent on $\{u_1, u_2\}$. The learned models $\hat{f}_2$ only finds $u_1$ as the relevant feature. The reason for this might be that $u_2$ is proportional to a very small constant $k_3=1.7\cdot 10^{-7}$. Variations in the line current $u_2$ are not big enough to be significant for the learning algorithm. The dynamics caused by $u_2$ are captured in a bias in the models.  
        
        \begin{figure}[H]
            \centering
            \includegraphics[width=0.8\linewidth]{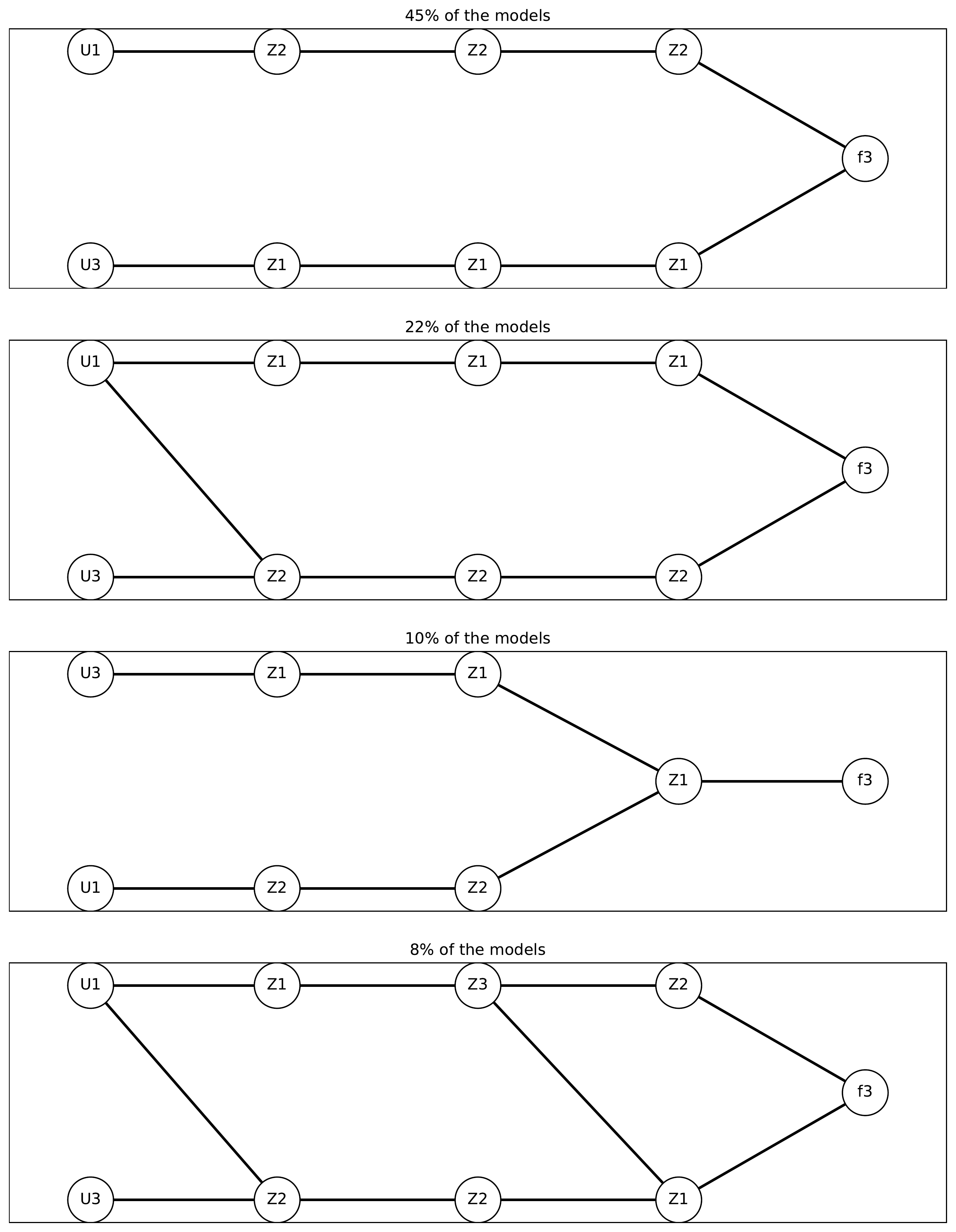}
            \caption{Most common learned structures for $\hat{f}_3(\mathbf{x}, \mathbf{u})$.}
            \label{fig:sparse_f3_most_learned_struct}
        \end{figure}
        
    \subsubsection{Model output $\hat{f}_3$}
        Fig.~\ref{fig:sparse_f3_most_learned_struct} shows the four most common learned structures of $\hat{f}_3$. $\hat{f}_3$ models the time derivative of the aluminum fluoride mass $\Dot{x}_3$ in the cell. The simulation model $\Dot{x}_3 = f_3$ in Eq.~(\ref{eq:xdot3}) is a linear model depending on the features $\{u_1, u_3\}$, and in all structures in Fig.~\ref{fig:sparse_f3_most_learned_struct}, only these two features are found. As shown in Table~\ref{table:freq_features}, these features are found in $100\%$ of the trained models. The structures found are mainly linear models. However, in the second and fourth structures, some weights connect the features in intermediate layers.
        \begin{figure}[H]
            \centering
            \includegraphics[width=0.8\linewidth]{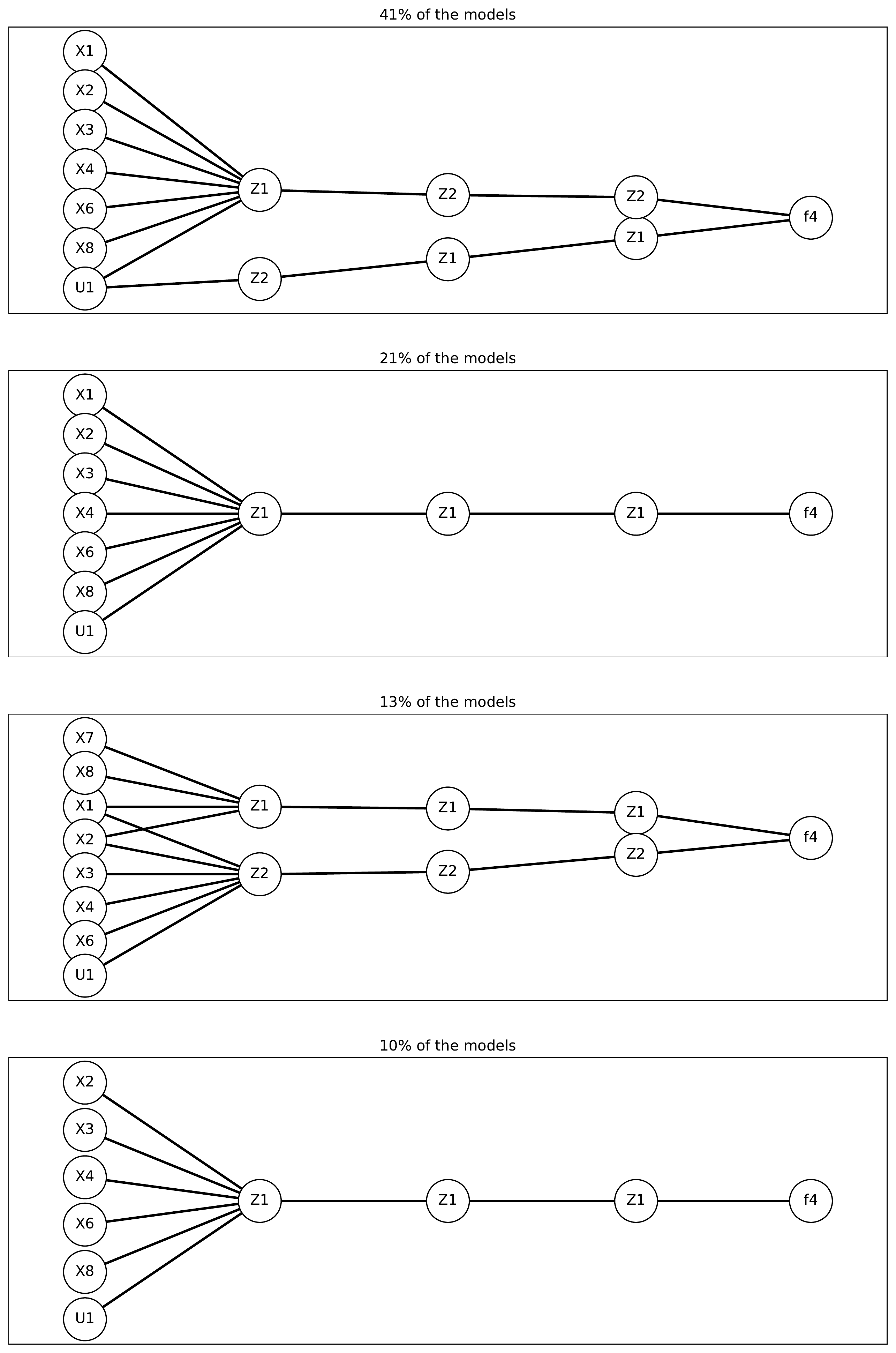}
            \caption{Most common learned structures for $\hat{f}_4(\mathbf{x}, \mathbf{u})$.}
            \label{fig:sparse_f4_most_learned_struct}
        \end{figure}
        
    \subsubsection{Model output $\hat{f}_4$}
        Fig.~\ref{fig:sparse_f4_most_learned_struct} shows the four most common learned structures among models $\hat{f}_4$ that models the mass rate of liquid cryolite $Na_3AlF_6$ in the bath, namely $\Dot{x}_4$. $\Dot{x}_4$ is simulated by the simulation model $\Dot{x}_4 = f_4$ in  Eq.~(\ref{eq:xdot3}). $f_4$ consist of the features $\{x_1, x_2, x_3, x_4, x_6, x_7, u_1\}$. Table~\ref{table:freq_features} show that $\{x_2, x_3, x_4, x_6, x_8, u_1\}$ are included in $100\%$ of the learned models, $x_1$ is included in $86\%$ of the models and $x_7$ is included in only $21\%$ of the models. As for $\hat{f}_1$, $x_8$ is erroneously included in the basis of the model. This might be explained by the fact that $x_7$ and $x_8$ highly correlate and that the wrong feature is included. The structures in Fig.~\ref{fig:sparse_f4_most_learned_struct} are all forming linear models. However, the simulation model $\Dot{x}_4 = f_4$ is partly nonlinear. Thus, the approximation $\hat{f}_4$ oversimplifies the dynamics. This might be caused by a high weighting of the sparse regularisation term. If the loss function is less penalized, there is room for more weights in the model and, therefore, more nonlinearities.  
        
    \subsubsection{Model output $\hat{f}_5$}
        \begin{figure}[H]
            \centering
            \includegraphics[width=0.8\linewidth]{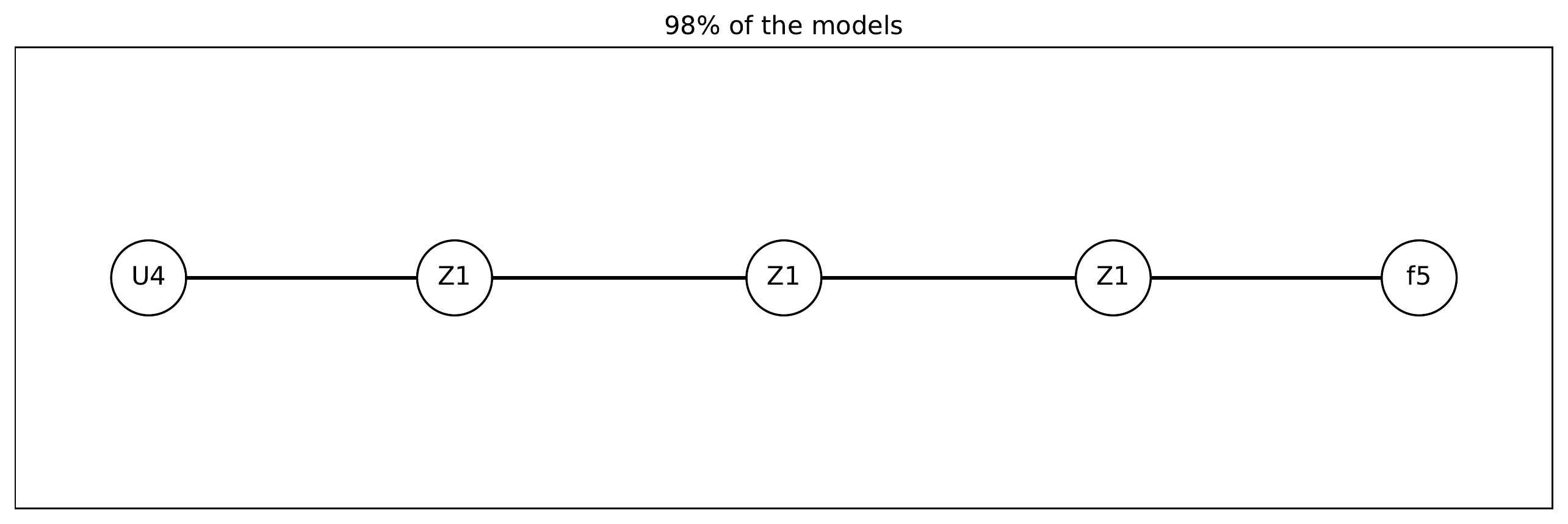}
            \caption{Most common learned structures for $\hat{f}_5(\mathbf{x}, \mathbf{u})$.}
            \label{fig:sparse_f5_most_learned_struct}
        \end{figure}
        Fig.~\ref{fig:sparse_f5_most_learned_struct} shows the most common learned structure among the models $\hat{f}_5$ and include $98\%$ of the learned model structures. $\hat{f}_5$ is modeling the mass rate of produced aluminum in the cell $\Dot{x}_5$. The $x_5$ time series are produced by the simulation model $\Dot{x}_5 = f_5$ in Eq.~\ref{eq:xdot5}. $f_5$ is a linear model dependent on the features $\{u_2, u_4\}$. However, most of the model structures are only depending on $u_4$. This can be caused by the fact that $u_2$ is proportional to a very small constant $k_6 = 4.43\cdot10^{-8}$. Thus, variations in $u_2$ might not be large enough for the learning algorithm to find $u_2$ significant as a basis for $\hat{f}_5$. 
        
    \subsubsection{Model output $\hat{f}_6$}
        \begin{figure}[H]
            \centering
            \includegraphics[width=0.68\linewidth]{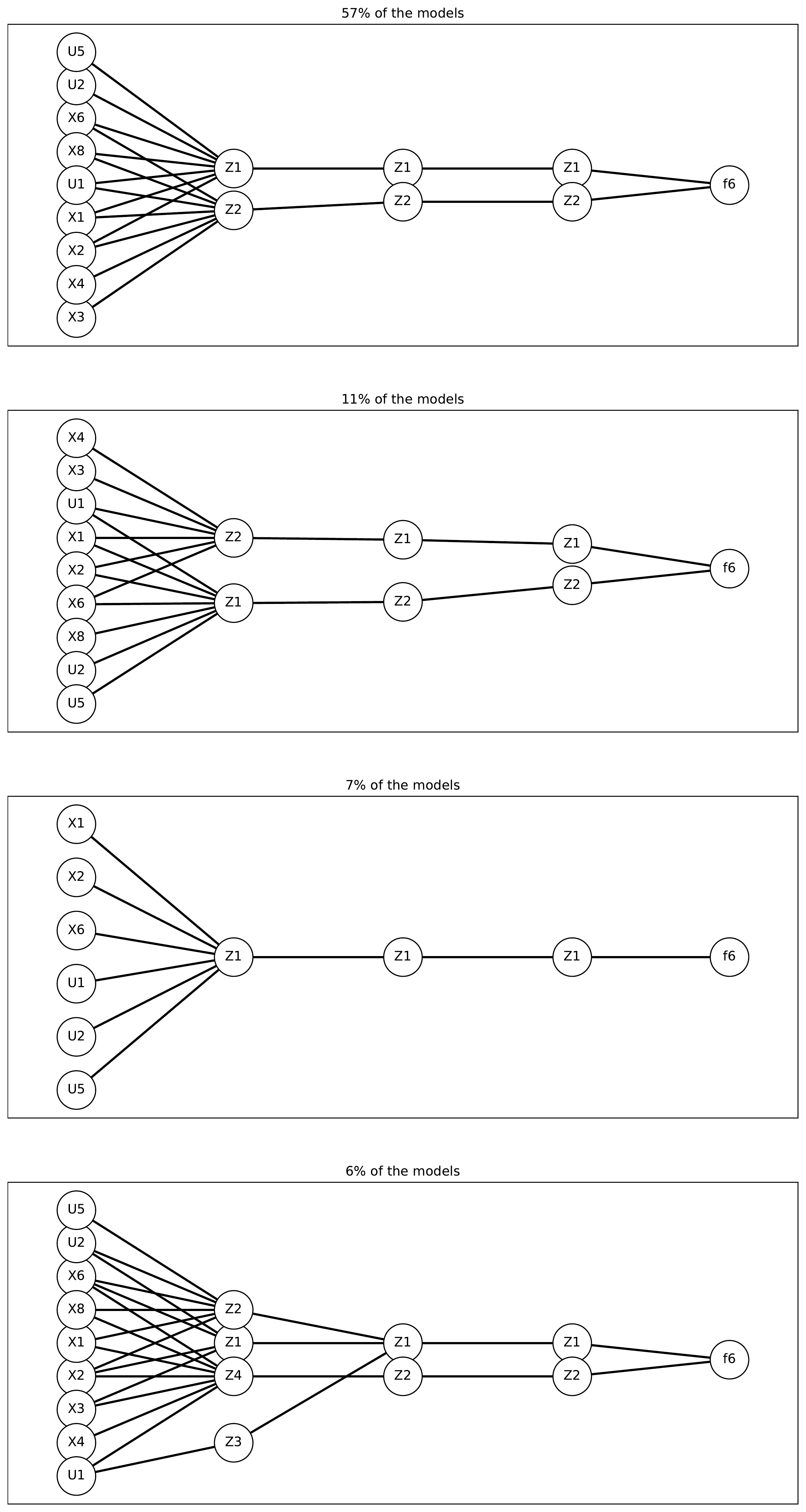}
            \caption{Most common learned structures for $\hat{f}_6(\mathbf{x}, \mathbf{u})$.}
            \label{fig:sparse_f6_most_learned_struct}
        \end{figure}
        Fig.~\ref{fig:sparse_f6_most_learned_struct} shows the most common model structures of $\hat{f}_6$. $\hat{f}_6$ models the bath temperature time derivative $\Dot{x}_6$. The bath temperature $x_6$ is simulated by the ODE in Eq.~(\ref{eq:xdot6}). It is a nonlinear equation depending on  $\{x_1, x_2, x_3, x_4, x_6, x_7, u_2, u_5\}$. The most common structure, learned by $57\%$ of the models $\hat{f}_6$ is illustrated in the top plot of Fig.~\ref{fig:sparse_f6_most_learned_struct}. This structure has the basis $\{x_1, x_2, x_3, x_4, x_6, x_8, u_1, u_2, u_5\}$. Hence, it finds $u_1$ and $x_8$, which is erroneously found in many of the structures above. A possible explanation for this trend is given above.
        The structure has two neurons in  the first layer, one with the basis $\{x_1,x_2,x_6, x_8, u_1, u_2, u_5\}$ and one with the basis $\{x_1, x_2, x_3,x_4, x_6, x_8, u_1\}$. Since the model collapses to a linear model, all terms are summed in the end. Thus, the separation of the basis is thus of little importance. 
        The second plot from above in Fig.~\ref{fig:sparse_f6_most_learned_struct} is the second most common structure, and accounts for $11\%$ of the models $\hat{f}_6$. It has the same feature basis as the most common structure, but they are arranged differently in the first layer. However, since both model structures are linear models, this arrangement is of minor importance. The third structure in Fig.~\ref{fig:sparse_f6_most_learned_struct} which account for $7\%$ of the structures of $\hat{f}_6$ has the basis $\{x_1, x_2, x_6, u_1, u_2, u_5\}$. Compared to the other structures, $\{x_3, x_4, x_8\}$ is not present. Since it only happens rarely, this is maybe partly caused by bad parameter initialization. The fourth most common structure, which accounts for $6\%$ of the structures, has the same feature basis as the first and the second most common structures. This structure is plotted in the bottom of Fig.~\ref{fig:sparse_f6_most_learned_struct}. While all other structures in Fig.~\ref{fig:sparse_f6_most_learned_struct} have one linear response region, the fourth most common model structure models some nonlinearities. That is,  neurons in the first hidden layer are connected in the second hidden layer. Hence, the input space must be divided into several linear response regions for this structure.
    
    \subsubsection{Model output $\hat{f}_7$}
        \begin{figure}[H]
            \centering
            \includegraphics[width=0.7\linewidth]{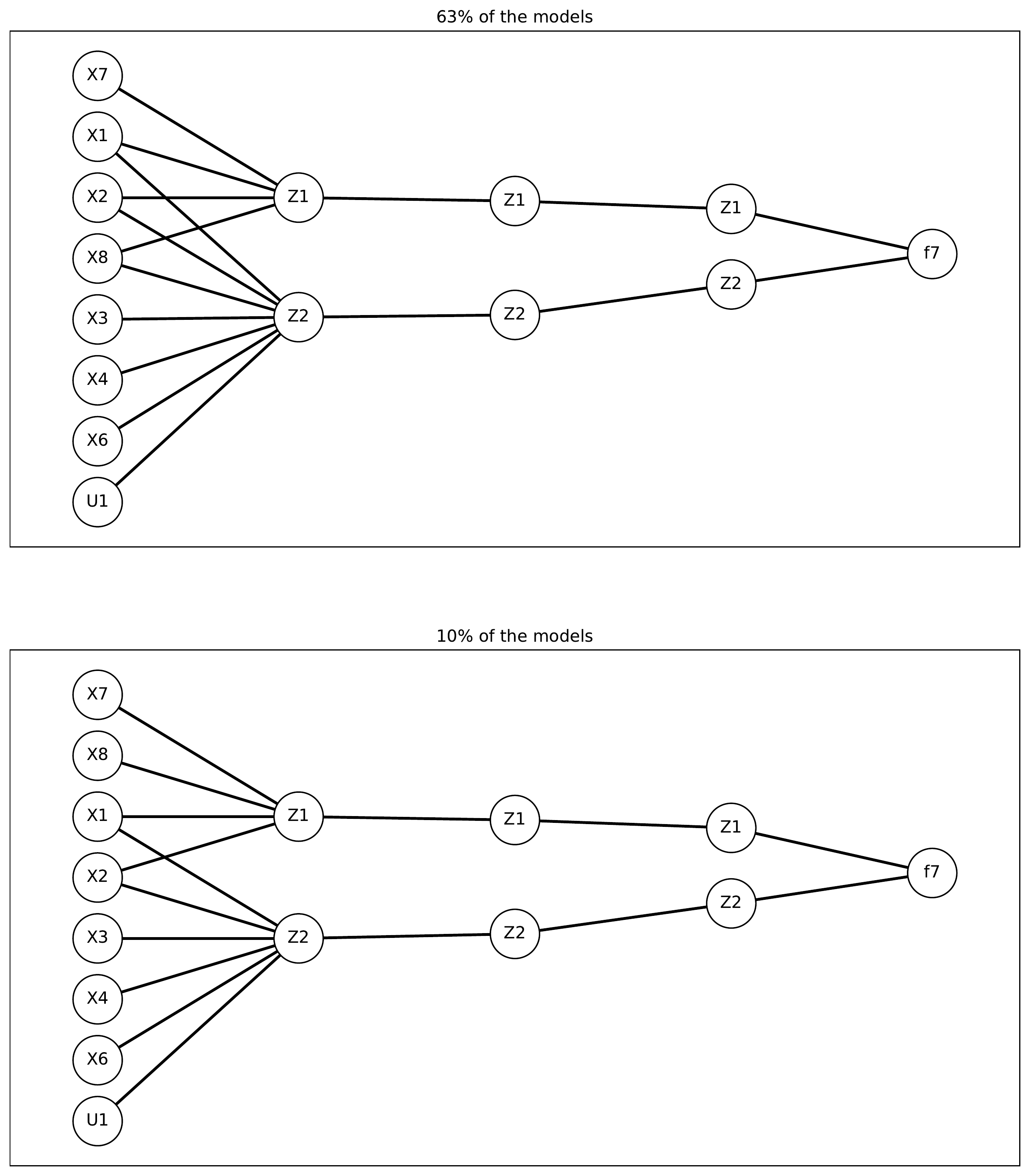}
            \caption{Most common learned structures for $\hat{f}_7(\mathbf{x}, \mathbf{u})$.}
            \label{fig:sparse_f7_most_learned_struct}
        \end{figure}
        Fig.~\ref{fig:sparse_f7_most_learned_struct} shows the two most common structures for the models $\hat{f}_7$. $\hat{f}_7$ models the time derivative of the side ledge temperature $\Dot{x}_7$. $\Dot{x}_7 = f_7$ is simulated by the ODE in Eq.~(\ref{eq:xdot7}). $\Dot{x}_7$ is depending on the feature basis $\{x_1, x_2, x_3, x_4, x_6, x_7, x_8\}$. Table~\ref{table:freq_features} states that the basis $\{x_2, x_3, x_4, x_6, x_8, u_1\}$ is present for $100\%$ of the models, $x_1$ is present for $99\%$ of the models and $x_7$ is present in $89\%$ of the models. The top plot in Fig.~\ref{fig:sparse_f7_most_learned_struct} show the structure that accounts for $63\%$ of the models. The bottom plot account for $10\%$ of the model structures of $\hat{f}_7$. These two structures collapse to linear models, and have the same feature basis $\{x_1, x_2, x_3, x_4, x_6, x_7, x_8, u_1\}$, but have minor differences in how weights are connected between input layer and first hidden layer. $u_1$ is also for this model output erroneously found as a basis, and a possible explanation is mentioned above.    
    
    \subsubsection{Model output $\hat{f}_8$}
        \begin{figure}[H]
            \centering
            \includegraphics[width=0.75\linewidth]{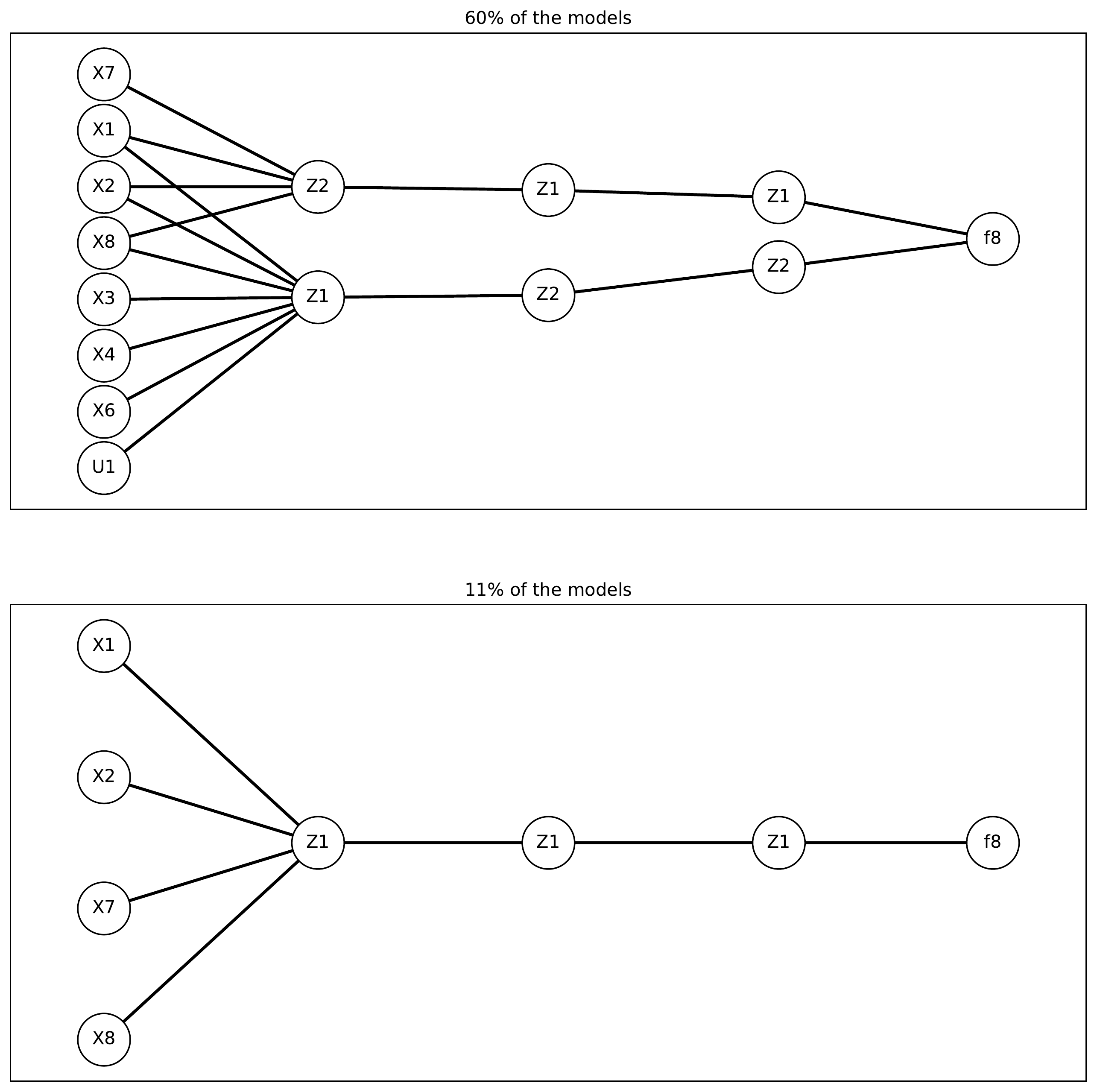}
            \caption{Most common learned structures for $\hat{f}_8(\mathbf{x}, \mathbf{u})$.}
            \label{fig:sparse_f8_most_learned_struct}
        \end{figure}
        Fig.~\ref{fig:sparse_f8_most_learned_struct} show the two most common model structures for the last model output $\hat{f}_8$. $\hat{f}_8$ models the time derivative of the wall temperature $\Dot{x}_8 = f_8$, which is simulated in Eq.~(\ref{eq:xdot8}). $\Dot{x}_8$ depends on the feature basis $\{x_1, x_7, x_8\}$. However, the most common learned structure for $\hat{f}_8$ has the basis $\{x_1, x_2, x_3, x_4, x_6, x_7, x_8, u_1\}$. This is the exact same structure as the most common learned structure for $\hat{f}_7$. Therefore, a possible explanation is that $\hat{f}_8$ adapts the same parameters as $\hat{f}_7$ in some cases as they highly correlates. The bottom plot in Fig.~\ref{fig:sparse_f8_most_learned_struct} show the second most common learned structure of the model output $\hat{f}_8$. This structure is learned by $11\%$ of the models $\hat{f}_8$. The feature basis for this structure is $\{x_1, x_2, x_7, x_8\}$, and reminds more of the actual basis. In this structure, there is only one erroneous learned feature, namely $x_2$.  
        Fig.~\ref{fig:sparse_f1_most_learned_struct}-\ref{fig:sparse_f8_most_learned_struct} and Table~\ref{table:freq_features} show that the sparse learning is quite consistent in finding the same feature basis and structure with similar characteristics. However, some differences that affect the models are present.

        It is clear that doing a similar analysis for models in Fig.~\ref{fig:dense_nn_structure} is impossible as all interconnections make the models a black box. On average, $93\%$ of the weights of dense DNNs are pruned in the sparse DNN models. For the outputs $\hat{f}_1,\; \hat{f}_4, \; \hat{f}_6, \; \hat{f}_7$ and $\hat{f}_8$, approximately $40\%$ of the input features are pruned of the model structures. For $\hat{f}_2,\; \hat{f}_3$ and $\hat{f}_5$, $85 - 95\%$ of the input features are pruned. For all outputs of the sparse DNN models, around $85-95\%$ of the neurons are pruned at each layer. In a neural network, the number of matrix operations only decreases if neurons are pruned. That is, removing a neuron in layer $j$ is equivalent to removing a row in weight matrix $\mathbf{W}_j$ and a column in weight matrix $\mathbf{W}_{j+1}$. The dense models have a compact model structure, where most weights are nonzero. The dense DNN models in the case study have the shapes 13-15-14-12-8. The first number is the number of features, the second, third, and fourth numbers are the number of neurons in hidden layers, and the last is the number of outputs. This shape gives $669$ matrix operations in a forward pass. An average sparse DNN, has the shape $13-6-6-6-8$. This gives $198$ matrix operations. Thus, the number of matrix operations in the forward pass of a sparse DNN model is reduced by approximately $70\%$.
    
    \subsection{Generalizability perspective}
    \label{subsec:Model_performance}
    This section focuses on the models' performance on test data in terms of accuracy and uncertainty. Furthermore, we also investigate the impact of the training data quantity and prediction horizon on the performance measured in terms of $\overline{\textrm{AN-RFMSE}}$.
    
    \subsubsection{Comparison of sparse and dense rolling forecast}
\begin{figure}[H]
        \centering
        \begin{subfigure}[t]{0.32\linewidth}
            \raisebox{-\height}{\includegraphics[width=\linewidth]{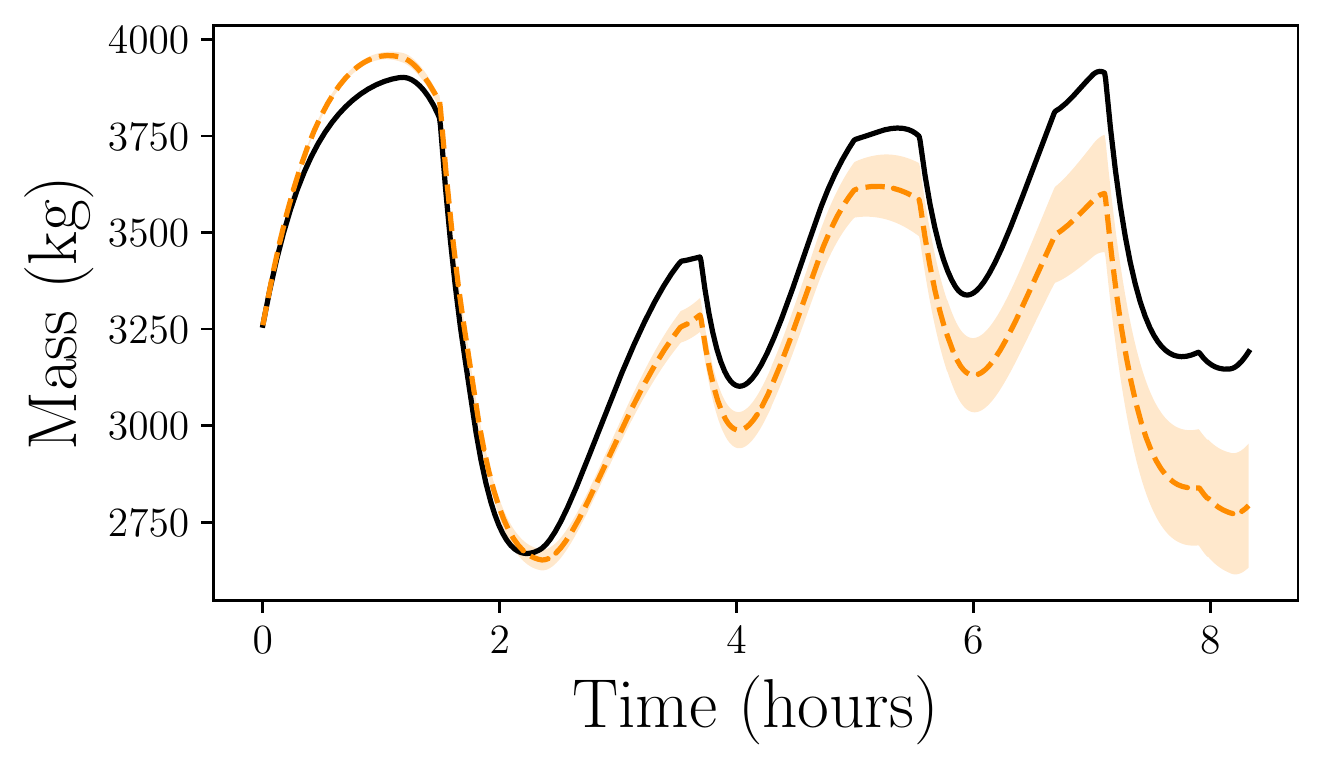}}
            \caption{Side ledge mass $x_1$}
            \label{subfig:Rolling_forecast_x1_sparse}
        \end{subfigure}
        \begin{subfigure}[t]{0.32\linewidth}
            \raisebox{-\height}{\includegraphics[width=\linewidth]{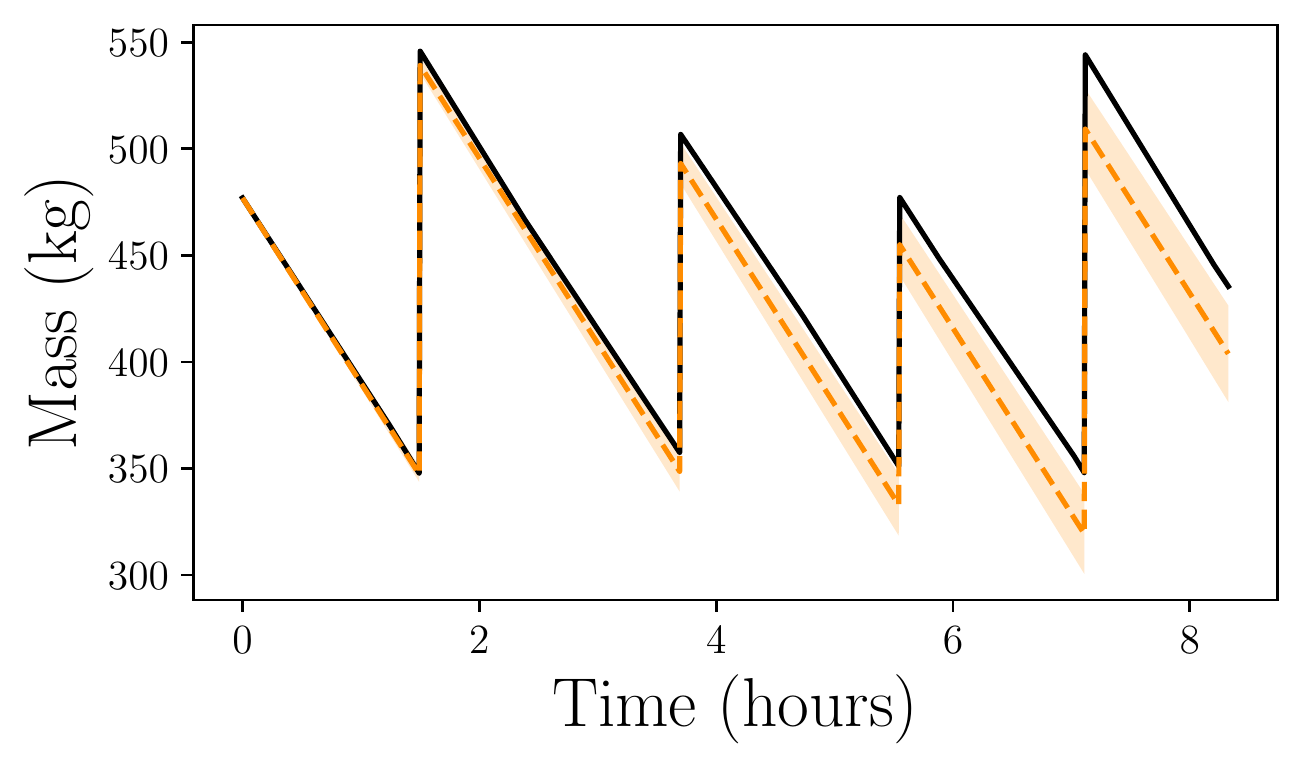}}
            \caption{Alumina mass $x_2$}
            \label{subfig:Rolling_forecast_x2_sparse}
        \end{subfigure}
        \begin{subfigure}[t]{0.32\linewidth}
            \raisebox{-\height}{\includegraphics[width=\linewidth]{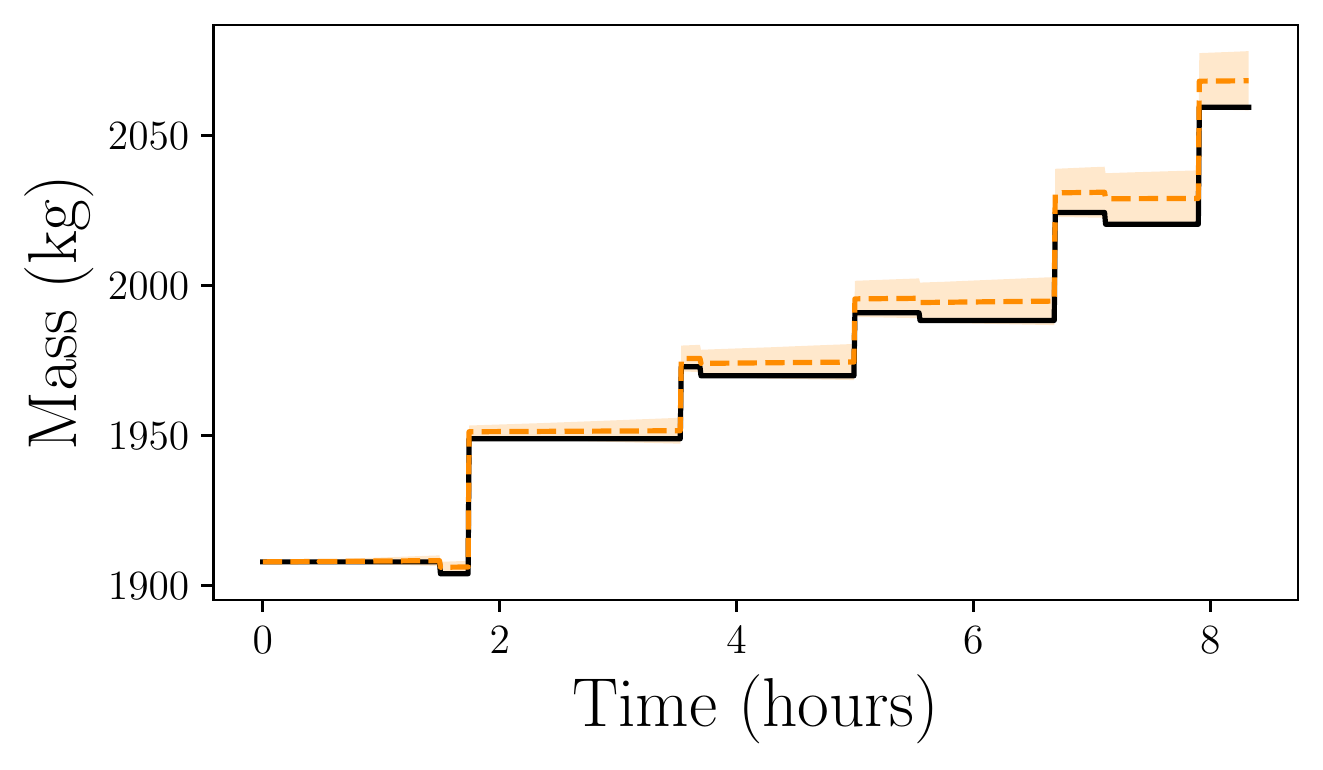}}
            \caption{Aluminum fluoride mass $x_3$}
            \label{subfig:Rolling_forecast_x3_sparse}
        \end{subfigure}
        \begin{subfigure}[t]{0.32\linewidth}
            \raisebox{-\height}{\includegraphics[width=\linewidth]{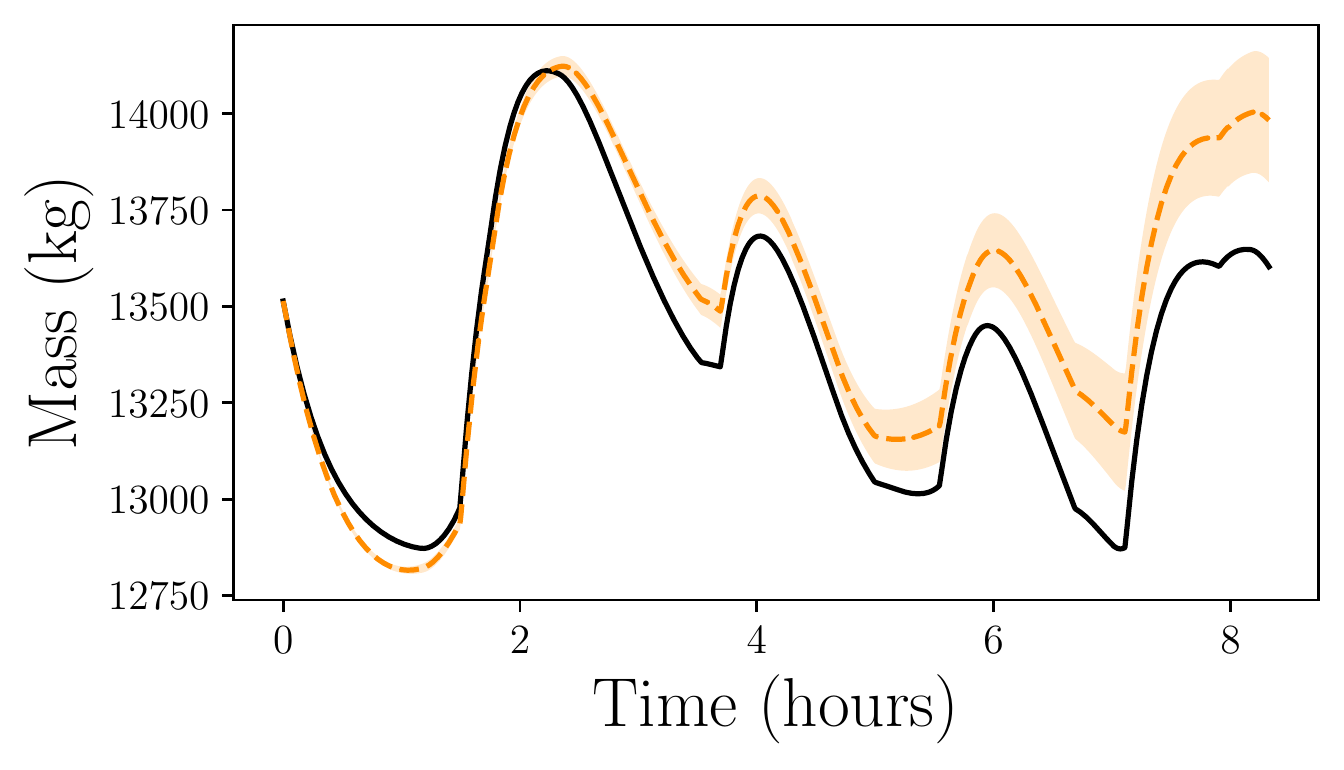}}
            \caption{Molten cryolite mass $x_4$}
            \label{subfig:Rolling_forecast_x4_sparse}
        \end{subfigure}
        \begin{subfigure}[t]{0.32\linewidth}
            \raisebox{-\height}{\includegraphics[width=\linewidth]{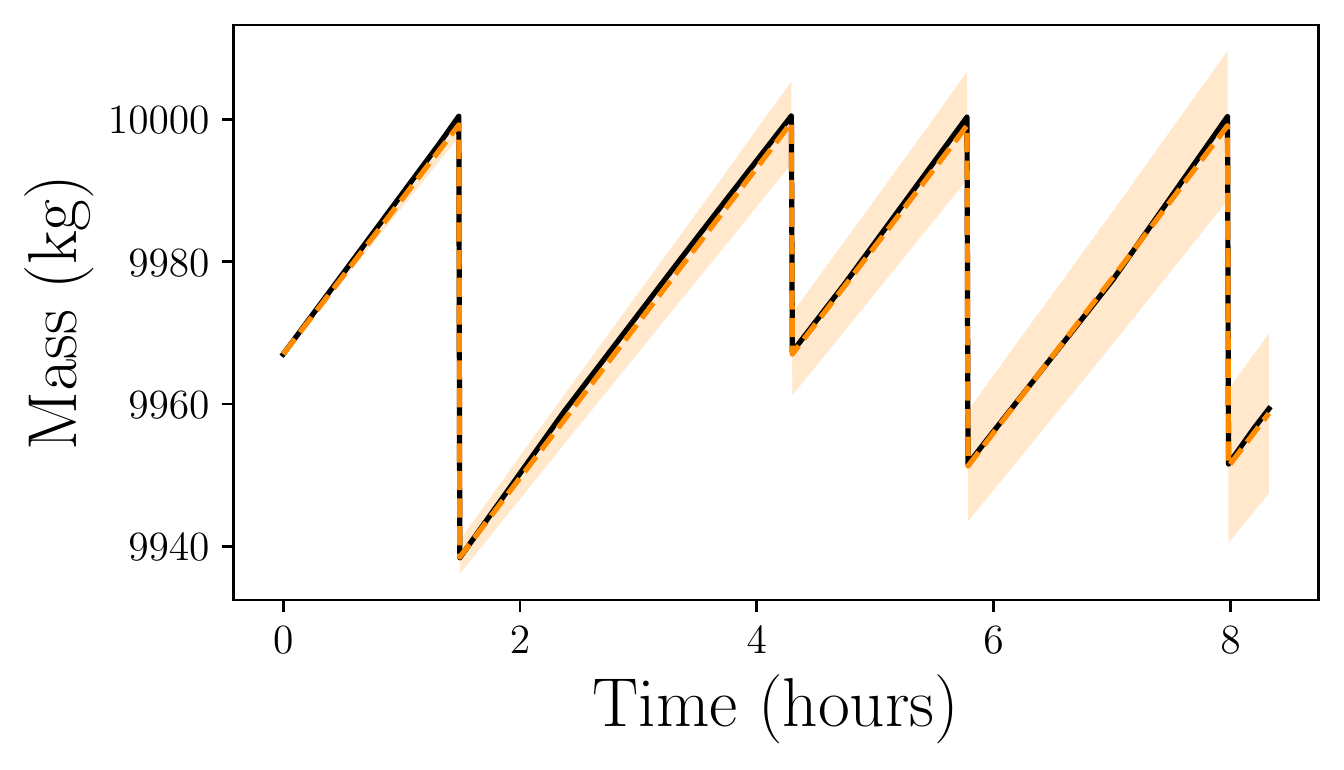}}
            \caption{Produced aluminum mass $x_5$}
            \label{subfig:Rolling_forecast_x5_sparse}
        \end{subfigure}
        \begin{subfigure}[t]{0.32\linewidth}
            \raisebox{-\height}{\includegraphics[width=\linewidth]{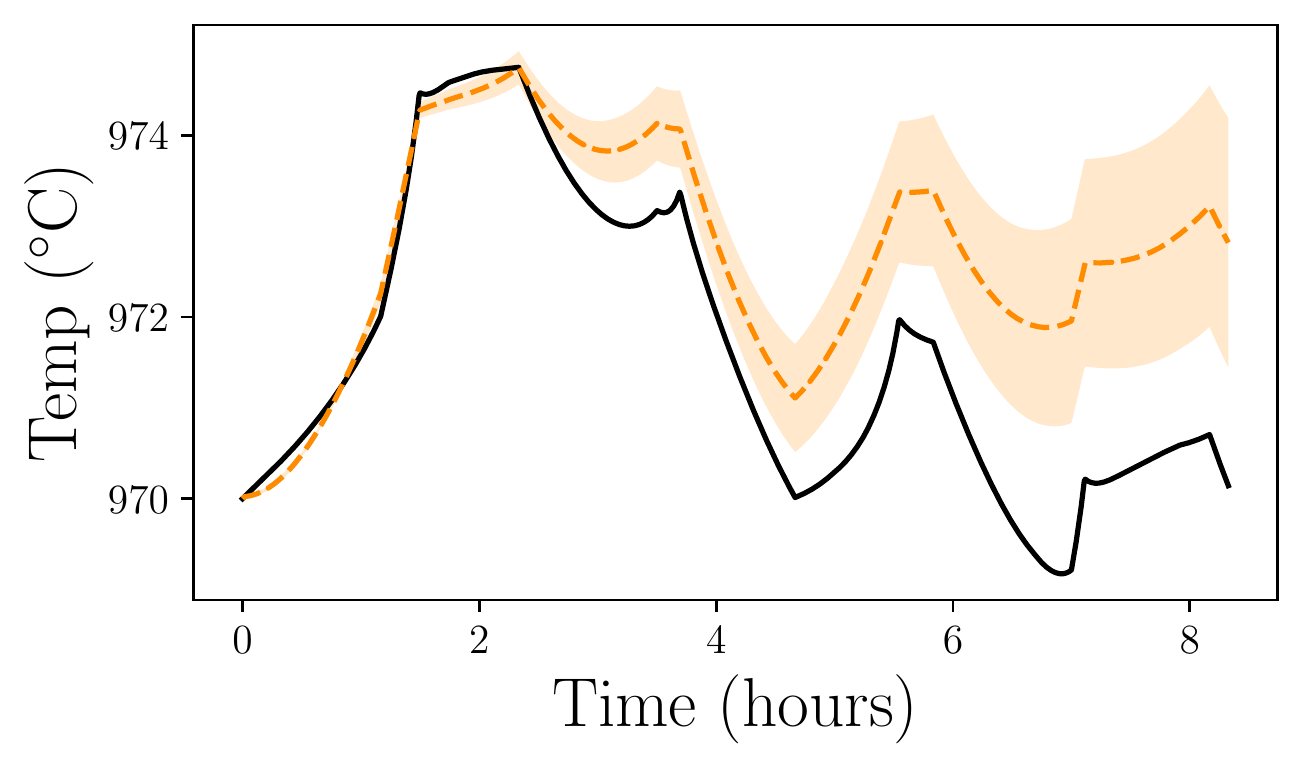}}
            \caption{Bath temperature $x_6$}
            \label{subfig:Rolling_forecast_x6_sparse}
        \end{subfigure}
        \begin{subfigure}[t]{0.32\linewidth}
            \raisebox{-\height}{\includegraphics[width=\linewidth]{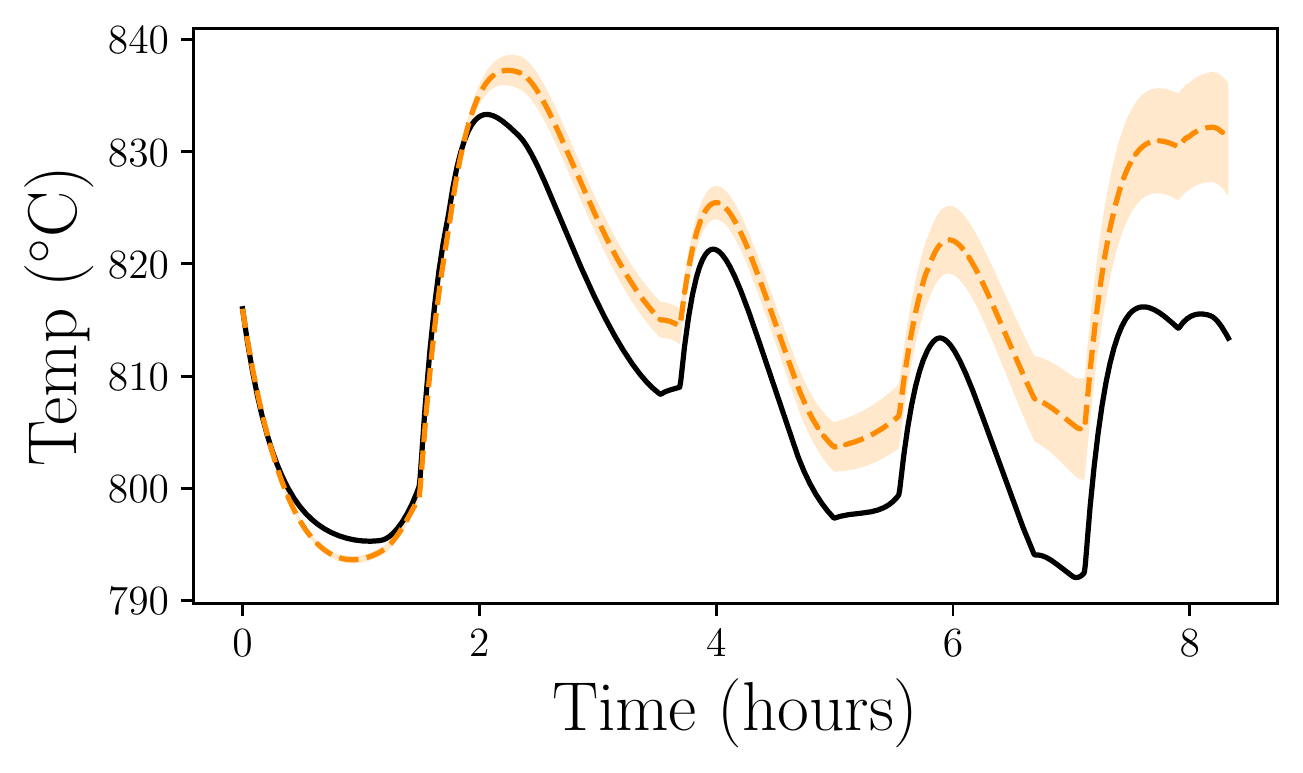}}
            \caption{Side ledge temperature $x_7$}
            \label{subfig:Rolling_forecast_x7_sparse}
        \end{subfigure}
        \begin{subfigure}[t]{0.32\linewidth}
            \raisebox{-\height}{\includegraphics[width=\linewidth]{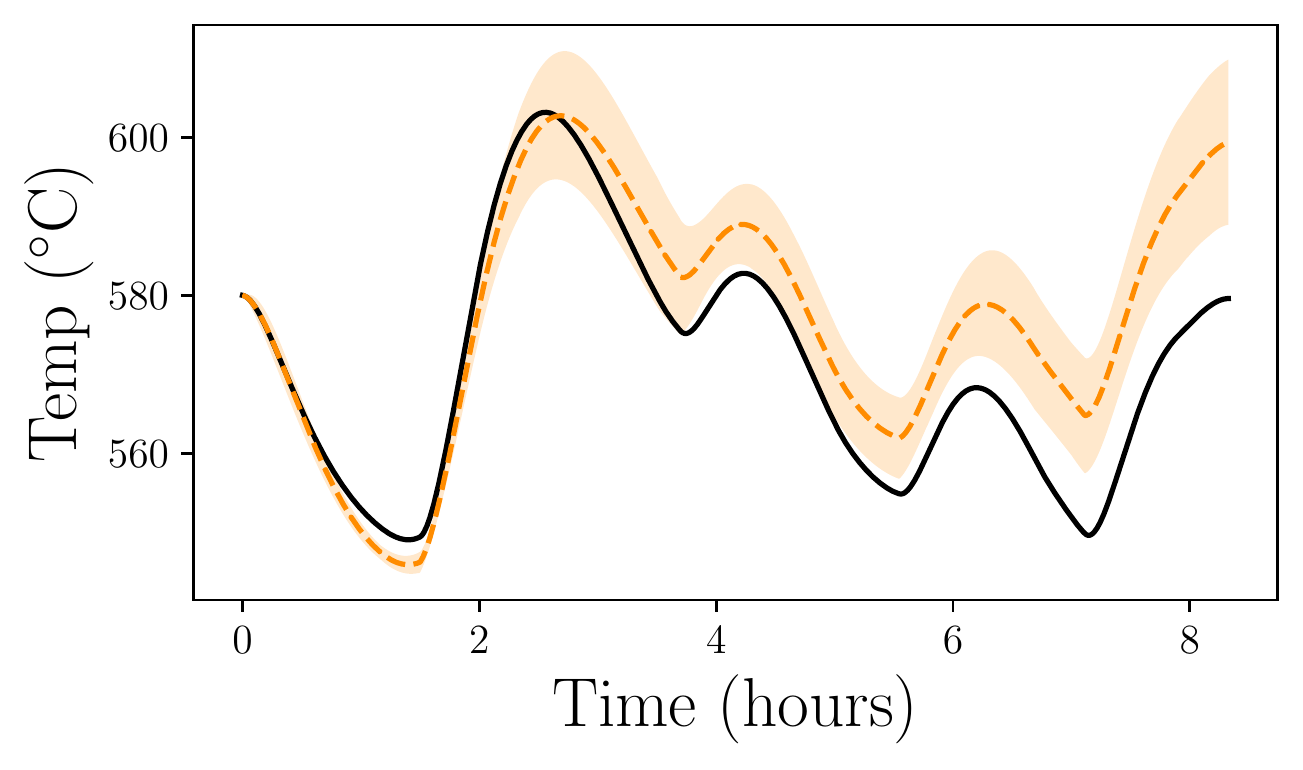}}
            \caption{Side wall temperature $x_8$}
            \label{subfig:Rolling_forecast_x8_sparse}
        \end{subfigure}
        \\
        \begin{subfigure}[t]{0.32\linewidth}
            \raisebox{-\height}{\includegraphics[width=\linewidth]{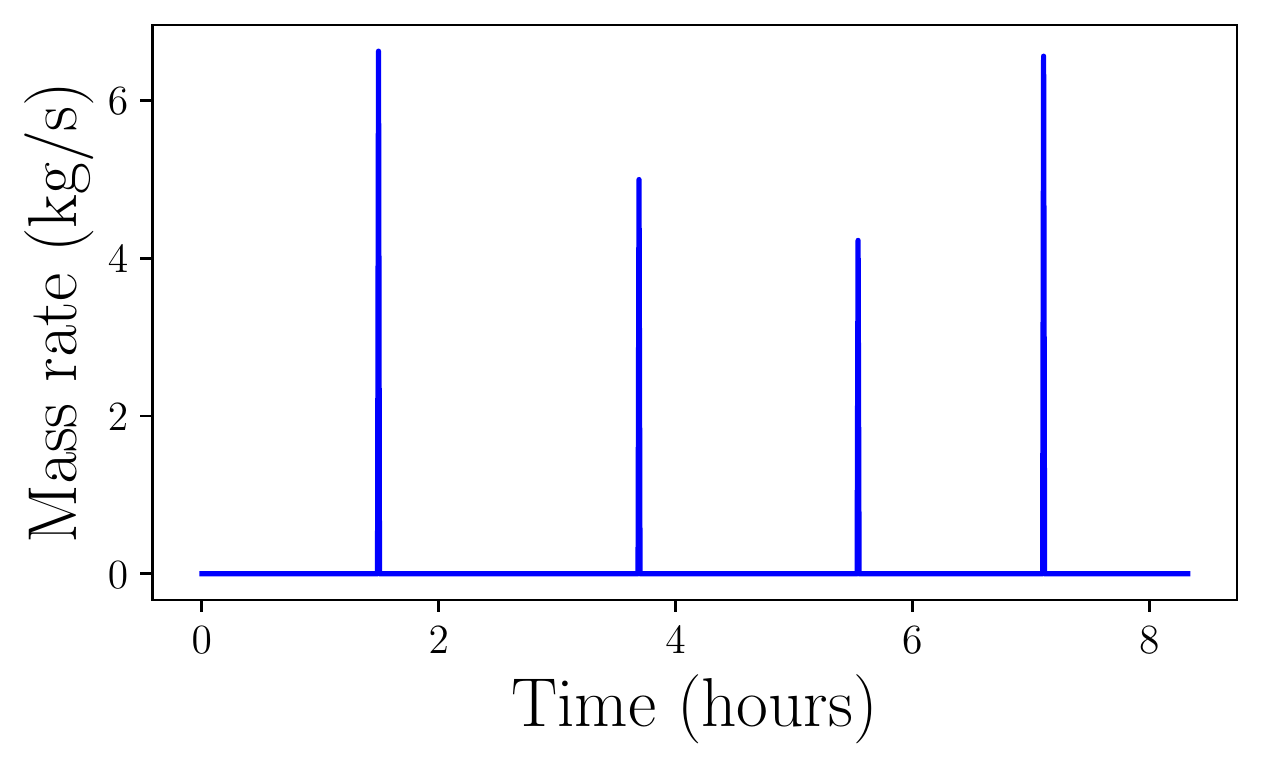}}
            \caption{Alumina feed $u_1$}
            \label{subfig:Rolling_forecast_u1_sparse}
        \end{subfigure}
        \begin{subfigure}[t]{0.32\linewidth}
            \raisebox{-\height}{\includegraphics[width=\linewidth]{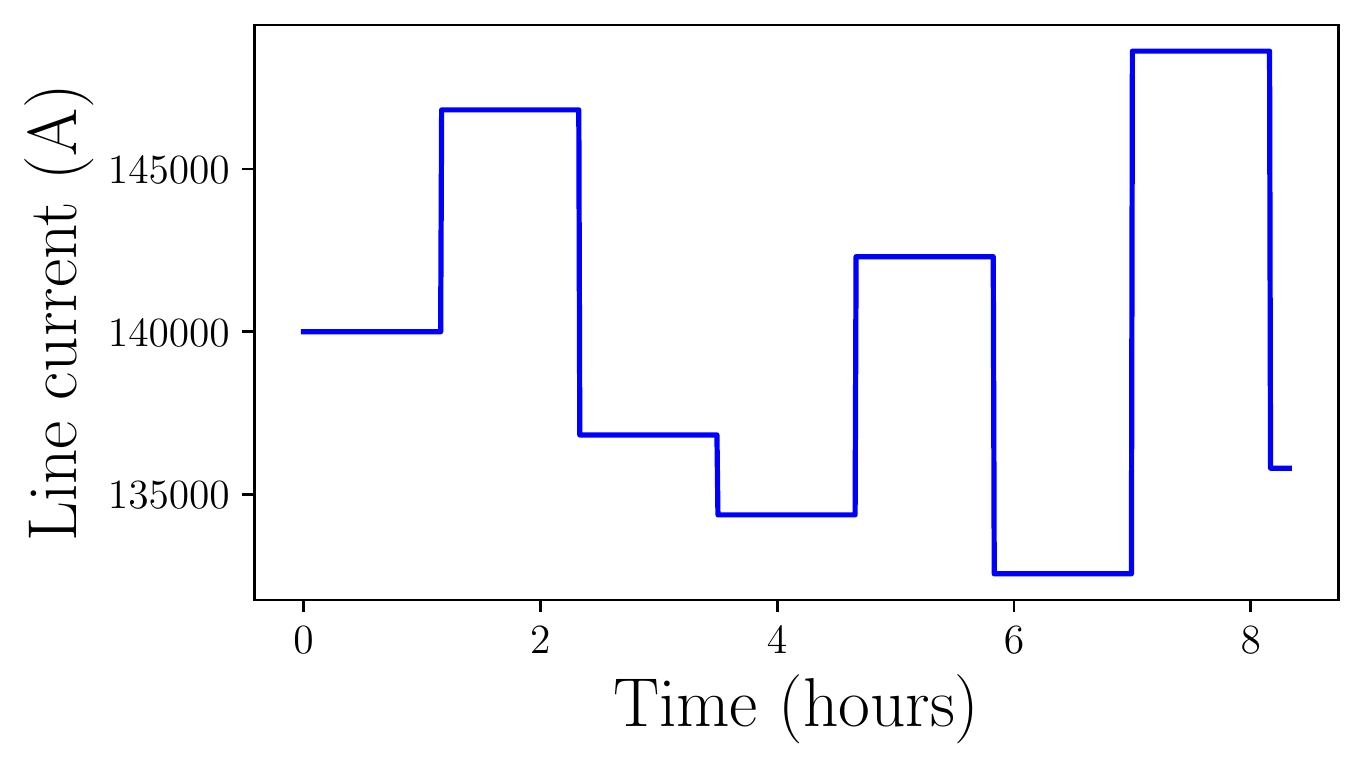}}
            \caption{Line current $u_2$}
            \label{subfig:Rolling_forecast_u2_sparse}
        \end{subfigure}
        \begin{subfigure}[t]{0.32\linewidth}
            \raisebox{-\height}{\includegraphics[width=\linewidth]{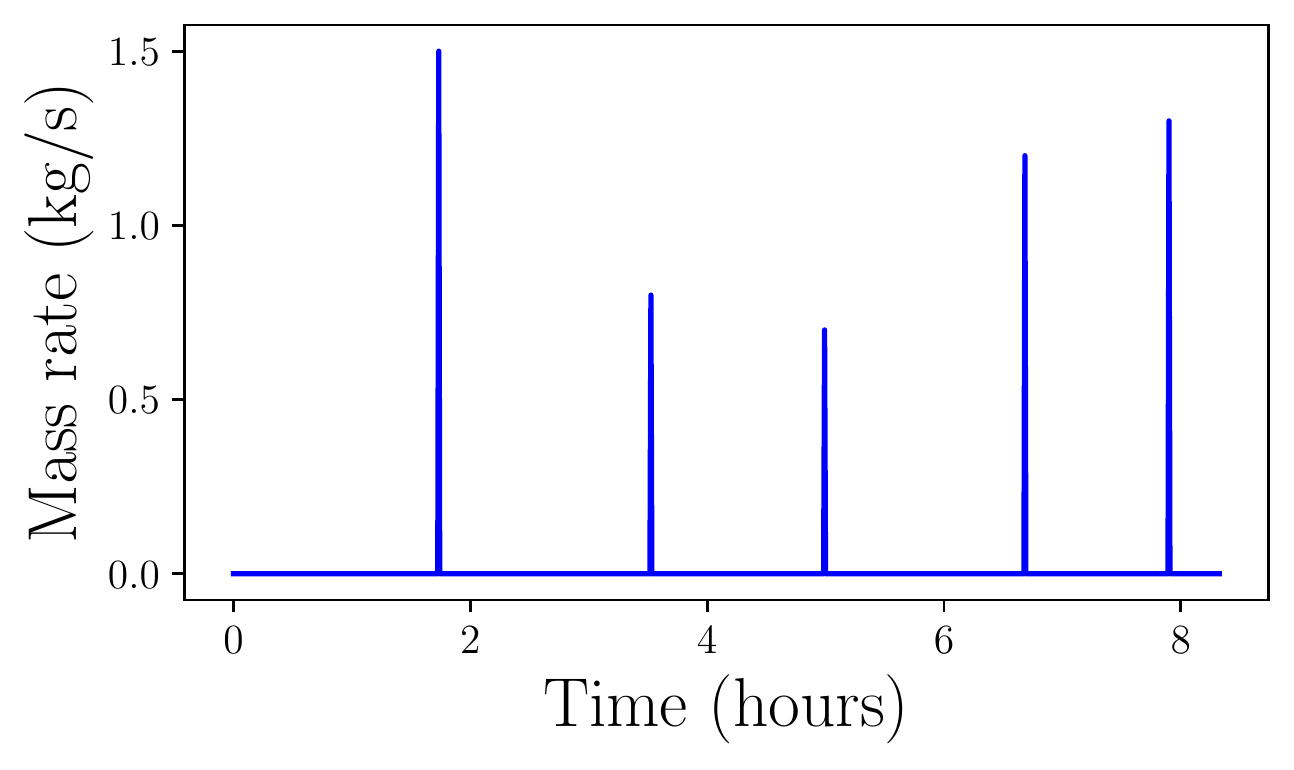}}
            \caption{Aluminum fluoride $u_3$}
            \label{subfig:Rolling_forecast_u3_sparse}
        \end{subfigure}
        \begin{subfigure}[t]{0.32\linewidth}
            \raisebox{-\height}{\includegraphics[width=\linewidth]{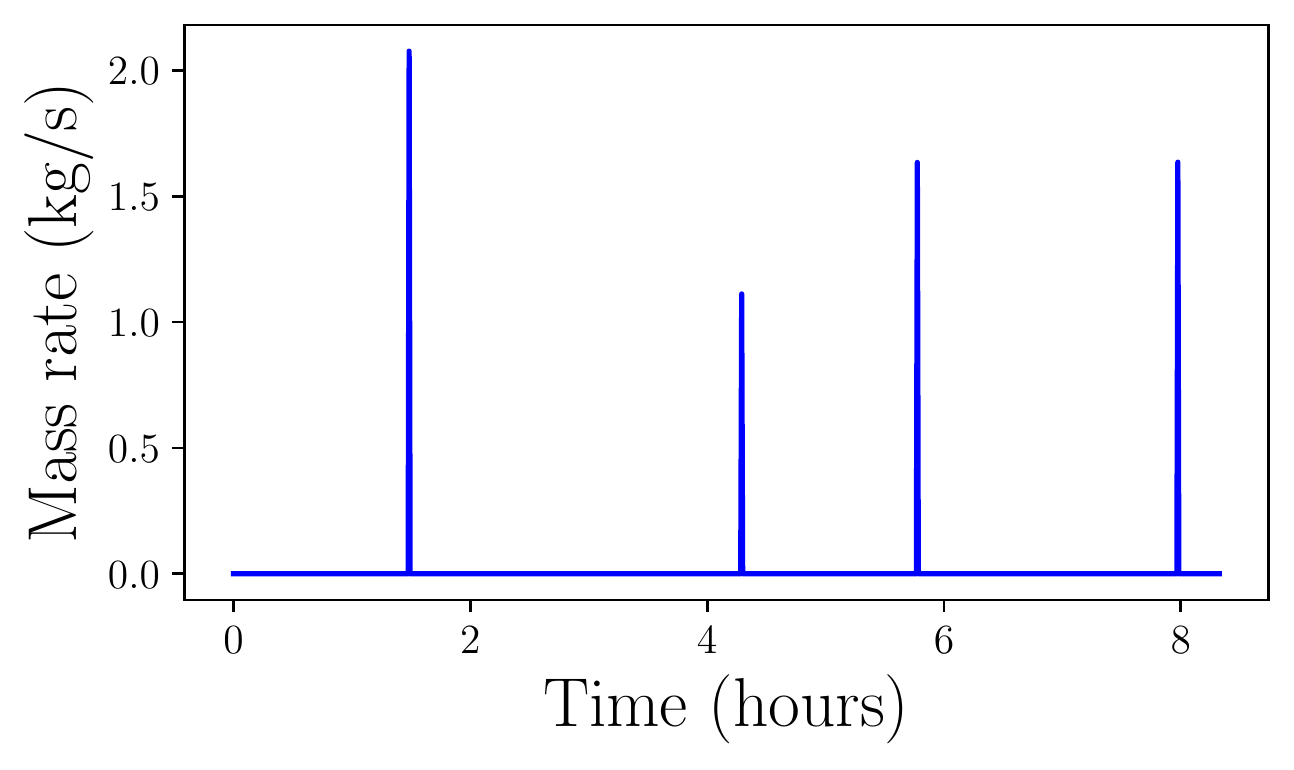}}
            \caption{Metal tapping $u_4$}
            \label{subfig:Rolling_forecast_u4_sparse}
        \end{subfigure}
        \begin{subfigure}[t]{0.32\linewidth}
            \raisebox{-\height}{\includegraphics[width=\linewidth]{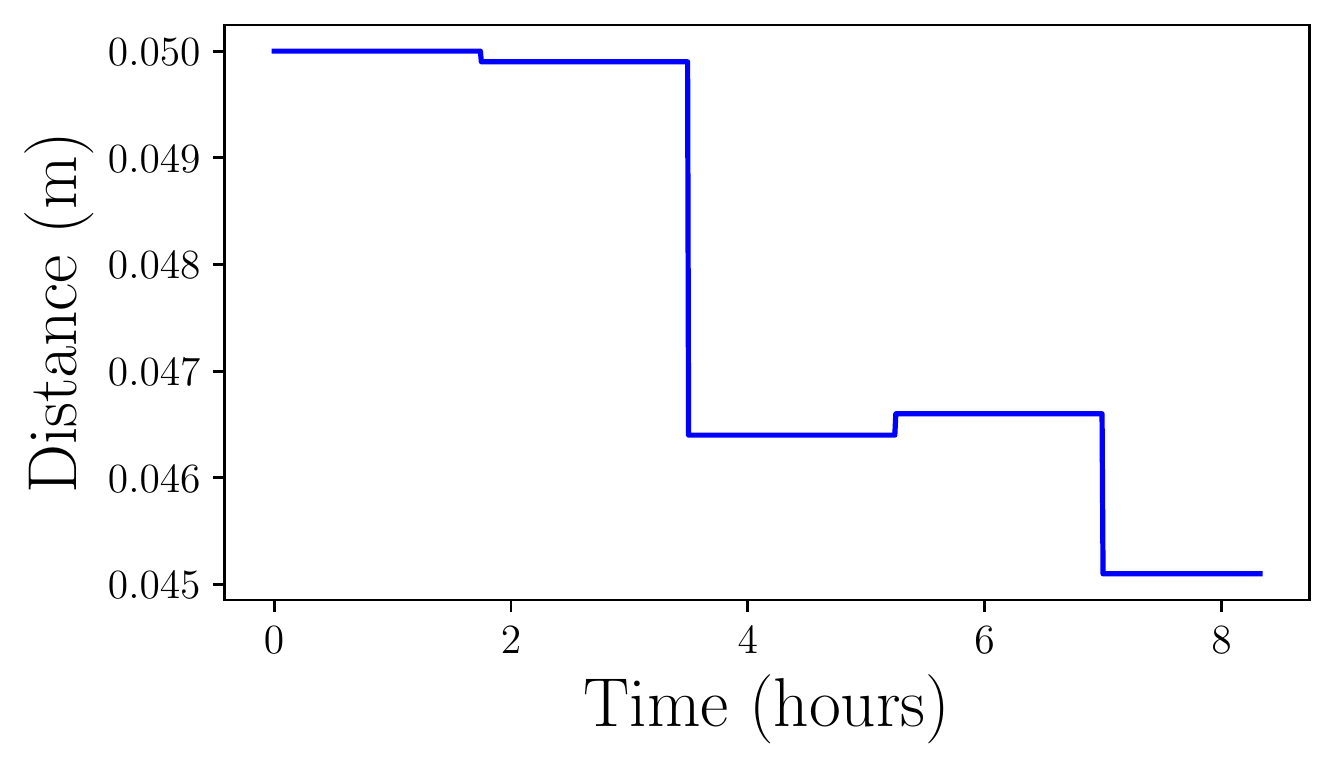}}
            \caption{Anode-cathode-distance $u_5$}
            \label{subfig:Rolling_forecast_u5_sparse}
        \end{subfigure}

        \begin{adjustbox}{max width=1\linewidth}
        \begin{tikzpicture}
            \begin{customlegend}[legend columns=6,legend style={draw=none, column sep=2ex},legend entries={Truth,Mean sparse, control input, std. dev. sparse}]
                \addlegendimage{black,thick, sharp plot}
                \addlegendimage{Torange,thick, dashed,sharp plot}
                \addlegendimage{defaultblue,thick, sharp plot}
                \addlegendimage{Torange!40, fill=Torange!40, area legend}
            \end{customlegend}
        \end{tikzpicture}
    \end{adjustbox}
        \caption{Sparse rolling forecast of state variables $\{x_1, .., x_8\}$ at each time instant. The true values of $\mathbf{x}$  and control inputs $\mathbf{u}$ are taken from one simulated set of test set trajectories $\mathbf{X}_i \in \mathbf{X}_{test}$. 
        The orange dotted line shows the average of 20 forecasts calculated by 20 sparse neural network models with different parameter initialization. The orange band shows the standard deviation of the same 20 forecasts calculated by 20 sparse models.}
        \label{fig:Rolling_Forecast_sparse}
    \end{figure}

\begin{figure}[H]
        \centering
        \begin{subfigure}[t]{0.32\linewidth}
            \raisebox{-\height}{\includegraphics[width=\linewidth]{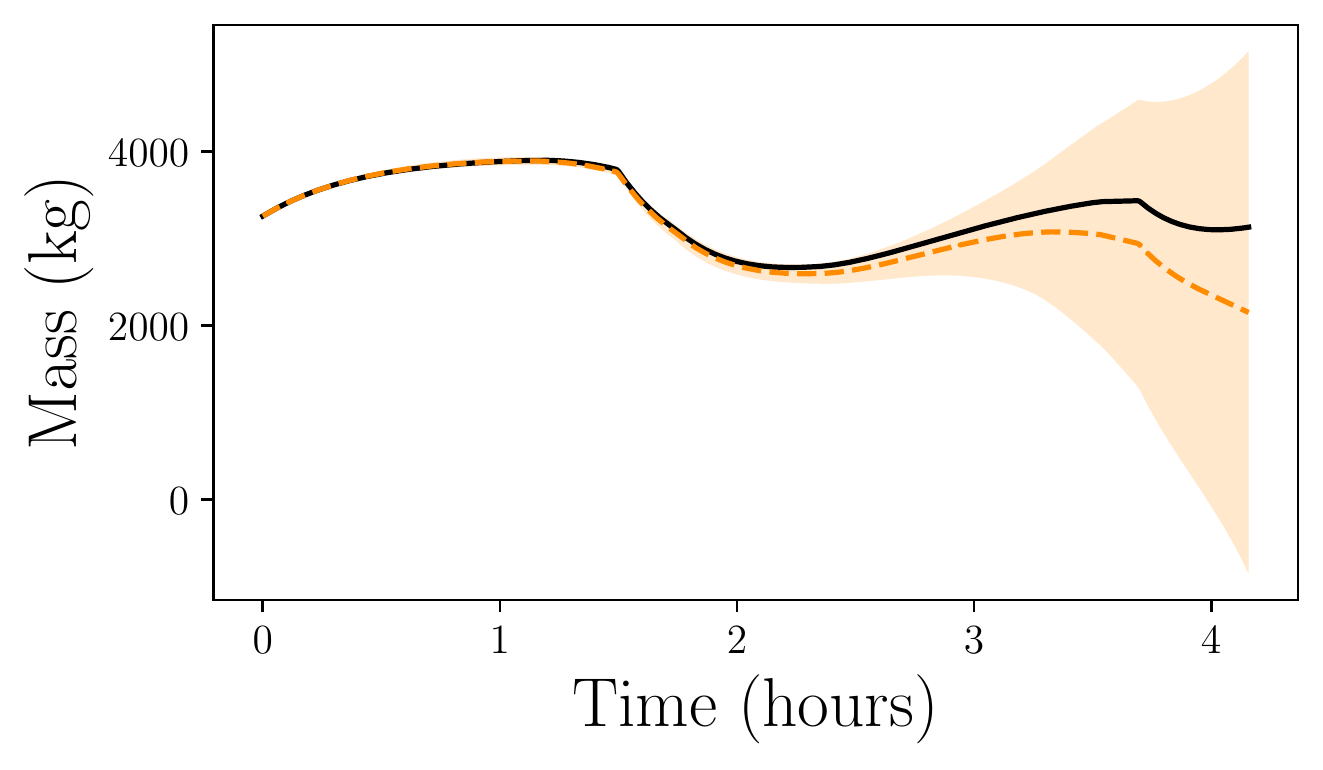}}
            \caption{Side ledge mass $x_1$}
            \label{subfig:Rolling_forecast_x1_dense}
        \end{subfigure}
        \begin{subfigure}[t]{0.32\linewidth}
            \raisebox{-\height}{\includegraphics[width=\linewidth]{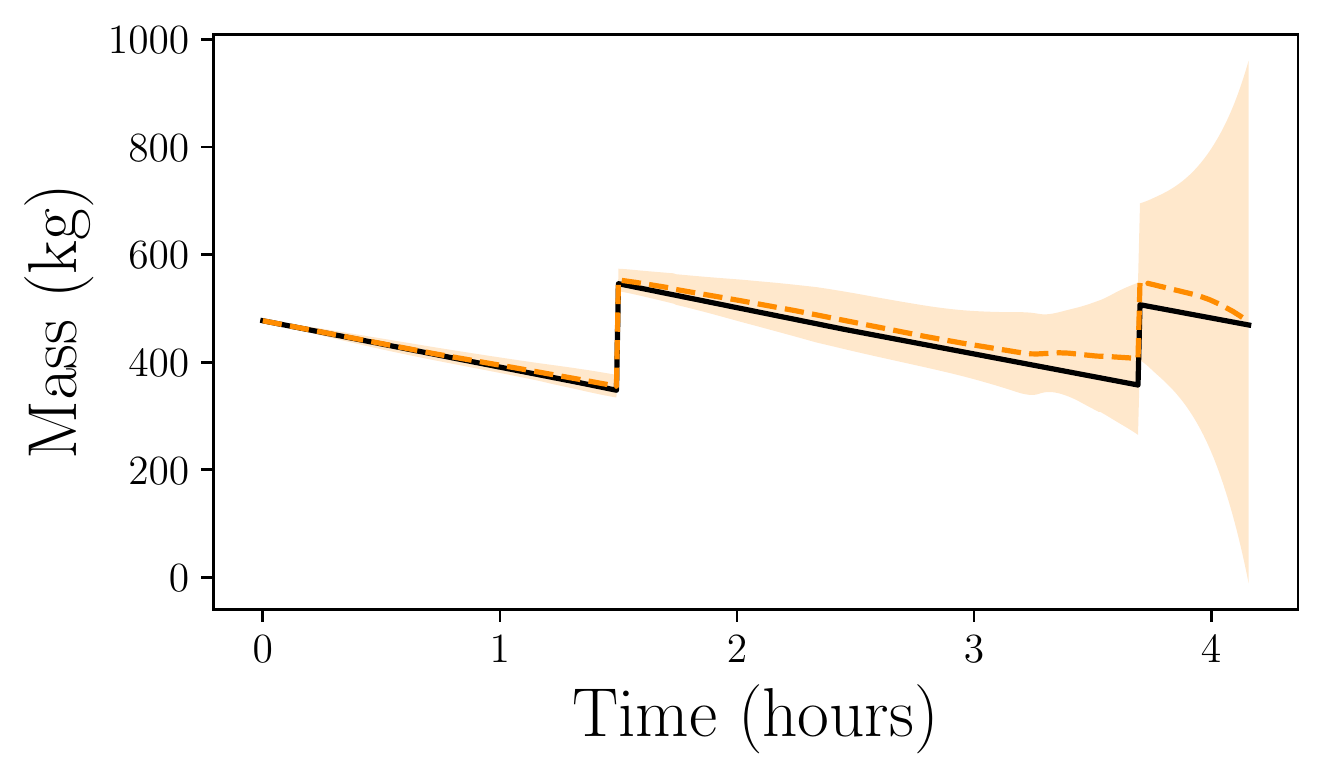}}
            \caption{Alumina mass $x_2$}
            \label{subfig:Rolling_forecast_x2_dense}
        \end{subfigure}
        \begin{subfigure}[t]{0.32\linewidth}
            \raisebox{-\height}{\includegraphics[width=\linewidth]{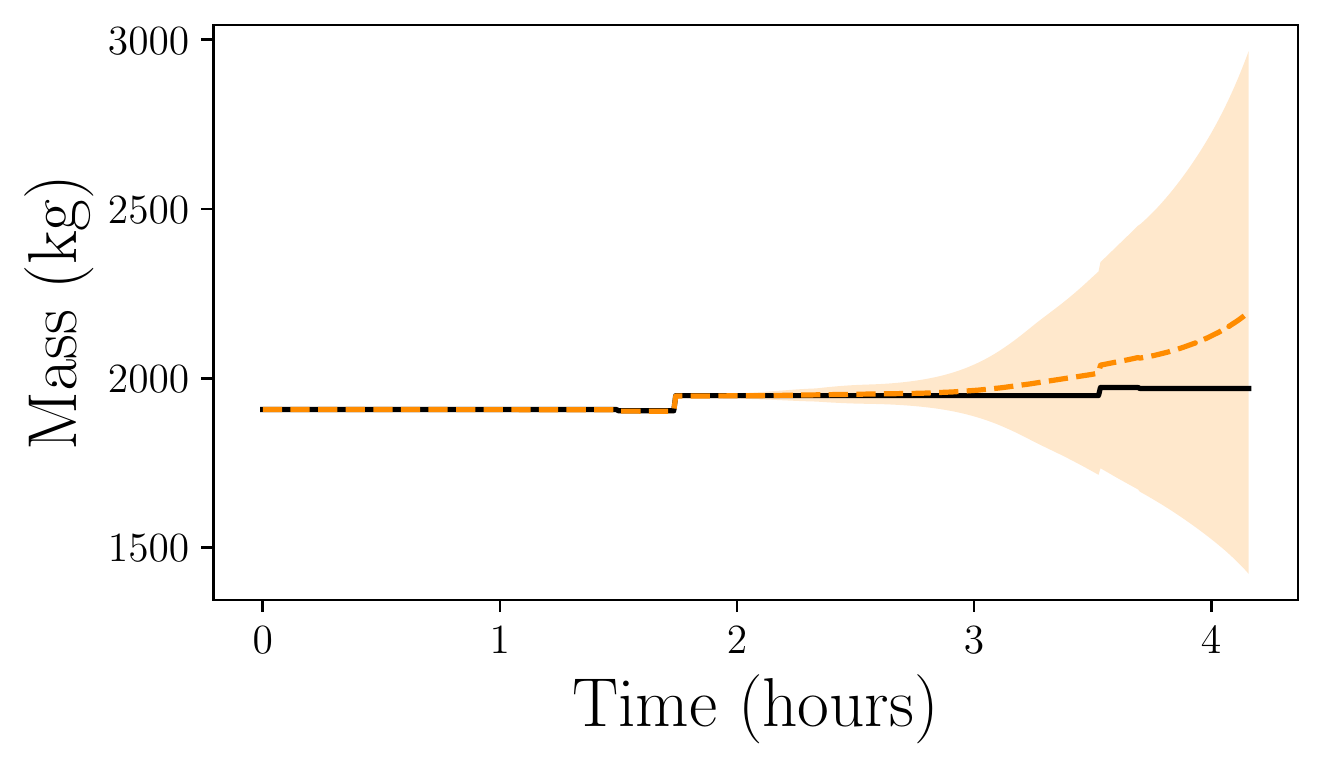}}
            \caption{Aluminum fluoride mass $x_3$}
            \label{subfig:Rolling_forecast_x3_dense}
        \end{subfigure}
        \begin{subfigure}[t]{0.32\linewidth}
            \raisebox{-\height}{\includegraphics[width=\linewidth]{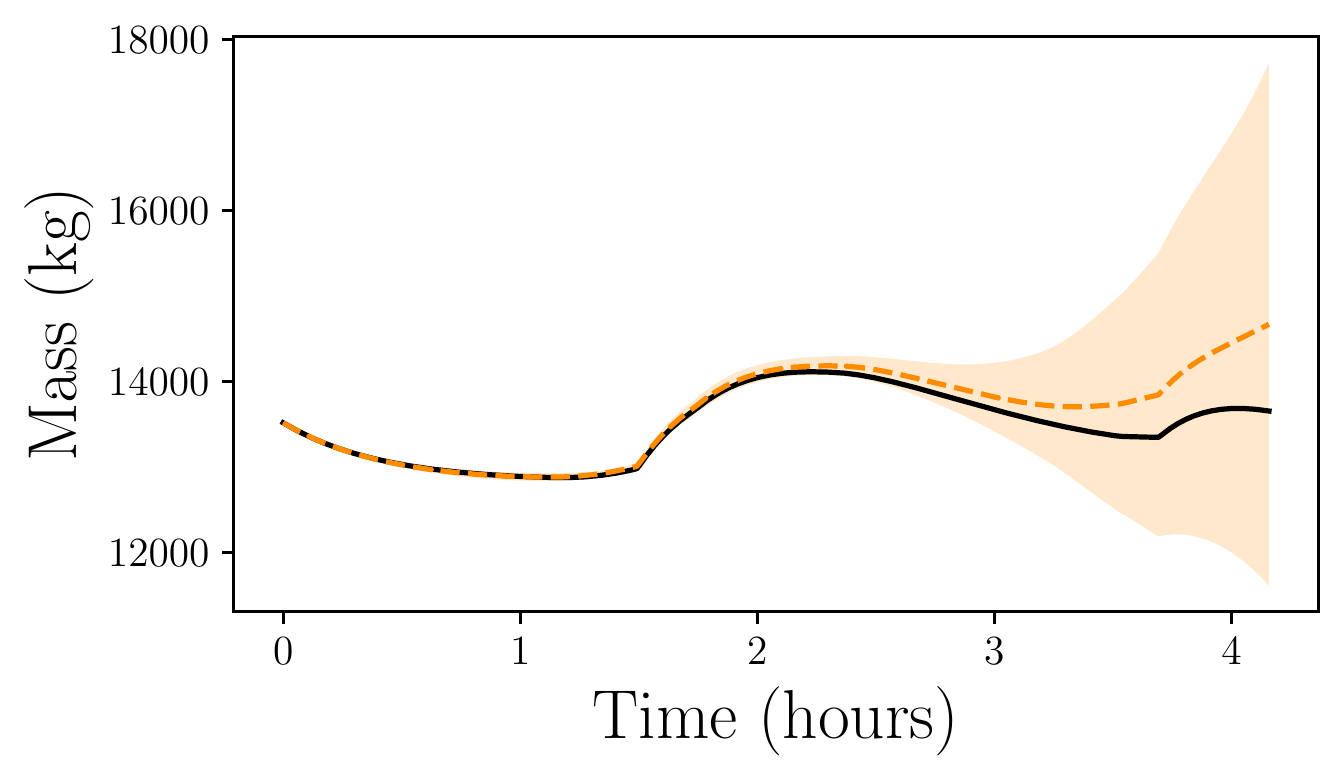}}
            \caption{Molten cryolite mass $x_4$}
            \label{subfig:Rolling_forecast_x4_dense}
        \end{subfigure}
        \begin{subfigure}[t]{0.32\linewidth}
            \raisebox{-\height}{\includegraphics[width=\linewidth]{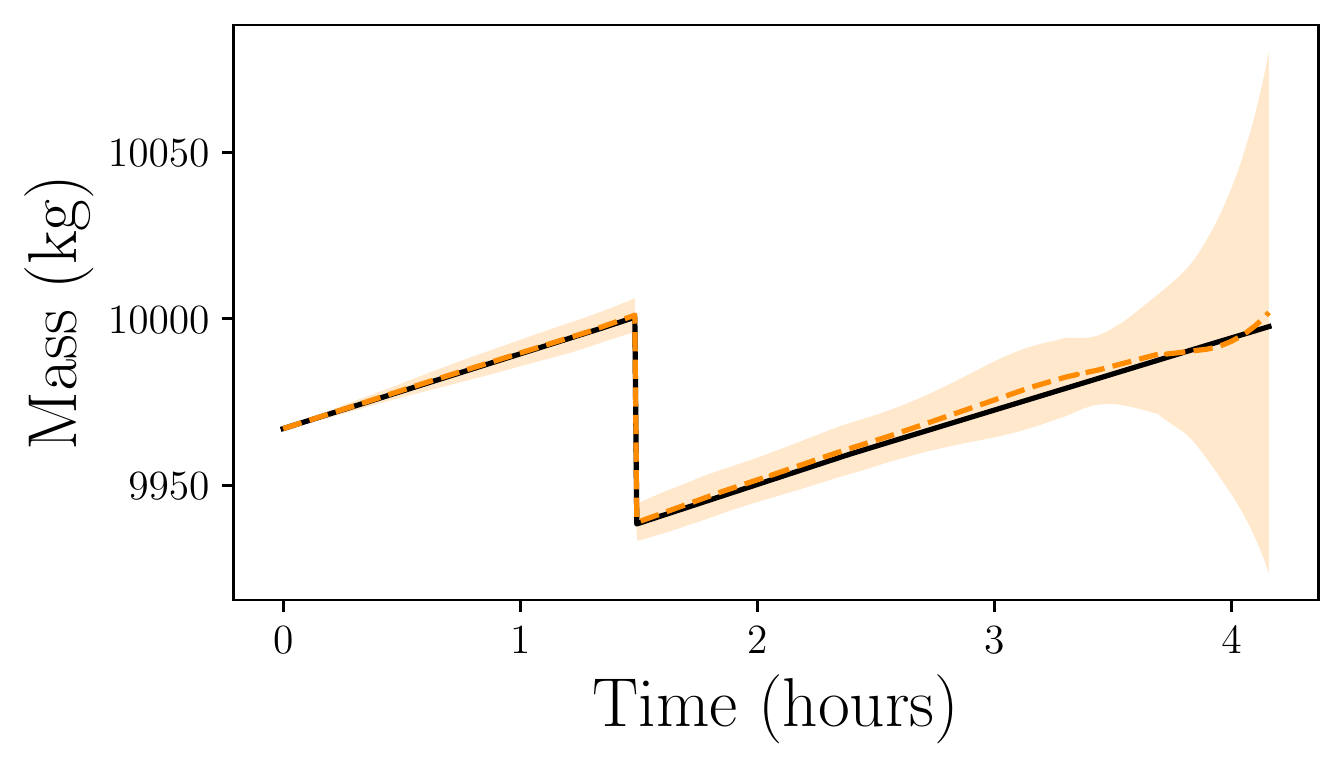}}
            \caption{Produced aluminum mass $x_5$}
            \label{subfig:Rolling_forecast_x5_dense}
        \end{subfigure}
        \begin{subfigure}[t]{0.32\linewidth}
            \raisebox{-\height}{\includegraphics[width=\linewidth]{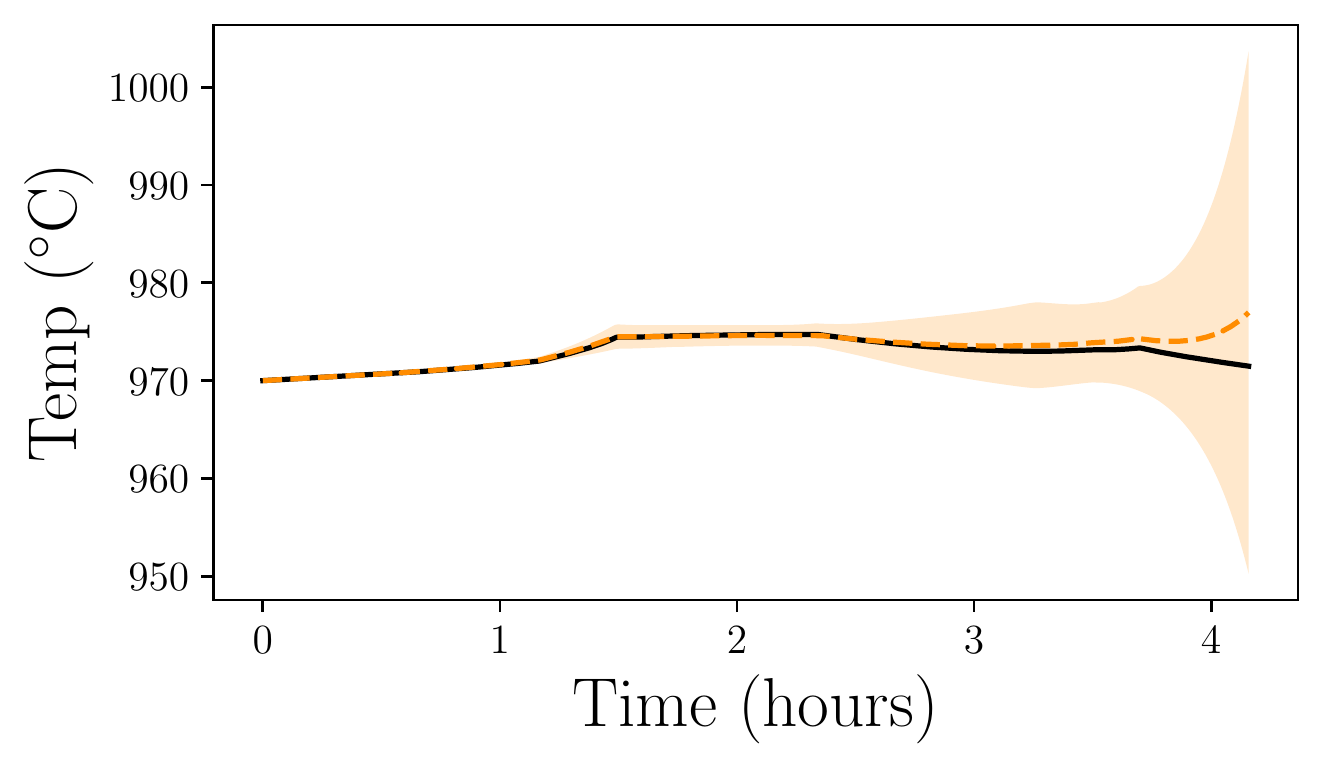}}
            \caption{Bath temperature $x_6$}
            \label{subfig:Rolling_forecast_x6_dense}
        \end{subfigure}
        \begin{subfigure}[t]{0.32\linewidth}
            \raisebox{-\height}{\includegraphics[width=\linewidth]{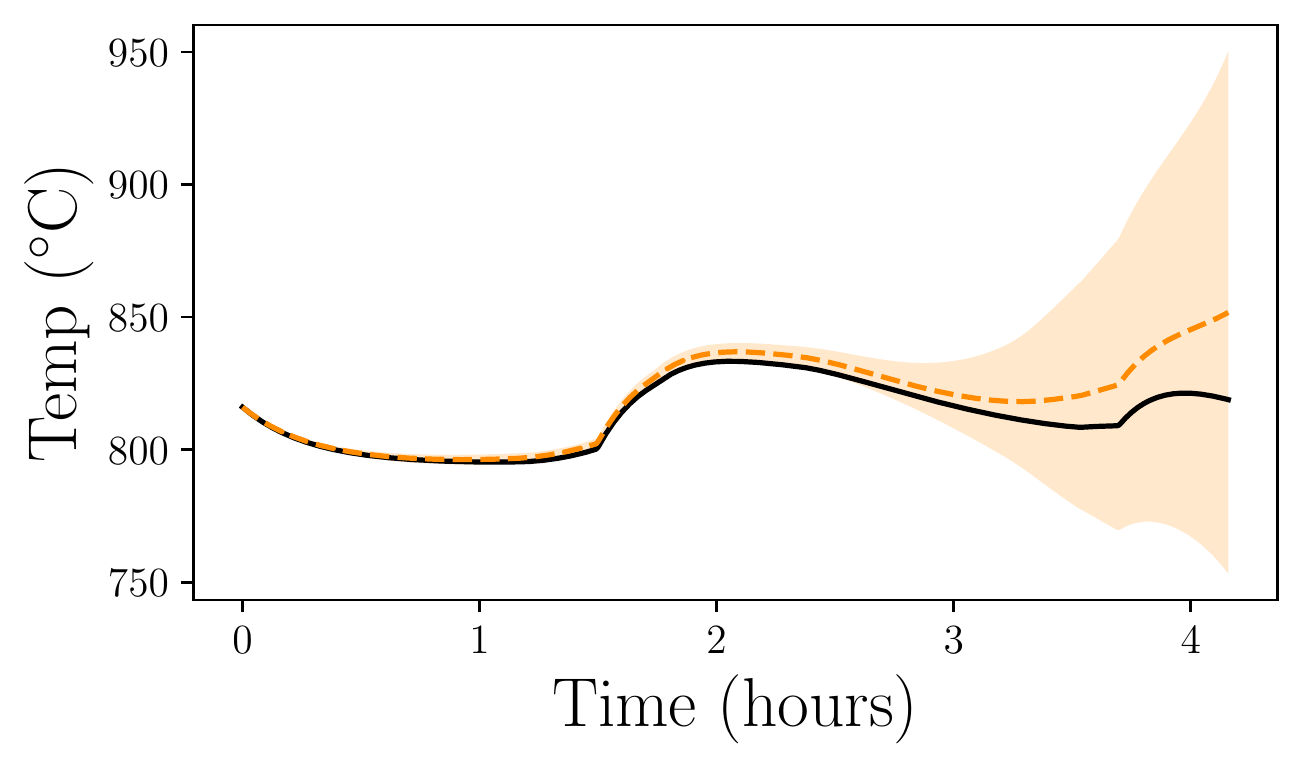}}
            \caption{Side ledge temperature $x_7$}
            \label{subfig:Rolling_forecast_x7_dense}
        \end{subfigure}
        \begin{subfigure}[t]{0.32\linewidth}
            \raisebox{-\height}{\includegraphics[width=\linewidth]{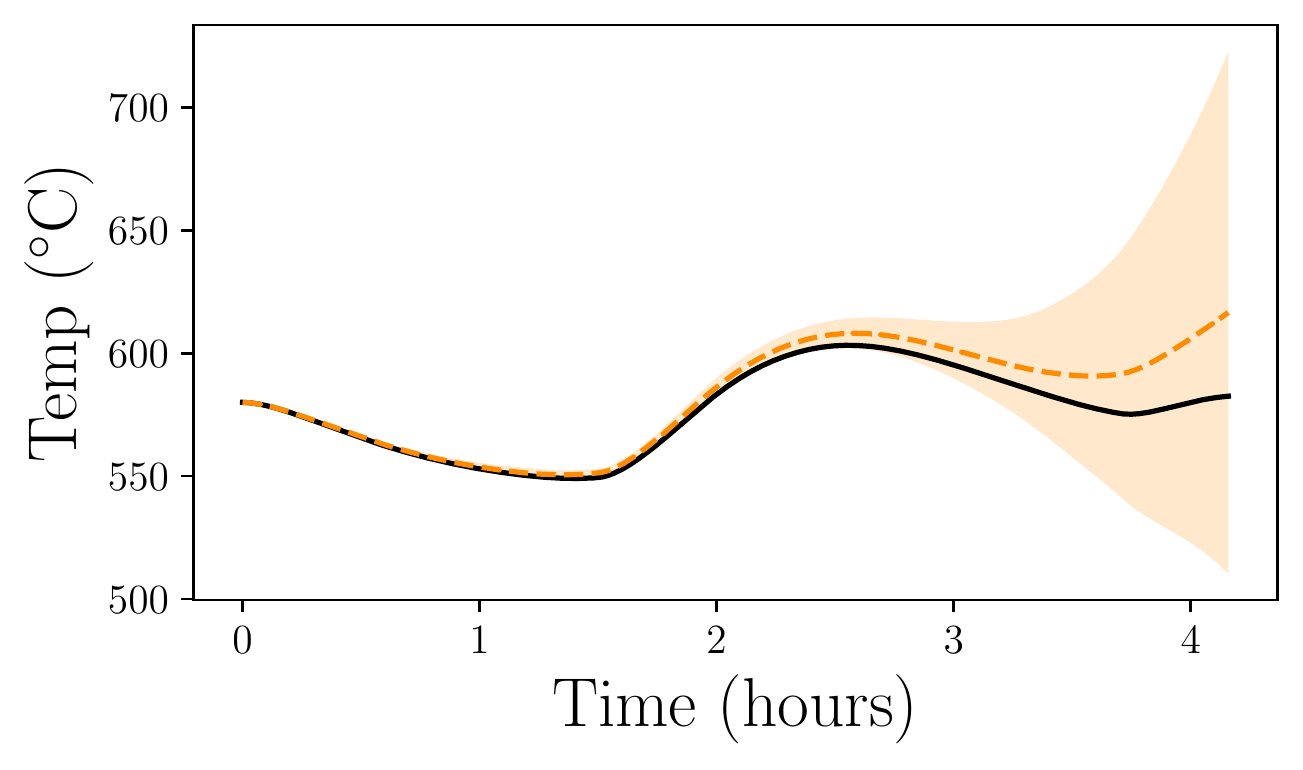}}
            \caption{Side wall temperature $x_8$}
            \label{subfig:Rolling_forecast_x8_dense}
        \end{subfigure}
        \\
        \begin{subfigure}[t]{0.32\linewidth}
            \raisebox{-\height}{\includegraphics[width=\linewidth]{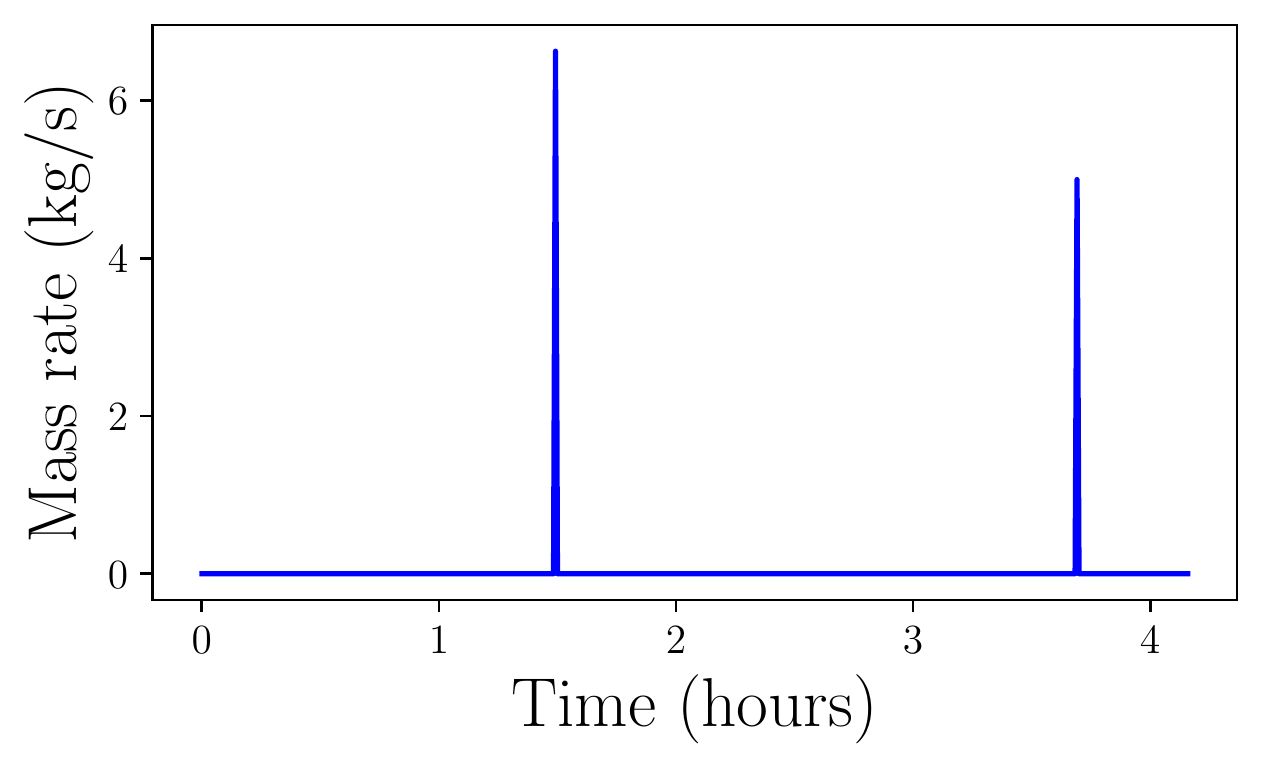}}
            \caption{Alumina feed $u_1$}
            \label{subfig:Rolling_forecast_u1_dense}
        \end{subfigure}
        \begin{subfigure}[t]{0.32\linewidth}
            \raisebox{-\height}{\includegraphics[width=\linewidth]{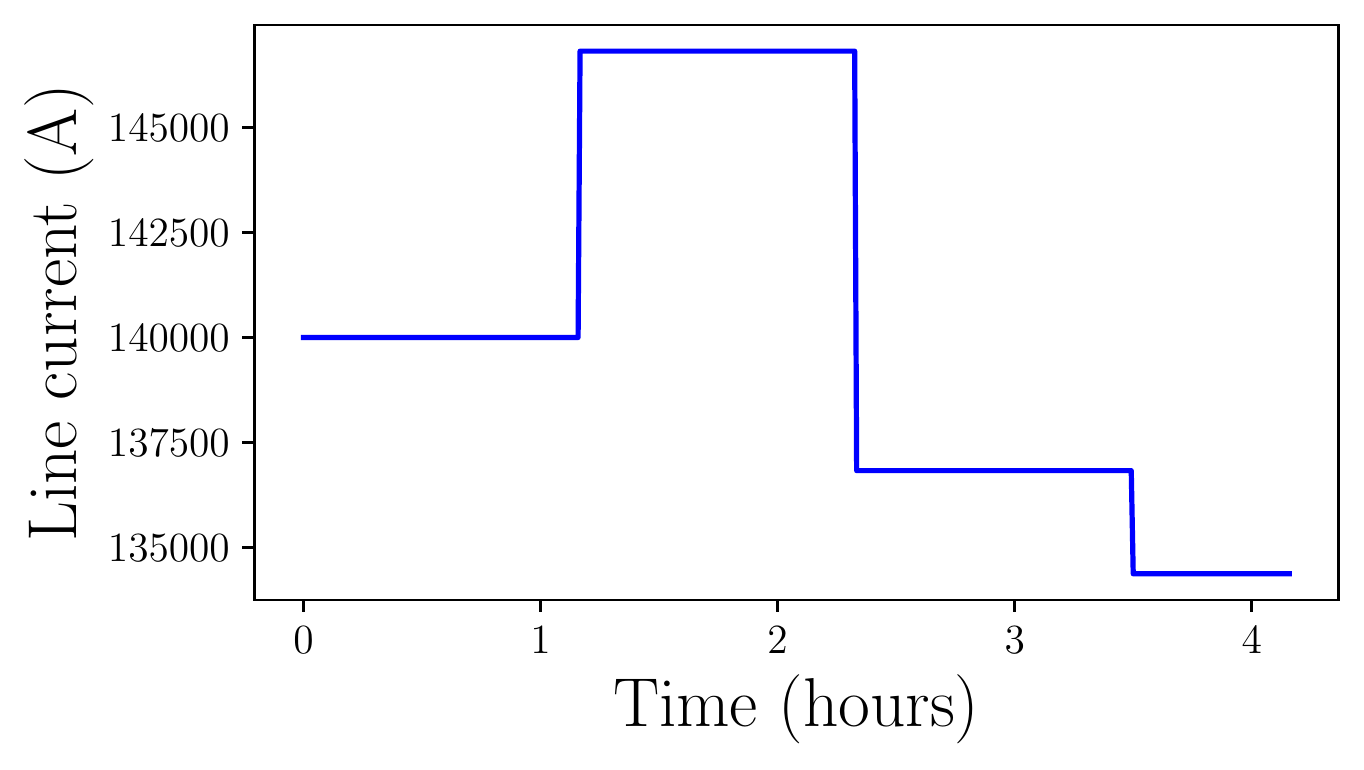}}
            \caption{Line current $u_2$}
            \label{subfig:Rolling_forecast_u2_dense}
        \end{subfigure}
        \begin{subfigure}[t]{0.32\linewidth}
            \raisebox{-\height}{\includegraphics[width=\linewidth]{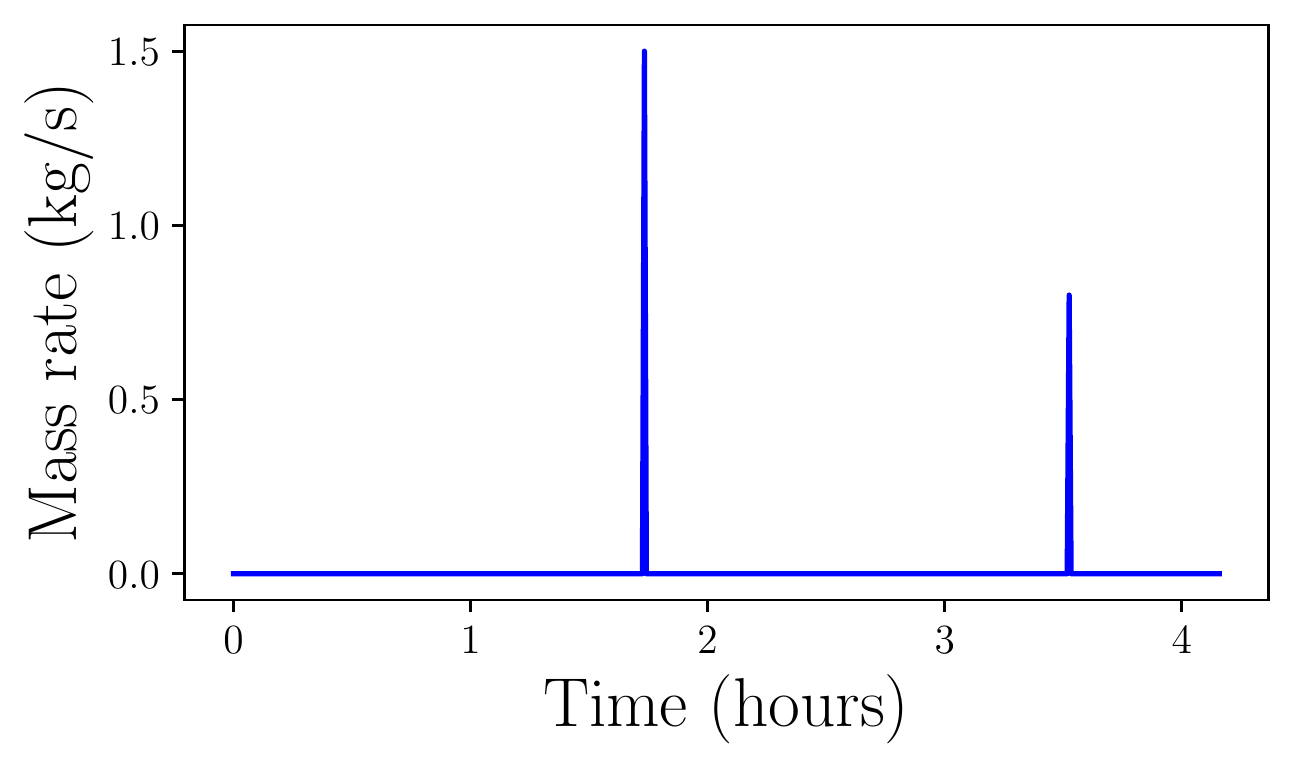}}
            \caption{Aluminum fluoride $u_3$}
            \label{subfig:Rolling_forecast_u3_dense}
        \end{subfigure}
        \begin{subfigure}[t]{0.32\linewidth}
            \raisebox{-\height}{\includegraphics[width=\linewidth]{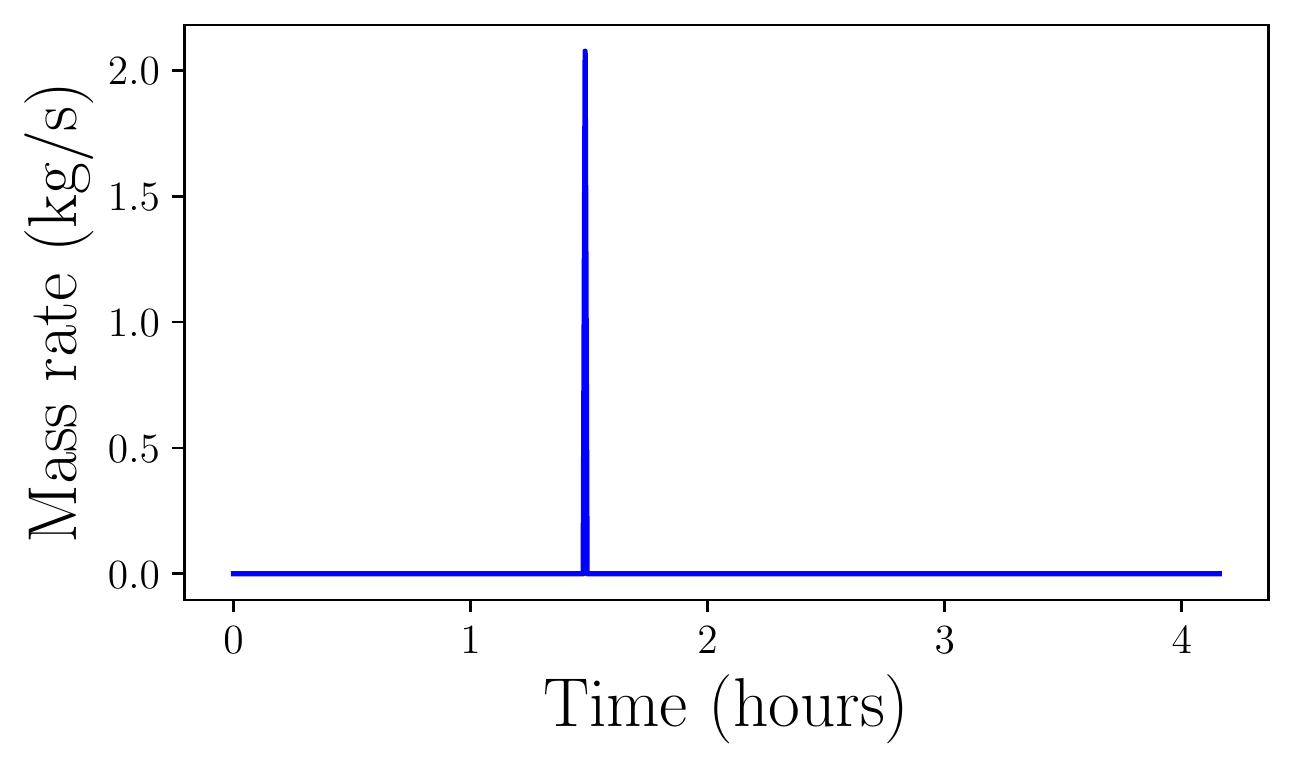}}
            \caption{Metal tapping $u_4$}
            \label{subfig:Rolling_forecast_u4_dense}
        \end{subfigure}
        \begin{subfigure}[t]{0.32\linewidth}
            \raisebox{-\height}{\includegraphics[width=\linewidth]{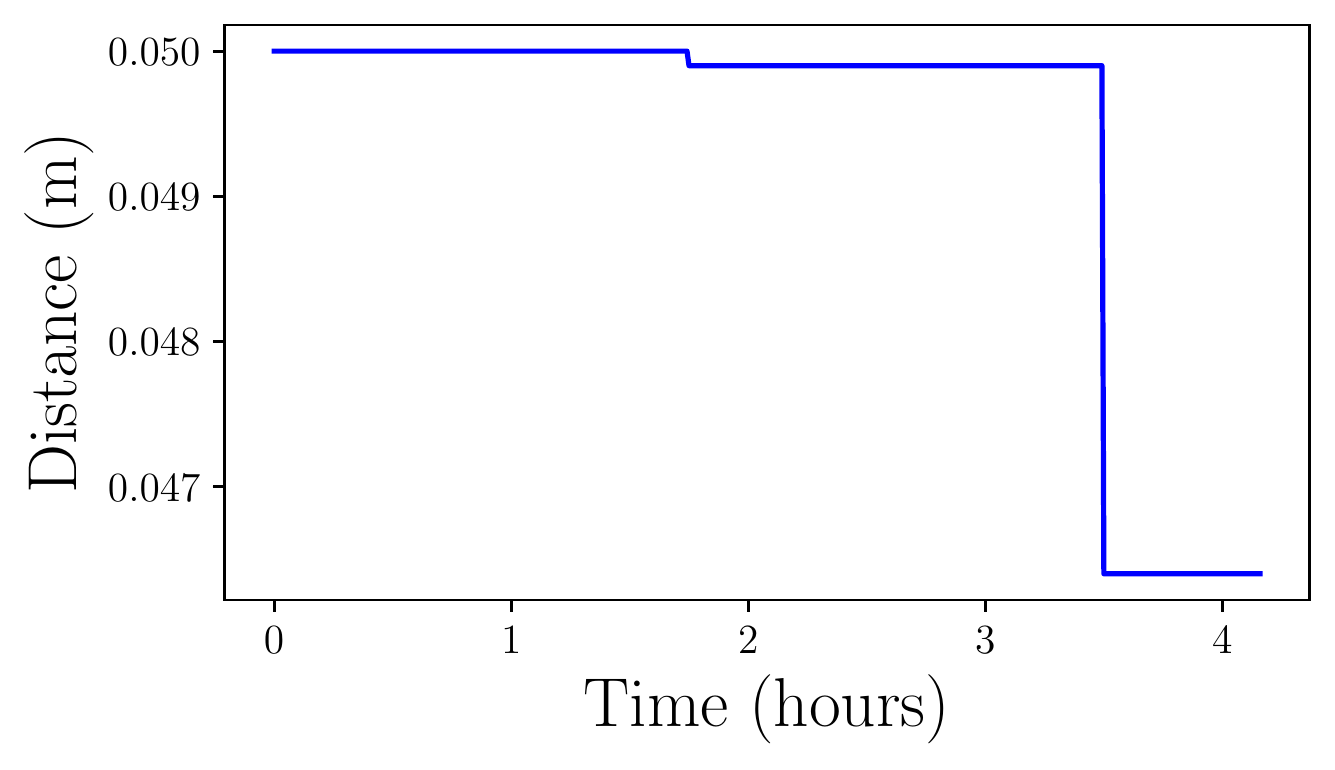}}
            \caption{Anode-cathode-distance $u_5$}
            \label{subfig:Rolling_forecast_u5_dense}
        \end{subfigure}

        \begin{adjustbox}{max width=1\linewidth}
        \begin{tikzpicture}
            \begin{customlegend}[legend columns=6,legend style={draw=none, column sep=2ex},legend entries={Truth,Mean sparse, control input, std. dev. sparse}]
                \addlegendimage{black,thick, sharp plot}
                \addlegendimage{Torange,thick, dashed,sharp plot}
                \addlegendimage{defaultblue,thick, sharp plot}
                \addlegendimage{Torange!40, fill=Torange!40, area legend}
            \end{customlegend}
        \end{tikzpicture}
    \end{adjustbox}
        \caption{Dense rolling forecast of state variables $\{x_1, .., x_8\}$ at each time instant. The true values of $\mathbf{x}$ and control inputs $\mathbf{u}$  are taken from one simulated set of test set trajectories $\mathbf{X}_i \in \mathbf{X}_{test}$. 
        The orange dashed line shows the average of 20 forecasts calculated by 20 dense neural network models with different parameter initialization. The orange band shows the standard deviation of the same 20 forecasts calculated by 20 dense models. The forecast is shown until some of the dense model estimates starts to diverge from true values.}
        \label{fig:Rolling_Forecast_dense}
    \end{figure}

    Fig.~\ref{fig:Rolling_Forecast_sparse} and Fig.~\ref{fig:Rolling_Forecast_dense} show the performance of the ensembles of 20 sparse and 20 dense DNN models with different parameter initialization forecasting the state variables $\mathbf{x}$  in one of the time series in the test set $\mathcal{S}_{test}(i) = \{\mathbf{X}_i\}$ as defined in Eq. (\ref{eq:test_set}). The models are trained on a data set $    \mathcal{S}_{train} = \left\{ \{\mathbf{X}_1,\;\mathbf{Y}_1\},\; \{\mathbf{X}_2,\;\mathbf{Y}_2\},\; ...,\;\{\mathbf{X}_{10},\;\mathbf{Y}_{10}\}\right\}$ consisting of ten time series $\{\mathbf{X}_1,\;..,\;\mathbf{X}_{10}\}$ with 999 time steps each.
    
    Fig.~\ref{fig:Rolling_Forecast_sparse} and Fig.~\ref{fig:Rolling_Forecast_dense} indicate that the forecasts of sparse and dense models are showing similar performance for the first time steps after they are given the initial conditions. However, while the forecasts calculated by sparse models show a consistently slow drift from the simulated values of $\mathbf{x}$, the mean and standard deviation of forecasts calculated by dense models suddenly drifts exponentially. The narrow banded standard deviation of sparse neural networks can indicate that these models converge to models with similar characteristics during training despite different parameter initialization. Furthermore, the consistently slow drift between the sparse DNN model forecast of $\mathbf{x}$ and the true values of $\mathbf{x}$ indicate that the sparse models are generalizing better as they are showing good forecasting capabilities in a broader region than the dense DNN models. 
     Fig.~\ref{fig:Rolling_Forecast_sparse} and Fig.~\ref{fig:Rolling_Forecast_dense} show some interesting results that indicate better generalization of sparse DNN models than dense DNN models and that the convergence of model parameters for sparse DNN models are more robust to random initialization than dense DNN models are. 
     
     \subsubsection{Impact of the training data quantity and prediction horizon}
     \label{subsubsec:impactofthetraining}
\begin{figure}[H]
    \centering
        	\begin{subfigure}{0.495\linewidth}
            	\centering 
        		\includegraphics[width=\linewidth]{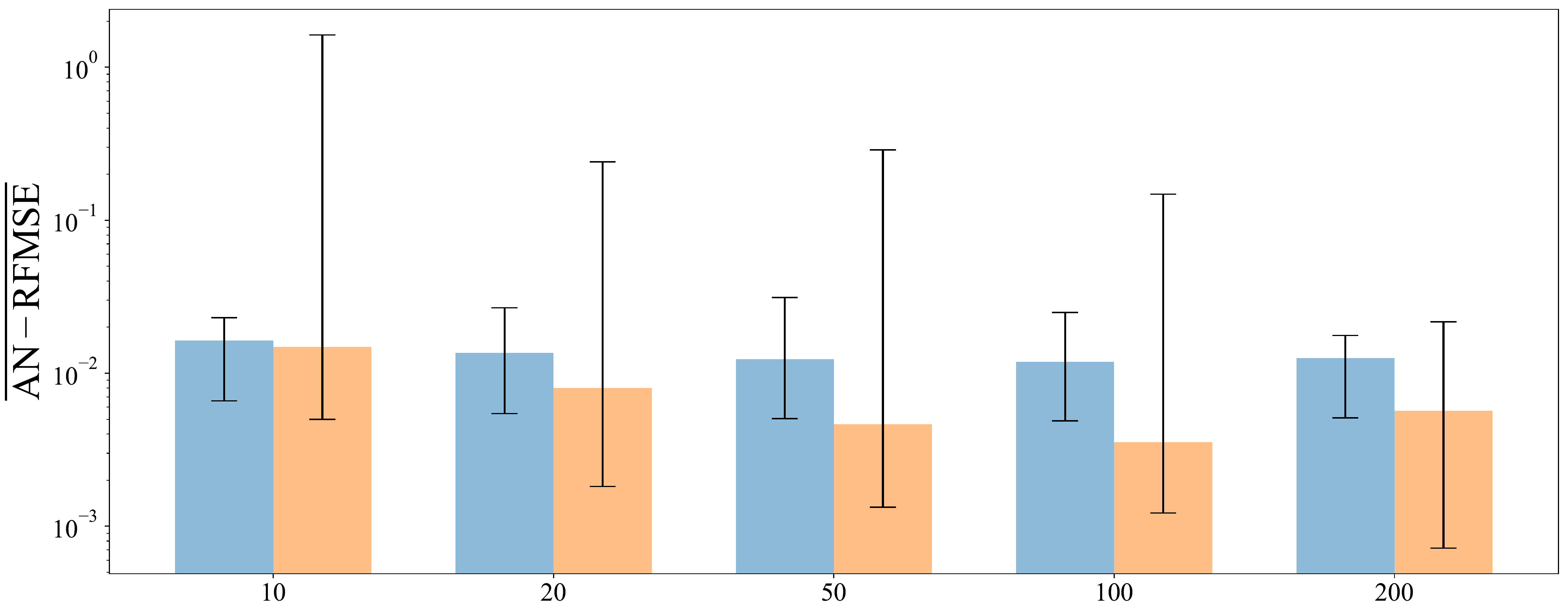}
        		\caption{200 $\Delta T$ of forecasting.}
        		\label{subfig:Norm_RFMSE200}
        	\end{subfigure}%
        	\begin{subfigure}{0.495\linewidth}
        		\centering 
        		\includegraphics[width=\linewidth]{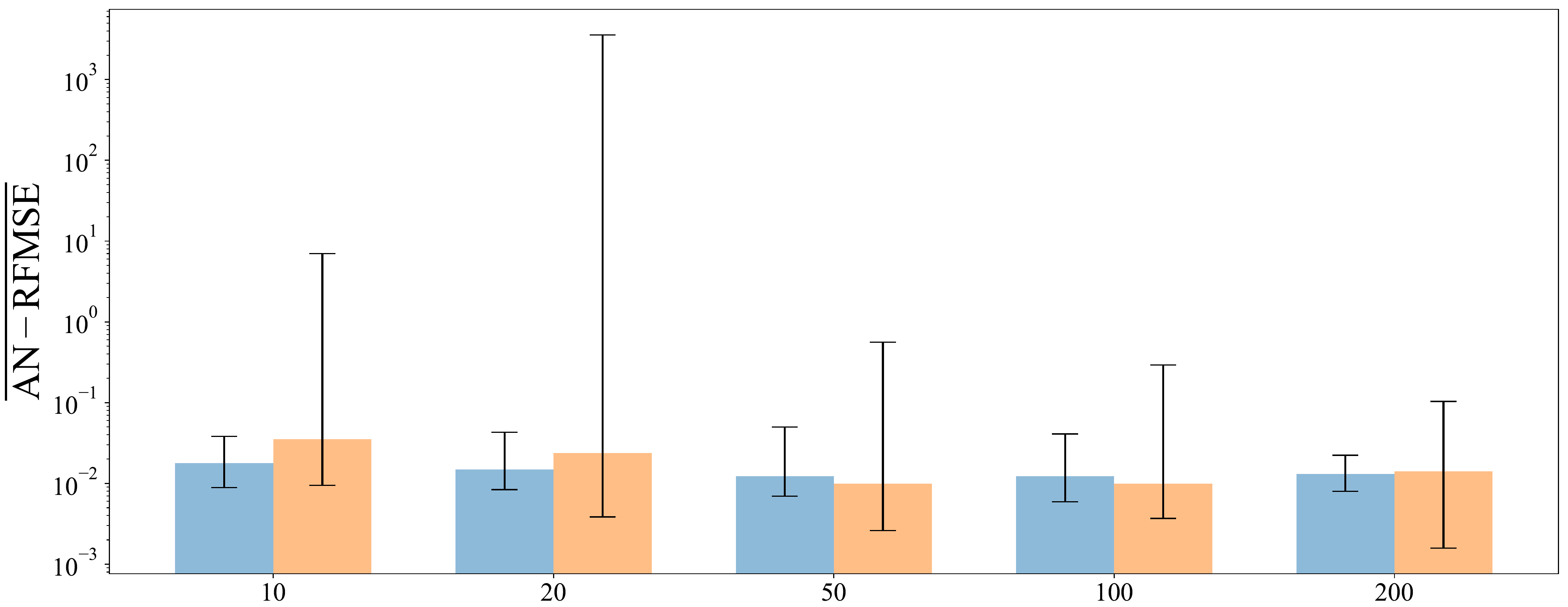}
        		\caption{300 $\Delta T$ of forecasting.}
        		\label{subfig:Norm_RFMSE300}
        	\end{subfigure}%
        	\\
        	\begin{subfigure}{0.495\linewidth}
        		\centering 
        		\includegraphics[width=\linewidth]{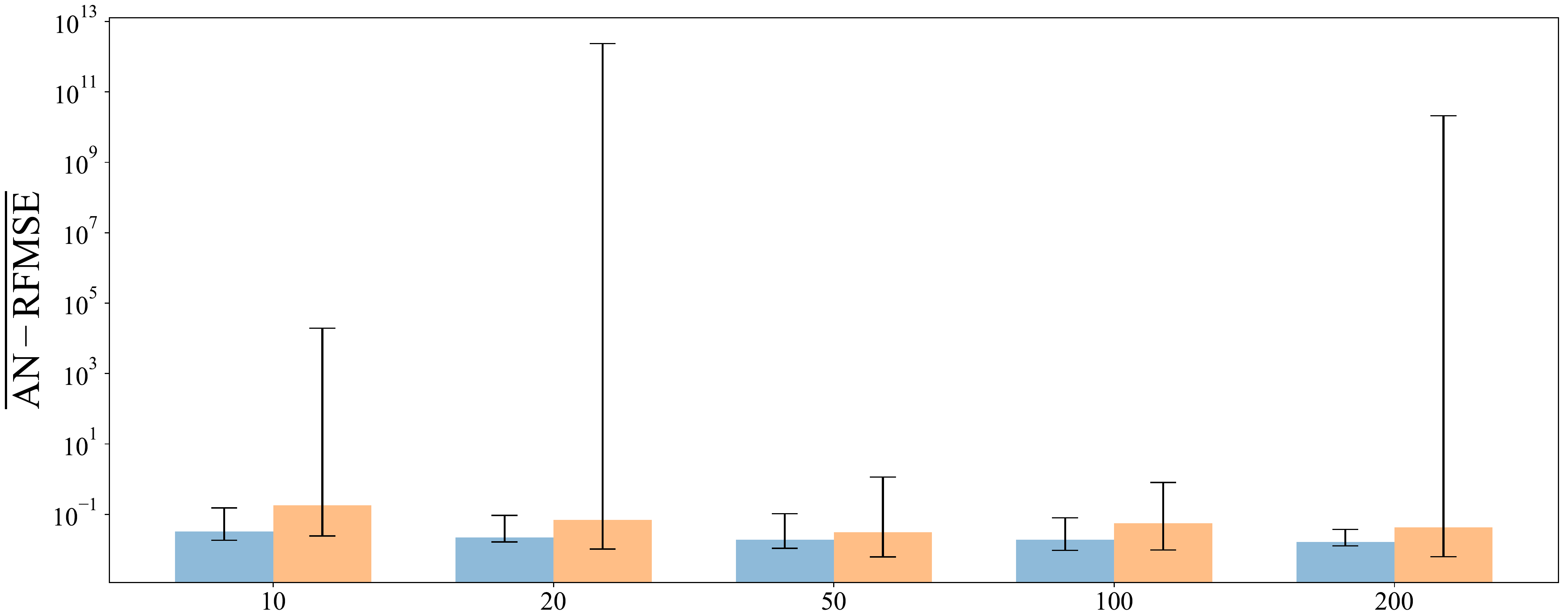}
        		\caption{500 $\Delta T$ of forecasting.}
        		\label{subfig:Norm_RFMSE500}
        	\end{subfigure}%
        	\begin{subfigure}{0.495\linewidth}
        		\centering 
        		\includegraphics[width=\linewidth]{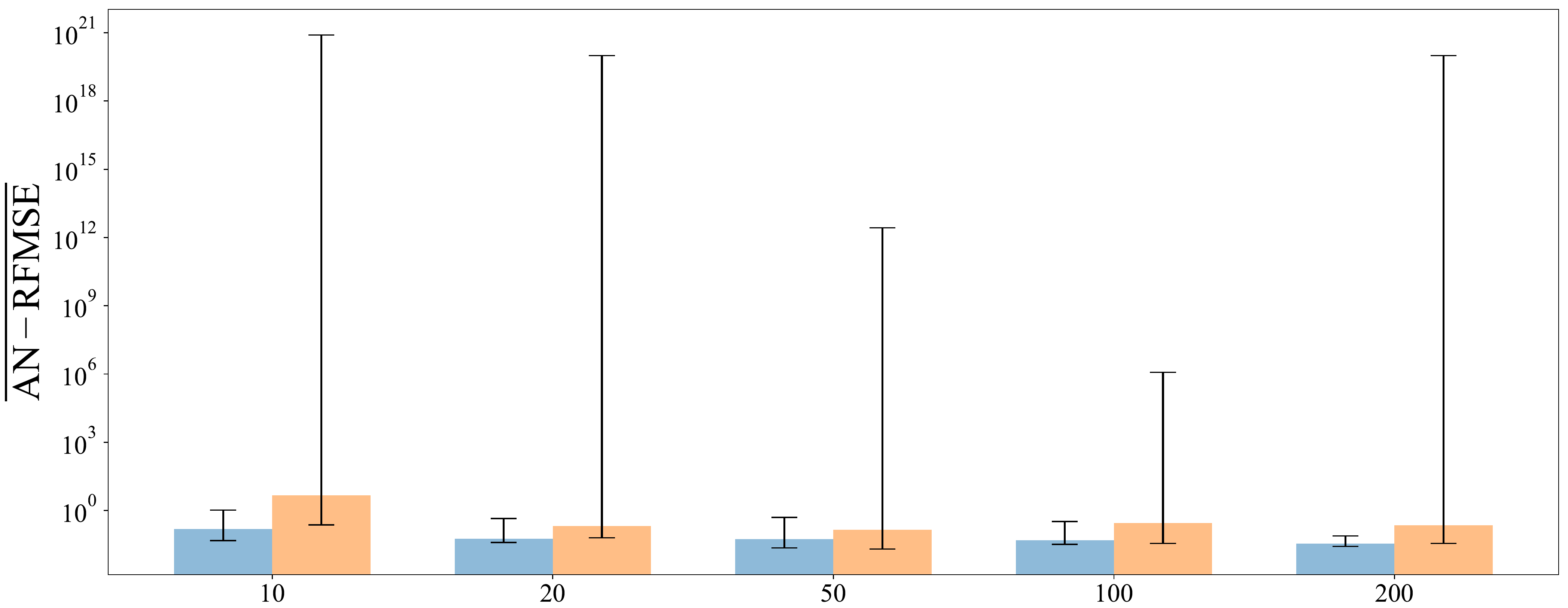}
        		\caption{1000 $\Delta T$ of forecasting.}
        		\label{subfig:Norm_RFMSE1000}
        	\end{subfigure}%
        	\\
        	 \begin{adjustbox}{max width=1\linewidth}
                    \begin{tikzpicture}
                        \begin{customlegend}[legend columns=6,legend style={draw=none, column sep=2ex},legend entries={Median $\overline{\textrm{AN-RFMSE}}$ sparse, Median $\overline{\textrm{AN-RFMSE}}$ dense}]
                            \addlegendimage{defaultblue!60, fill=defaultblue!60, area legend}
                            \addlegendimage{defaultorange!80, fill=defaultorange!80, area legend}
                        \end{customlegend}
                    \end{tikzpicture}
                \end{adjustbox}
        	\caption{Median, maximum and minimum $\overline{\textrm{AN-RFMSE}}$. There are five ensembles of models of both sparse and dense DNNs trained on data sets with different sizes. For each ensemble of models, there is a corresponding colored bar indicating median values and an error bar indicating maximum and minimum values of the vector $\overline{\textrm{AN-RFMSE}}_{vec}$. Moreover, the size of the training sets are indicated on the x-axis of the subplots. The blue bar shows the median $\overline{\textrm{AN-RFMSE}}$ among 20 sparse DNN models over 20 test sets. The orange bar shows the median $\overline{\textrm{AN-RFMSE}}$ among 20 dense DNN models over 20 test sets. For each subfigure, the $\overline{\textrm{AN-RFMSE}}$ are calculated for a given number of timesteps reported in the captions of each subfigure (\ref{subfig:Norm_RFMSE200} - \ref{subfig:Norm_RFMSE1000}). Notice the logarithmic scale of the y-axis. The x-axis is in thousand.
        	}
        	\label{fig:RFMSE_histogram} 
        \end{figure}

         Fig.~\ref{fig:RFMSE_histogram} shows the median, maximum and minimum elements of the $\overline{\textrm{AN-RFMSE}}_{vec}$ vector. 
        Fig.~\ref{fig:RFMSE_histogram} contains a good amount of information about the model performance of dense DNN models and  the sparse DNN models with weight penalty $\lambda_{\ell_1}=1e^{-2}$. Figs. \ref{subfig:Norm_RFMSE200} to \ref{subfig:Norm_RFMSE1000} report median and extreme values of $\overline{\textrm{AN-RFMSE}}_{vec}$ over four different time horizons for five ensembles of models trained on five data sets with different sizes. Hence, the results in Fig.~\ref{fig:RFMSE_histogram} show how dense and sparse DNN models perform with varying amounts of training data over varying time horizons. Figs.~\ref{subfig:Norm_RFMSE200} to \ref{subfig:Norm_RFMSE1000} show that the ensembles of \textbf{sparse models}  trained on data sets with varying size show similar results, both in terms of median $\overline{\textrm{AN-RFMSE}}$ and extreme values. However, as Fig.~\ref{subfig:Norm_RFMSE1000} shows, there seems to be a small trend that model ensembles trained with more data perform slightly better over longer forecasting horizons. Furthermore, the band between the minimum and maximum values of $\overline{\textrm{AN-RFMSE}}$ is overall relatively small for all ensembles of models and all forecast horizons for sparse models. The converging behavior of the performance of ensembles of sparse models as a function of the amount of data in the training set indicates that only small amounts of data are required to gain significance for the model parameters. While the sparse models show stable performance across ensembles of models with different amounts of training data and slowly increasing values of $\overline{\textrm{AN-RFMSE}}$ proportional to the length of the forecasting horizon, the same cannot be said about the performance of the dense models. When considering the \textbf{dense models}, Figs.~\ref{subfig:Norm_RFMSE200} to \ref{subfig:Norm_RFMSE1000} indicate that there is a trend where both median, minimum and maximum values of $\overline{\textrm{AN-RFMSE}}$ decreases significantly as sizes of training set decreases. This expected trend indicates that the performance improves with increasing dataset size. However, the trend is not consistent for all ensembles of dense DNN models for all forecasting horizons. Furthermore, the maximum values of $\overline{\textrm{AN-RFMSE}}$ for ensembles of dense DNN models for longer forecasting horizons such as in Fig.~\ref{subfig:Norm_RFMSE500} and Fig.~\ref{subfig:Norm_RFMSE1000} show that $\overline{\textrm{AN-RFMSE}}$ exponentially increases for some of the models within the ensemble. This indicates that the dense DNN models are likely to have some input regions where the model output is not sound. If the model estimate enters a region with poorly modeled dynamics, the model estimate might drift exponentially. For short-term prediction, that is in Fig.~\ref{subfig:Norm_RFMSE200} and Fig.~\ref{subfig:Norm_RFMSE300}, the trend is that dense models show better performance for median and minimum values than sparse models, especially within the ensembles with large training sets. This may be because dense models have more flexibility in terms of more parameters. However, it is important to state that the weights of the sparse models evaluated in Fig.~\ref{fig:RFMSE_histogram} are especially hard penalized, indicating that the flexibility of these models are limited.  For longer forecasting horizons (Fig.~\ref{subfig:Norm_RFMSE500} and Fig.~\ref{subfig:Norm_RFMSE1000}), sparse models are always showing better performance than dense models in terms of median $\overline{\textrm{AN-RFMSE}}$. This is a typical example of a bias-variance trade-off. For all forecasting horizons and within all groups of training set sizes, sparse models are always showing a smaller maximum value of $\overline{\textrm{AN-RFMSE}}$.  
        \\
        While it is valuable to have models that can give reasonable estimates in the long term, short-term prediction accuracy can be given extra attention since the models typically perform best on shorter horizons, and can therefore be used more aggressively for optimal control. As observed above, dense models tend to give better median accuracy than sparse models with hard $\ell_1$ regularization in shorter prediction horizons if trained on larger datasets. We, therefore, conducted a study where we compared dense models with sparse models trained with different levels of weight regularization with different sizes of the training set.
        \begin{figure}
            \centering
            \begin{subfigure}{0.495\linewidth}
            \centering
            \includegraphics[width=\linewidth]{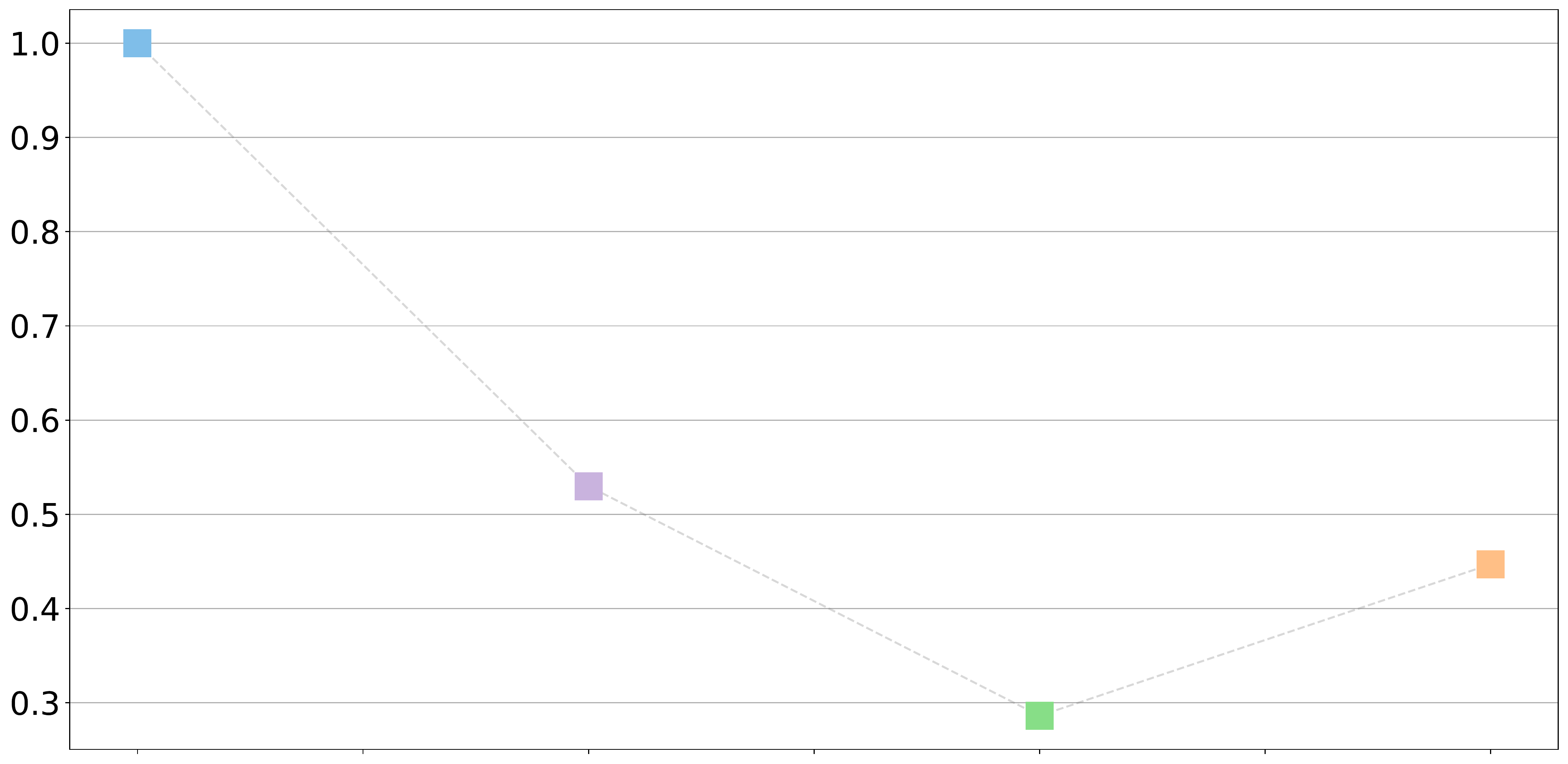}
            \caption{25 000 training datapoints}
            \label{subfig:short_pred_hor_Dset1}
            \end{subfigure}
            \begin{subfigure}{0.495\linewidth}
            \centering
            \includegraphics[width=\linewidth]{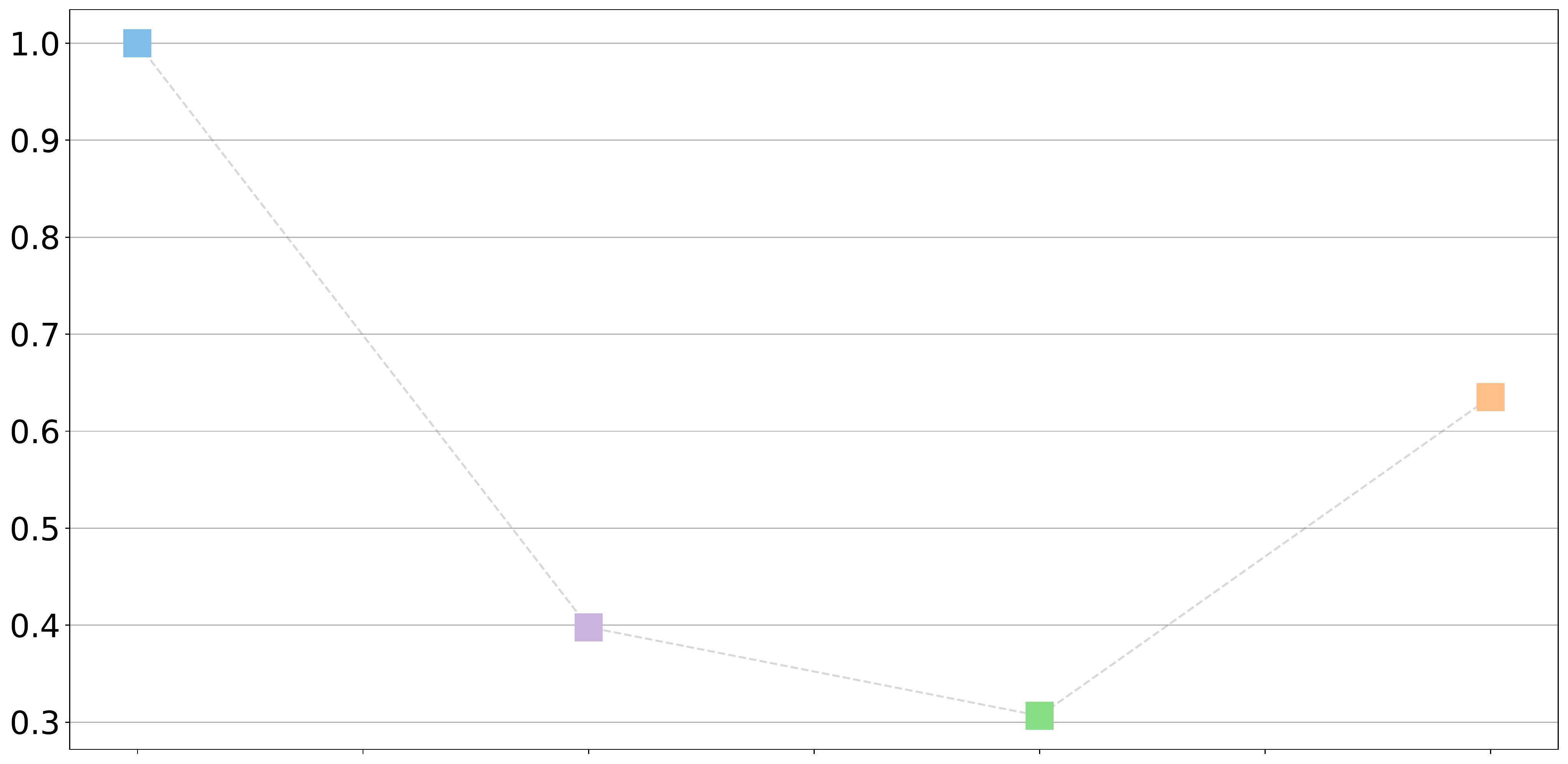}
            \caption{50 000 training datapoints}
            \label{subfig:short_pred_hor_Dset2}
            \end{subfigure}
            \\
            \begin{subfigure}{0.495\linewidth}
            \centering
            \includegraphics[width=\linewidth]{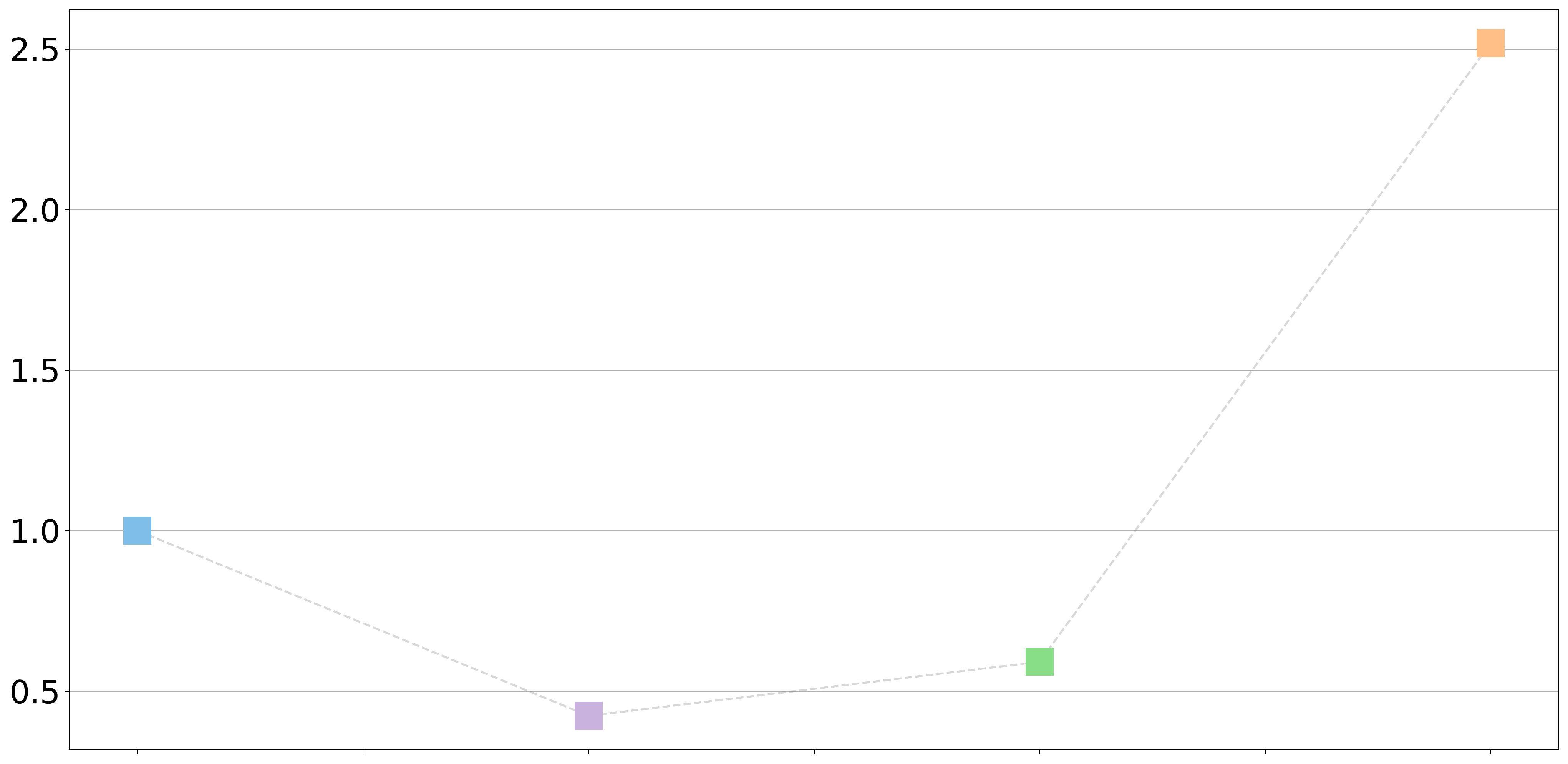}
            \caption{100 000 training datapoints}
            \label{subfig:short_pred_hor_Dset3}
            \end{subfigure}
            \begin{subfigure}{0.495\linewidth}
            \centering
            \includegraphics[width=\linewidth]{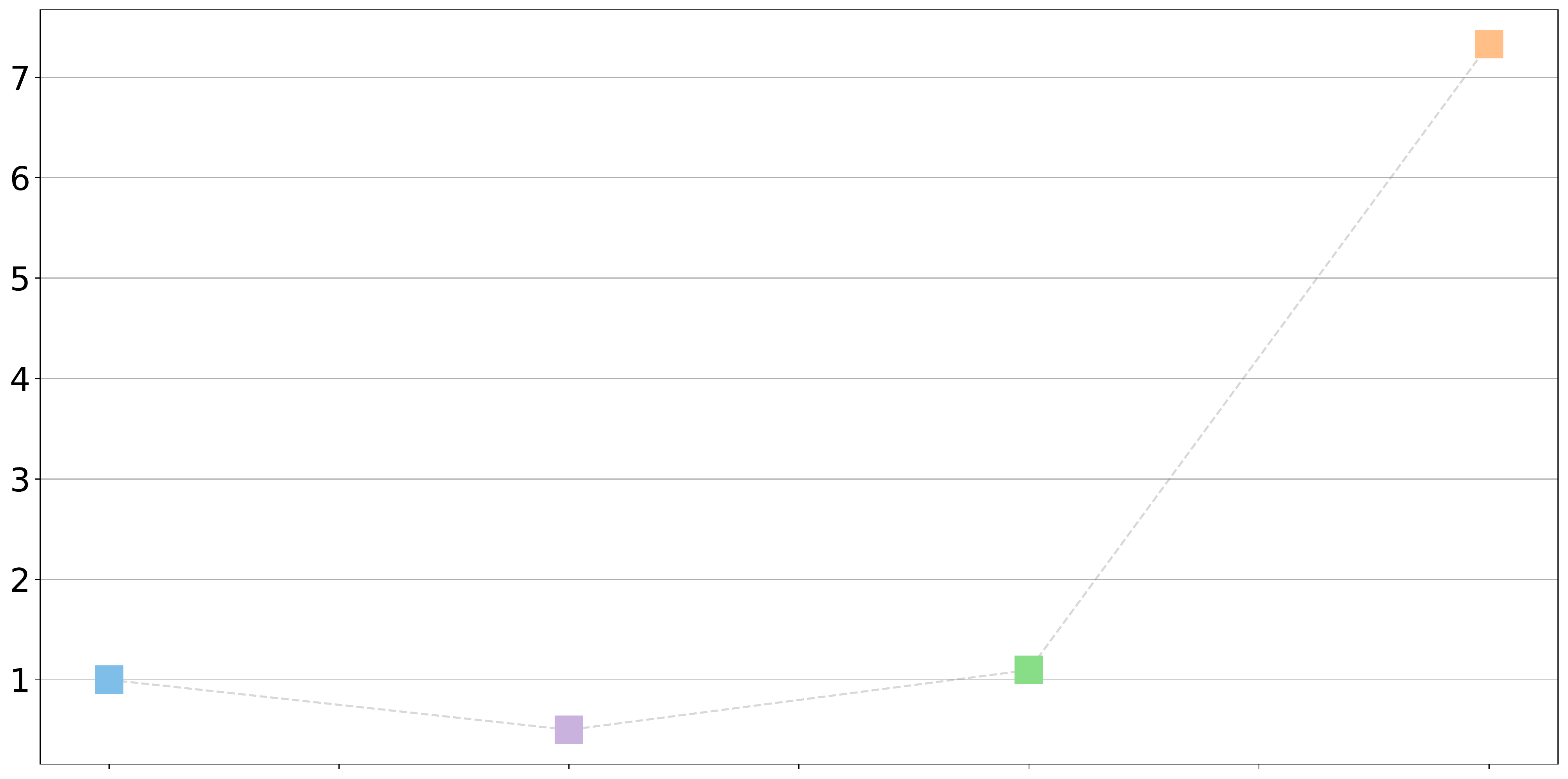}
            \caption{200 000 training datapoints}
            \label{subfig:short_pred_hor_Dset4}
            \end{subfigure}
            
            \begin{adjustbox}{max width=\linewidth}
            \begin{tikzpicture}
            \begin{customlegend}[legend columns=4,legend style={draw=none, column sep=1ex},legend entries={$\lambda_{\ell_1}=0$ (Dense),$\lambda_{\ell_1}=1e^{-4}$,$\lambda_{\ell_1}=1e^{-3}$,$\lambda_{\ell_1}=1e^{-2}$}]
                \addlegendimage{Tblue!50, fill=Tblue!50,area legend}
                \addlegendimage{Tpurple!50, fill=Tpurple!50,area legend}
                \addlegendimage{Tgreen!50, fill=Tgreen!50,area legend}
                \addlegendimage{Torange!50, fill=Torange!50,area legend}
            \end{customlegend}
            \end{tikzpicture}
            \end{adjustbox}
            \caption{Median prediction error ratio for short prediction horizon. All median prediction errors are divided by the median prediction error of dense models. Thus, dense models will always have a prediction error ratio of 1.}
            \label{fig:Short_pred_horizon}
        \end{figure}
        Fig.~\ref{fig:Short_pred_horizon} show the mean prediction accuracy of an ensemble of dense models and three different ensembles of sparse models with different degree of weight penalization (namely $\lambda_{\ell_1}=1e^{-4}$, $\lambda_{\ell_1}=1e^{-3}$ and $\lambda_{\ell_1}=1e^{-2}$). Each ensemble consists of 20 models, and each of the models forecast 200 timesteps of 50 different test trajectories.
        The median accuracy is expressed in terms a ratio called median prediction error ratio. This ratio is given by of the ratio between the median prediction error of the given ensemble of models and the median prediction error of the ensemble of dense models. This means that the median prediction error ratio of dense models is always one, and that smaller ratio indicates higher median accuracy. 
        For low and medium data sizes (Fig.~\ref{subfig:short_pred_hor_Dset1} and \ref{subfig:short_pred_hor_Dset2}), the median accuracy of all ensembles of sparsely regularized models are more accurate then dense models also in the short term. In this data regime, the different sparse models show similar prediction capabilities. In Fig.~\ref{subfig:short_pred_hor_Dset3}, the median prediction error ratio for models trained on 100 000 data points are plotted. In this data regime, the median accuracy ratio between medium sparse models ($\lambda_{\ell_1}=1e^{-3}$ and  $\lambda_{\ell_1}=1e^{-4}$) and dense models remains quite similar to the median accuracy ratio in the between dense and medium sparse models in the low data regime. Moreover, the prediction error ratio of very sparse models ($\lambda_{\ell_1}=1e^{-2}$) and the other models increase significantly. This is probably because the model accuracy of very sparse models converges at low data limits. In contrast, medium sparse and dense models can exploit their nonlinear prediction capabilities when large amounts of training data are available. In Fig.~\ref{subfig:short_pred_hor_Dset4}, median prediction error ratios for models trained on 200 000 data samples are presented. At this point, sparse models trained with $\lambda_{\ell_1}=1e^{-4}$ has a median model prediction error ratio of 0.5, sparse models trained with $\lambda_{\ell_1}=1e^{-3}$ has a prediction error ratio of 1.1 and sparse models trained with $\lambda_{\ell_1}=1e^{-2}$ has a prediction error ratio of approximately 7. In this large data regime, it seems clear that the medium sparse, and dense models are very accurate. In contrast, the accuracy of very sparse models converges for small training datasets. Furthermore, the sparse model trained with the smallest sparsity regularization parameter ($\lambda_{\ell_1}=1e^{-4}$) still outperforms the dense models. This study illustrates that dense models require enormous amounts of data to outperform sparse models.
        
        To quantify the exponential drift of model estimates for different levels of sparsity, we run a test on ensembles of models trained on datasets with different data sizes.
        
        \begin{figure}[H]
    \centering
        	\begin{subfigure}{0.495\linewidth}
            	\centering 
        		\includegraphics[width=\linewidth]{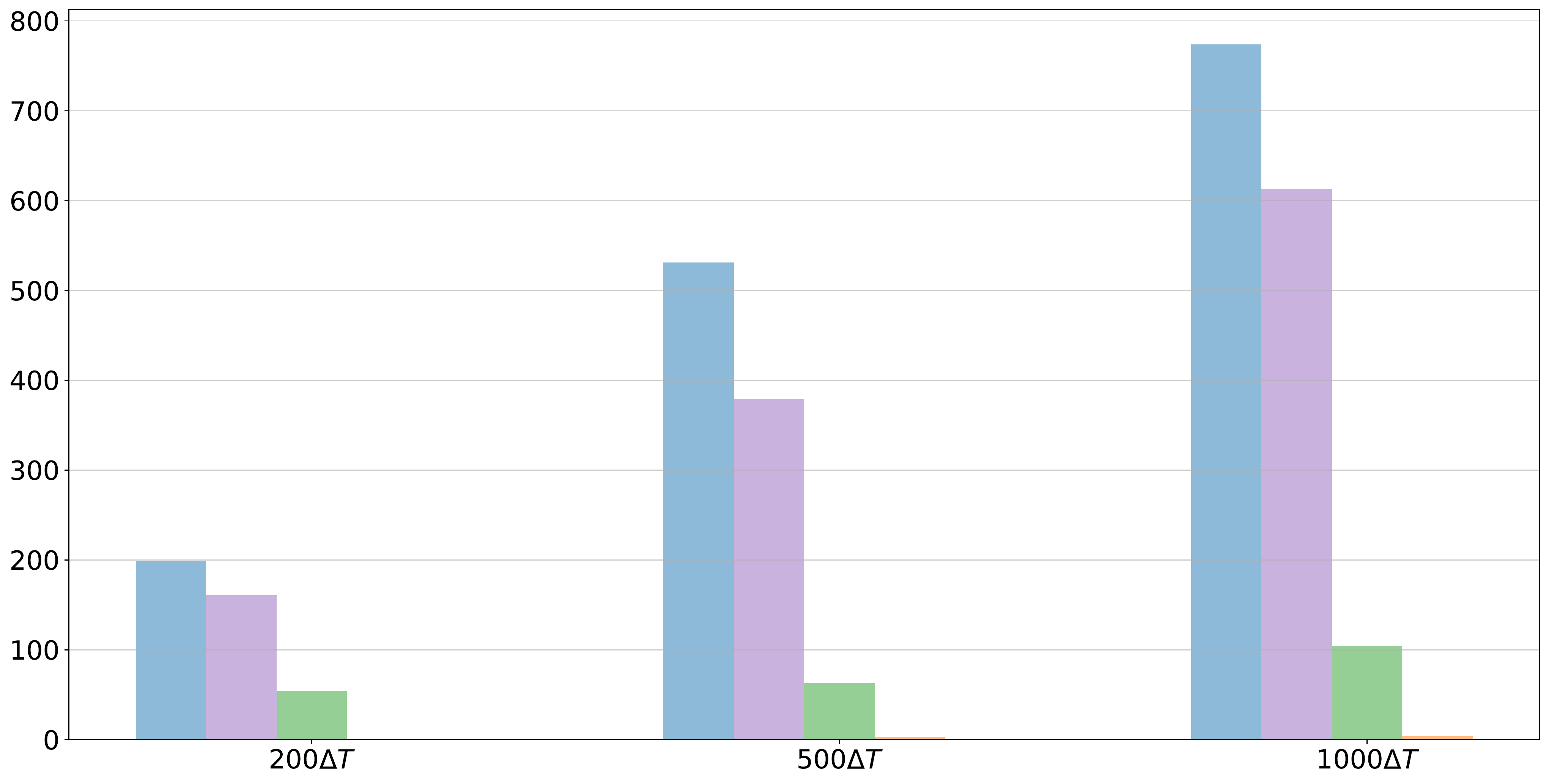}
        		\caption{25 000 training data}
        		\label{subfig:BlowUpData1}
        	\end{subfigure}%
        	\begin{subfigure}{0.495\linewidth}
        		\centering 
        		\includegraphics[width=\linewidth]{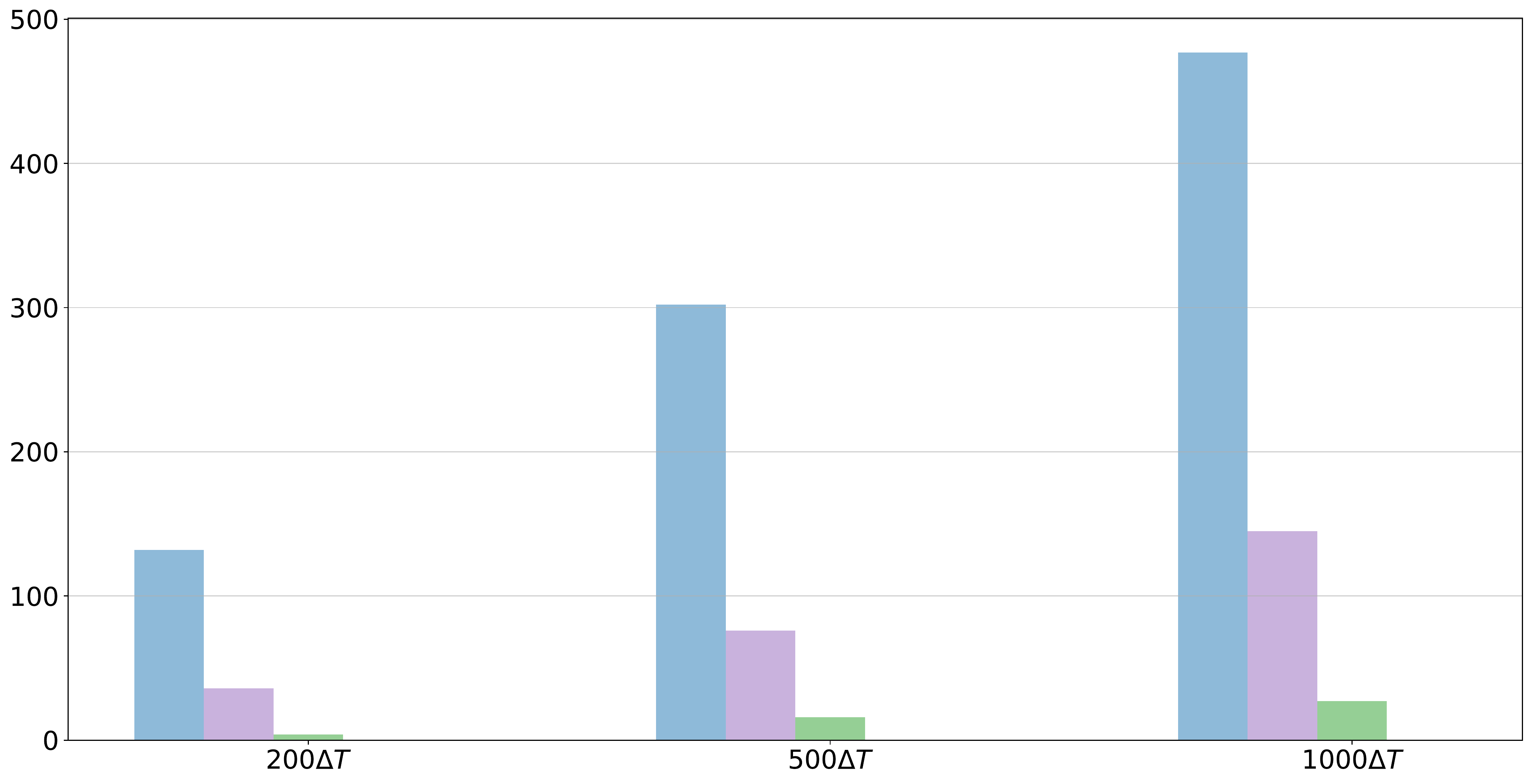}
        		\caption{50 000 trainig data}
        		\label{subfig:BlowUpData2}
        	\end{subfigure}%
        		\\
        	\begin{subfigure}{0.495\linewidth}
        		\centering 
        		\includegraphics[width=\linewidth]{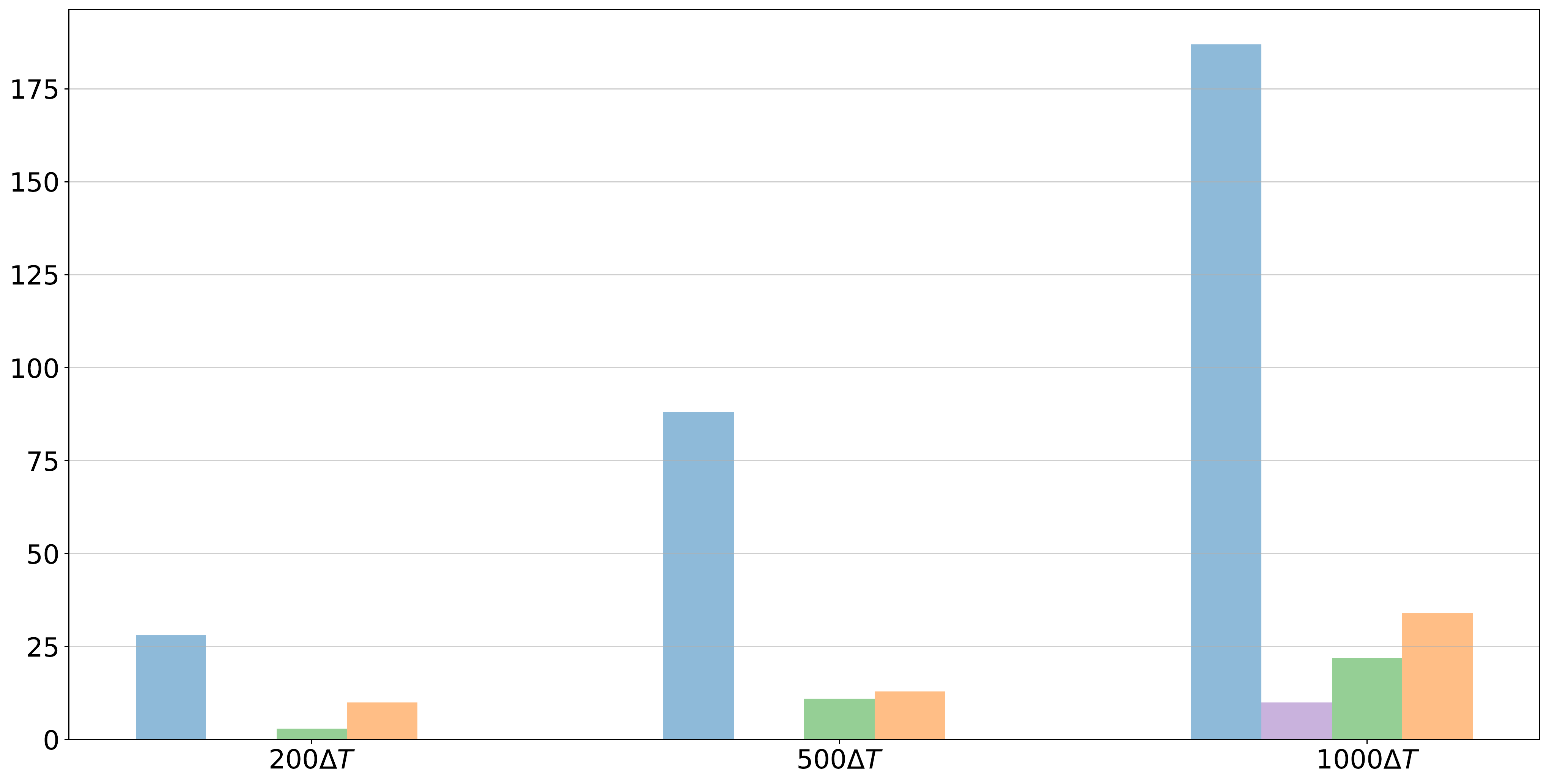}
        		\caption{100 000 training data}
        		\label{subfig:BlowUpData3}
        	\end{subfigure}%
        	\begin{subfigure}{0.495\linewidth}
        		\centering 
        		\includegraphics[width=\linewidth]{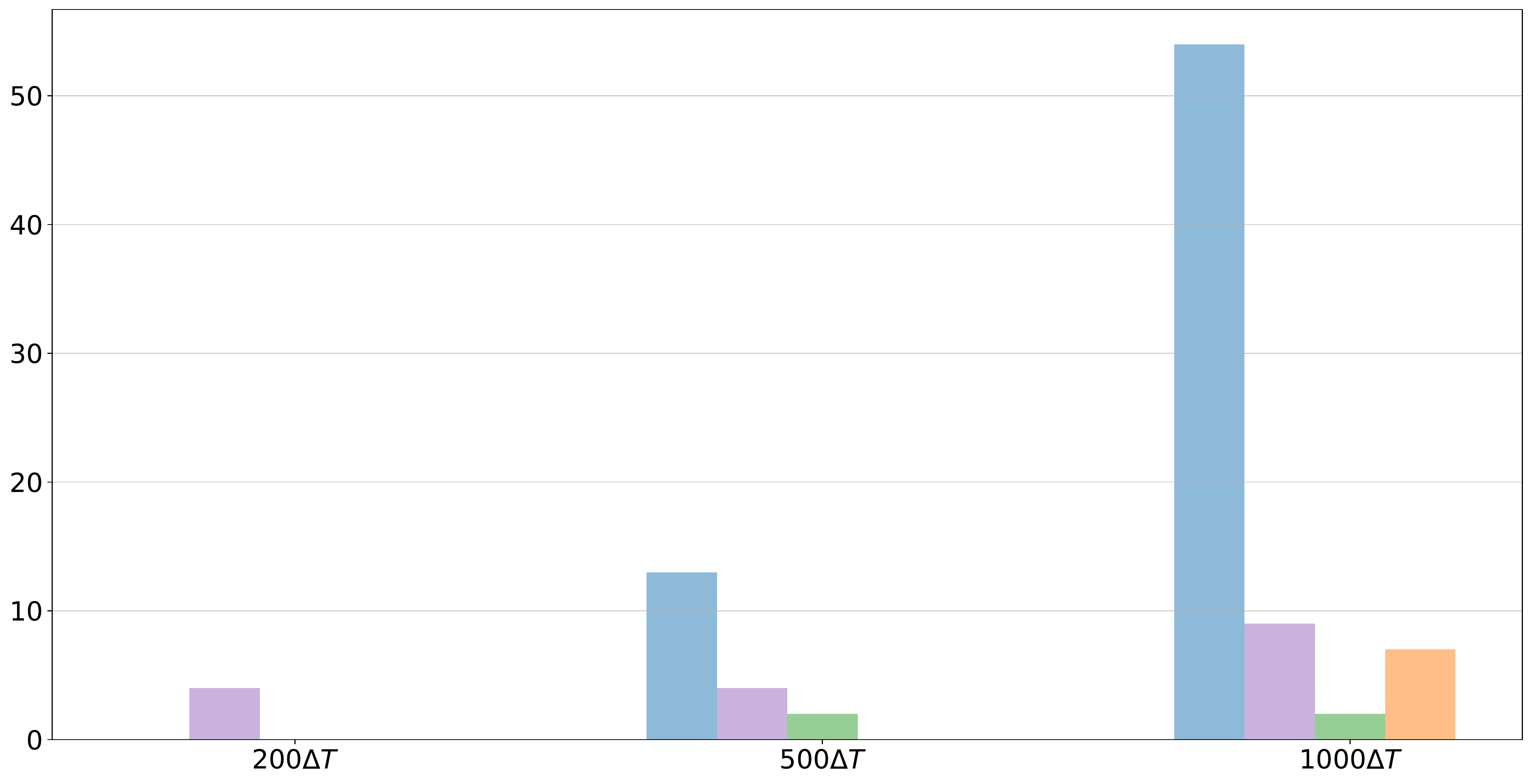}
        		\caption{200 000 training data}
        		\label{subfig:BlowUpData4}
        	\end{subfigure}\\
        \begin{adjustbox}{max width=\linewidth}
        \begin{tikzpicture}
            \begin{customlegend}[legend columns=4,legend style={draw=none, column sep=1ex},legend entries={$\lambda_{\ell_1}=0$ (Dense),$\lambda_{\ell_1}=1e^{-4}$,$\lambda_{\ell_1}=1e^{-3}$,$\lambda_{\ell_1}=1e^{-2}$}]
                \addlegendimage{Tblue!50, fill=Tblue!50,area legend}
                \addlegendimage{Tpurple!50, fill=Tpurple!50,area legend}
                \addlegendimage{Tgreen!50, fill=Tgreen!50,area legend}
                \addlegendimage{Torange!50, fill=Torange!50,area legend}
            \end{customlegend}
        \end{tikzpicture}
        \end{adjustbox}
        	\caption{Effect of regularization on the number of blowups for different amount of training data
        	}
        	\label{fig:BlowUpHist} 
        \end{figure}
    Fig.~\ref{fig:BlowUpHist} shows the frequency of blowups for models with different degree of $\ell_1$ regularization ranging from $\lambda_{\ell_1}=0$ giving dense DNNs to $\lambda_{\ell_1}=1e^{-2}$ giving very sparse DNNs. The models are trained on four different datasets, and for each dataset and each value of regularization parameter $\lambda_{\ell_1}$, an ensemble of 20 DNNs are trained. The models are then tested on 50 different simulated test trajectories, yielding 1000 possible blowups for each ensemble of models. The training datasets consists of respectively 25 000, 50 000, 100 000 and 200 000 datapoints to show the effect sparsfication has for preventing blow ups for a range of training set sizes. In each subfigure of Fig.~\ref{fig:BlowUpHist}, the number of blow ups are given after three different prediction horizons. For the models trained on the smallest training set (show in Fig.~\ref{subfig:BlowUpData1} and Fig.~\ref{subfig:BlowUpData2}) the trend is clearly that the higher degree of sparsity, the lesser the blow ups. For models trained on larger datasets (Fig.~\ref{subfig:BlowUpData3} and Fig.~\ref{subfig:BlowUpData4}) the dense models are still having significantly more blow ups than all the sparse models. However, when we compare very sparse models ($\lambda_{\ell_1}=1e^{-2}$) with medium-, and little sparse models ($\lambda_{\ell_1}=1e^{-3}$ and $\lambda_{\ell_1}=1e^{-3}$ respectively), the difference becomes less significant. This can maybe be explained by the fact that the amount of data converges to a sufficient level also for the models trained with smaller sparsity promoting regularization on the parameters. 
    
    \subsection{Training stability perspective}
    \label{subsec:trainingstabilityperspective}
        Sparsification on a large ensemble of neural networks gives similar sparse structures. This has been shown in the structure plots in the Figs.\ref{fig:sparse_f1_most_learned_struct}-\ref{fig:sparse_f8_most_learned_struct}. Furthermore, Table~\ref{table:freq_features} show that sparse models to a large extent finds the same feature basis for each of the model outputs $\{\Hat{f}_1,\; ..., \;\Hat{f}_8\}$.
        
        Moreover, Fig.~\ref{fig:RFMSE_histogram} clearly indicates for all forecasting horizons that the uncertainty bounds for dense models are much larger than those for sparse models. For the longer horizons, some of the dense models tend to blow up. This is probably due to that the model enters a region of the input space where it overfits. This can be seen as poor generalization to that specific area. Fig.~\ref{fig:BlowUpHist} quantify this by showing that state estimates calculated by dense models tend to blow up much more frequently than sparse models for all forecasting horizons and all training dataset sizes used in the case study. This indicates that the risk of finding bad minimas is higher for dense models. Furthermore, sparse models are more likely to converge to reasonable minimas with smaller training data than dense models. Hence, the study shows that sparse models have better training stability than dense models with limited data. 
        
\section{Conclusions and future work}
\label{sec:conclusions}
This article presents a sparse neural network model that approximates a set of nonlinear ODEs based on time series data sampled from the system variables included in the ODEs. The set of nonlinear ODEs represents an aluminum electrolysis simulator based on the mass and energy balance of the process. This includes nonlinear and interrelated models of electrochemical and thermal subprocesses. The sparsity in the model is achieved by imposing sparsity using $\ell_1$ regularization on the weights. The main conclusions from the study can be itemized as follows:
\begin{itemize}
    \item $\ell_1$ regularization drastically reduces the number of parameters in the deep neural network (DNN). In our case we witnessed a 93\% reduction in the parameters compared to the corresponding dense DNN for a regularization parameter of $\lambda_{\ell_1} = 1e^{-2}$.
    \item The sparse neural network was more interpretable using the domain knowledge of the aluminium electrolysis process. In contrast, the dense neural networks were completely black-box.
    \item Sparse neural networks were consistently more stable than their dense counterparts. This was reflected in the model uncertainty estimates based on a large ensemble of models. Furthermore, dense model estimates tend to diverge from the states that they are estimating way more often than sparse models. This means that the parameters of sparse neural networks are more likely to end up at a reasonable minima than the parameters of dense DNN with limited training data.
    \item For small to medium amounts of data, even the most sparse models have better median prediction accuracy than the dense models for a short prediction horizon.
    \item For medium to large amounts of training data, sparse models with low weight penalization still has better prediction accuracy than dense models in the short term. However, dense models outperform very sparse models in terms of median prediction accuracy in short prediction horizons. For longer prediction horizons, sparse models outperform dense models both in terms of higher median accuracy.
\end{itemize}
While the sparse models show promising results within interpretability and generalizability, there is still a high potential for improvement. There is a desire to increase prediction accuracy and decrease the bias of the sparse models. This might be addressed by investigating other sparsity structures at different layers that better compromise the bias-variance trade-off. One possible direction is to inject simplified theories known from first principle into the neural network to possibly increase accuracy. Lastly, we have not addressed the additional challenges related to the presence of noise.  

\section{Acknowledgments}
This work was supported by the industry partners Borregaard, Elkem, Hydro, Yara and the Research Council of Norway through the project TAPI: Towards Autonomy in Process Industries, project number 294544.

\bibliographystyle{elsarticle-num} 
\bibliography{ref.bib}

\begin{thebibliography}{10}
\expandafter\ifx\csname url\endcsname\relax
  \def\url#1{\texttt{#1}}\fi
\expandafter\ifx\csname urlprefix\endcsname\relax\def\urlprefix{URL }\fi
\expandafter\ifx\csname href\endcsname\relax
  \def\href#1#2{#2} \def\path#1{#1}\fi

\bibitem{hedrea2021tensor}
E.-L. Hedrea, R.-E. Precup, R.-C. Roman, E.~M. Petriu, Tensor product-based
  model transformation approach to tower crane systems modeling, Asian Journal
  of Control 23~(3) (2021) 1313--1323.

\bibitem{QuocNguyenARMAX2022}
V.-H.~V. Quoc-Cuong~Nguyen, M.~Thomas, Optimal armax model order identification
  of dynamic systems, London Journal of Engineering Research 22 (2022) 1--22.

\bibitem{brunton2016discovering}
S.~L. Brunton, J.~L. Proctor, J.~N. Kutz, Discovering governing equations from
  data by sparse identification of nonlinear dynamical systems, Proceedings of
  the national academy of sciences 113~(15) (2016) 3932--3937.

\bibitem{Naimi2020100}
A.~Naimi, J.~Deng, A.~Abdulrahman, V.~Vajpayee, V.~Becerra, N.~Bausch, Dynamic
  neural network-based system identification of a pressurized water reactor,
  2020, p. 100 – 104.

\bibitem{9723038}
E.~M. Rentería-Vargas, C.~J. Zúñiga~Aguilar, J.~Y. Rumbo~Morales, F.~D.
  J.~S. Vázquez, M.~De-La-Torre, J.~A. Cervantes, E.~S. Bustos,
  M.~Calixto~Rodríguez, Neural network-based identification of a psa process
  for production and purification of bioethanol, IEEE Access 10 (2022)
  27771--27782.

\bibitem{C7ME00107J}
D.~Fooshee, A.~Mood, E.~Gutman, M.~Tavakoli, G.~Urban, F.~Liu, N.~Huynh,
  D.~Van~Vranken, P.~Baldi, Deep learning for chemical reaction prediction,
  Mol. Syst. Des. Eng. 3 (2018) 442--452.

\bibitem{PAPADOPOULOS20115155}
V.~D. Papadopoulos, G.~N. Beligiannis, D.~G. Hela, Combining experimental
  design and artificial neural networks for the determination of chlorinated
  compounds in fish using matrix solid-phase dispersion, Applied Soft Computing
  11~(8) (2011) 5155--5164.

\bibitem{blakseth2022dnn}
S.~S. Blakseth, A.~Rasheed, T.~Kvamsdal, O.~San, Deep neural network enabled
  corrective source term approach to hybrid analysis and modeling, Neural
  Networks 146 (2022) 181--199.

\bibitem{san2021hybrid}
O.~San, A.~Rasheed, T.~Kvamsdal, Hybrid analysis and modeling, eclecticism, and
  multifidelity computing toward digital twin revolution (2021).
\newblock \href {http://arxiv.org/abs/2103.14629} {\path{arXiv:2103.14629}}.

\bibitem{zhang2021survey}
Y.~Zhang, P.~Tiňo, A.~Leonardis, K.~Tang, A survey on neural network
  interpretability (2021).
\newblock \href {http://arxiv.org/abs/2012.14261} {\path{arXiv:2012.14261}}.

\bibitem{liu2019improving}
S.~Liu, D.~C. Mocanu, M.~Pechenizkiy, On improving deep learning generalization
  with adaptive sparse connectivity, arXiv preprint arXiv:1906.11626 (2019).

\bibitem{blaksethcpb2022}
S.~S. Blakseth, A.~Rasheed, T.~Kvamsdal, O.~San, Combining physics-based and
  data-driven techniques for reliable hybrid analysis and modeling using the
  corrective source term approach, Applied Soft Computing 128 (2022) 109533.
\newblock \href {https://doi.org/https://doi.org/10.1016/j.asoc.2022.109533}
  {\path{doi:https://doi.org/10.1016/j.asoc.2022.109533}}.

\bibitem{robinson2022anc}
H.~Robinson, E.~Lundby, A.~Rasheed, J.~T. Gravdahl,
  \href{https://arxiv.org/abs/2209.10861}{A novel corrective-source term
  approach to modeling unknown physics in aluminum extraction process} (2022).
\newblock \href {https://doi.org/10.48550/ARXIV.2209.10861}
  {\path{doi:10.48550/ARXIV.2209.10861}}.
\newline\urlprefix\url{https://arxiv.org/abs/2209.10861}

\bibitem{LUNDBY202162}
E.~T.~B. Lundby, A.~Rasheed, J.~T. Gravdahl, I.~J. Halvorsen, A novel hybrid
  analysis and modeling approach applied to aluminum electrolysis process,
  Journal of Process Control 105 (2021) 62--77.

\bibitem{RAISSI2019686}
M.~Raissi, P.~Perdikaris, G.~Karniadakis, Physics-informed neural networks: A
  deep learning framework for solving forward and inverse problems involving
  nonlinear partial differential equations, Journal of Computational Physics
  378 (2019) 686--707.

\bibitem{pawar2021physics}
S.~Pawar, O.~San, B.~Aksoylu, A.~Rasheed, T.~Kvamsdal, Physics guided machine
  learning using simplified theories, Physics of Fluids 33~(1) (2021) 011701.

\bibitem{NIPS1988_07e1cd7d}
M.~C. Mozer, P.~Smolensky, Skeletonization: A technique for trimming the fat
  from a network via relevance assessment, in: D.~Touretzky (Ed.), Advances in
  Neural Information Processing Systems, Vol.~1, Morgan-Kaufmann, 1988.

\bibitem{ZHOU2022110489}
H.~Zhou, C.~Ibrahim, W.~X. Zheng, W.~Pan, Sparse bayesian deep learning for
  dynamic system identification, Automatica 144 (2022) 110489.

\bibitem{schoukens2016cascaded}
M.~Schoukens, P.~Mattson, T.~Wigren, J.-P. Noel, Cascaded tanks benchmark
  combining soft and hard nonlinearities, in: Workshop on nonlinear system
  identification benchmarks, 2016, pp. 20--23.

\bibitem{wigren2017coupled}
T.~Wigren, M.~Schoukens, Coupled electric drives data set and reference models,
  Department of Information Technology, Uppsala Universitet, 2017.

\bibitem{grjotheim1993introduction}
K.~Grotheim, H.~Kvande, Introduction to Aluminium Electrolysis-Understanding
  the Hall-Heroult Process, Aluminium-Verlag, Dusseldorf, Germany, 1993.

\bibitem{gusberti2012modeling}
V.~Gusberti, D.~S. Severo, B.~J. Welch, M.~Skyllas-Kazacos, Modeling the mass
  and energy balance of different aluminium smelting cell technologies, in:
  Light Metals 2012, Springer, 2012, pp. 929--934.

\bibitem{EINARSRUD20173}
K.~E. Einarsrud, I.~Eick, W.~Bai, Y.~Feng, J.~Hua, P.~J. Witt, Towards a
  coupled multi-scale, multi-physics simulation framework for aluminium
  electrolysis, Applied Mathematical Modelling 44 (2017) 3--24.

\bibitem{mandin2009industrial}
P.~Mandin, R.~W{\"u}thrich, H.~Roustan, Industrial aluminium production: the
  hall-heroult process modelling, ECS Transactions 19~(26) (2009) 1.

\bibitem{Chermont2016}
P.~R. Chermont, F.~M. Soares, R.~C. {De Oliveira}, {Simulations on the bath
  chemistry variables using neural networks}, in: TMS Light Metals, Vol.
  2016-January, 2016.

\bibitem{souza2019soft}
A.~M.~F. de~Souza, F.~M. Soares, M.~A.~G. de~Castro, N.~F. Nagem, A.~H. d.~J.
  Bitencourt, C.~d.~M. Affonso, R.~C.~L. de~Oliveira, {Soft sensors in the
  primary aluminum production process based on neural networks using clustering
  methods}, Sensors 19~(23) (2019).

\bibitem{bhattacharyay2017artificial}
D.~Bhattacharyay, D.~Kocaefe, Y.~Kocaefe, B.~Morais, An artificial neural
  network model for predicting the co 2 reactivity of carbon anodes used in the
  primary aluminum production, Neural computing and Applications 28~(3) (2017)
  553--563.

\bibitem{goodfellow2016deep}
I.~Goodfellow, Y.~Bengio, A.~Courville, Deep learning, MIT press, 2016.

\bibitem{frankle2018lottery}
J.~Frankle, M.~Carbin, The lottery ticket hypothesis: Finding sparse, trainable
  neural networks, arXiv preprint arXiv:1803.03635 (2018).

\bibitem{https://doi.org/10.48550/arxiv.1710.01878}
M.~Zhu, S.~Gupta, To prune, or not to prune: exploring the efficacy of pruning
  for model compression (2017).

\bibitem{zeng2006hidden}
X.~Zeng, D.~S. Yeung, Hidden neuron pruning of multilayer perceptrons using a
  quantified sensitivity measure, Neurocomputing 69~(7-9) (2006) 825--837.

\bibitem{hoefler2021sparsity}
T.~Hoefler, D.~Alistarh, T.~Ben-Nun, N.~Dryden, A.~Peste, Sparsity in deep
  learning: Pruning and growth for efficient inference and training in neural
  networks, arXiv preprint arXiv:2102.00554 (2021).

\bibitem{natarajan1995sparse}
B.~K. Natarajan, Sparse approximate solutions to linear systems, SIAM journal
  on computing 24~(2) (1995) 227--234.

\bibitem{montufar2014LinReg}
G.~F. Montufar, R.~Pascanu, K.~Cho, Y.~Bengio, On the number of linear regions
  of deep neural networks, in: Z.~Ghahramani, M.~Welling, C.~Cortes,
  N.~Lawrence, K.~Q. Weinberger (Eds.), Advances in Neural Information
  Processing Systems, Vol.~27, Curran Associates, Inc., 2014.

\bibitem{pascanu2014numberResponsReg}
R.~Pascanu, G.~Montufar, Y.~Bengio, On the number of response regions of deep
  feed forward networks with piece-wise linear activations, ICLR, 2014.

\bibitem{SerraBoundingLinear2018}
T.~Serra, C.~Tjandraatmadja, S.~Ramalingam, Bounding and counting linear
  regions of deep neural networks, CoRR abs/1711.02114 (2017).
\newblock \href {http://arxiv.org/abs/1711.02114} {\path{arXiv:1711.02114}}.

\bibitem{gale2003smithells}
14 - general physical properties, in: W.~Gale, T.~Totemeier (Eds.), Smithells
  Metals Reference Book (Eighth Edition), eighth edition Edition,
  Butterworth-Heinemann, Oxford, 2004, pp. 14--1--14--45.

\bibitem{dincer2014advancedpower}
I.~Dincer, C.~Zamfirescu, Chapter 1 - fundamentals of thermodynamics, in:
  I.~Dincer, C.~Zamfirescu (Eds.), Advanced Power Generation Systems, Elsevier,
  Boston, 2014, pp. 1--53.

\bibitem{skogestad2008chemical}
S.~Skogestad, Chemical and energy process engineering, CRC press, 2008.

\bibitem{TAYLOR1996913}
A dynamic model for the energy balance of an electrolysis cell, Chemical
  Engineering Research and Design 74~(8) (1996) 913--933.

\bibitem{drengstig1997process}
T.~Drengstig, On process model representation and alf3 dynamics of aluminum
  electrolysis cells, Ph.D. thesis, Ph. D. thesis, Norwegian Univ. of Science
  and Tech.(NUST) (1997).

\bibitem{jessen2008mathematical}
S.~W. Jessen, Mathematical modeling of a hall h{\'e}roult aluminium reduction
  cell, Master's thesis, Technical University of Denmark, DTU, DK-2800 Kgs.
  Lyngby, Denmark (2008).

\bibitem{hyde1997gas}
T.~M. Hyde, B.~J. Welch, The gas under anodes in aluminium smelting cells. part
  i: Measuring and modelling bubble resistance under horizontally oriented
  electrodes, LIGHT METALS-WARRENDALE- (1997) 333--340.

\bibitem{solheim2013liquidus}
A.~Solheim, S.~Rolseth, E.~Skybakmoen, L.~St{\o}en, {\AA}.~Sterten,
  T.~St{\o}re, Liquidus temperature and alumina solubility in the system na 3
  alf 6-alf 3-lif-caf 2-mgf 2, in: Essential Readings in Light Metals: Aluminum
  Reduction Technology, Volume 2, John Wiley \& Sons, 2013, pp. 73--82.

\end{thebibliography}

\clearpage
\newpage
\appendix
\section{Simulation model}
\label{appendix:a}
In this section we will follow a purely physics-based approach to deriving the equations. At appropriate places we will highlight the challenges and assumption 

The dynamical system simulated in this study is generated by the set of ordinary differential equations (ODE's) in Eqs. (\ref{eq:xdot1}) - (\ref{eq:xdot8}). This system of equations is derived from a simplified model of an aluminum electrolysis cell. This model comprises simplified energy and mass balance of the electrolysis cell. The model consider an energy balance based on sideways heat transfer, energy transfer between side ledge and bath due to melting and freezing of cryolite, and energy input as a function of electrical resistance in the electrolyte and voltage drop due to bubbles. The mass balance includes mass transfer between side ledge and bath, input of $Al_2O_3$ and $AlF_3$, production of metal and consumption of the raw material $Al_2O_3$ and tapping of metal from the electrolysis cell.  In Table \ref{table:states_inputs}, the system states and inputs are described. The purpose of the simulation model is not to mimic the exact dynamics of an aluminum electrolysis cell, but rather to generate nonlinear dynamics similar to what occurs in a real aluminum electrolysis.

\subsection{Heat capacity}
Heat capacity is a measure of the amount of thermal energy a body of a certain material can store for a given temperature and volume and is given by the definition \cite{gale2003smithells}:
\begin{equation}
    C = \frac{\delta q}{\delta T}.
\end{equation}
$C [J/^\circ C]$ is the heat capacity, $\delta q [J]$ is an infinitesimal heat quantity and $\delta T [^\circ C]$. Specific heat capacity $c_p$ is heat capacity at constant pressure per unit of mass:
\begin{equation}
    c_p = \left(\frac{d h}{d T}\right)_p,
\end{equation}
where $c_p \; [J/(kg ^\circ C]$ is the specific heat capacity, $h \; [J/kg]$ is specific enthalpy and $T \; [^\circ C]$ is temperature. The subscript $p$ indicates constant pressure. In the process of aluminum electrolysis, the pressure can be assumed to be constant at $p = 1 \;[atm]$. 

\subsection{Energy and mass balance}
The first law of thermodynamics known as the energy conservation principle states the following \cite{dincer2014advancedpower}:
\begin{equation}
    \frac{d E_{i}}{d t} = \Dot{E}_{in, \; i} - \Dot{E}_{out, \; i}.
\end{equation}
$\frac{d E_{i}}{d t} \; [W]$ is the change of energy of species $i$ in the system, $\Dot{E}_{in, \; i} \; [W]$ is the energy input rate and $\Dot{E}_{out, \; i} \; [W]$ is the energy output rate of species $i$ the system. System is here used synonymous to control volume. The energy of the system can be transferred through heat, work or through the energy associated with the mass crossing the system boundary. This can be expressed as follows:
\begin{eqnarray}
    \Dot{E}_{in, \; i} &=&  \Dot{Q}_{in, \; i} +  \Dot{W}_{in, \; i} + \Dot{m}_{in, \; i}e_{in, \; i}\\
    \Dot{E}_{out, \; i} &=&  \Dot{Q}_{out, \; i} + \Dot{W}_{out, \; i} + \Dot{m}_{out, \; i}e_{out, \; i}
\end{eqnarray}
    where $\Dot{Q}_{in, \; i} \; [W]$ and $\Dot{Q}_{out, \; i}\; [W]$ are the rates of heat in and out of the system and $\Dot{W}_{in, \; i} \; [W]$ and $\Dot{W}_{out, \; i} \; [W]$ is the rate of work generated on the system. $\Dot{m}_{in, \; i} \; [kg/s]$ and $\Dot{m}_{out, \; i}\; [kg/s]$ is the mass rate into the system and out of the system respectively, whereas $e_{in, \; i}\; [J/kg]$ and $e_{out, \; i}\; [J/kg]$ is the specific energy of the mass entering and leaving the system. The specific energy can be formulated as:
    \begin{equation}
        e = u + \frac{1}{2}v^2 + gz,
    \end{equation}
    where $u \; [J/kg]$ is the specific internal energy, $\frac{1}{2}v^2 \; [J/kg]$ is the specific energy related to velocity $v \; [m/s]$, and $gz \; [J/kg]$ is the specific energy related to elevation difference $z \; [m]$. 
    
    The change of system energy $ \frac{d E_{i}}{d t} = \Dot{E}_{i}$ can be written as:
    \begin{equation}
        \frac{d E_{i}}{d t} = \frac{d (m_{i}e_{i})}{d t},
    \end{equation}
where $m \; [kg]$ is the mass of the system and $e_{i} \; [J/kg]$ is the specific energy of the system. Since the relevant control volumes are related to an aluminum electrolysis, it is reasonable to neglect the terms $\frac{1}{2}v^2$ and $gz$. In this work, $\Dot{Q} = Q_{in, \; i} - Q_{out, \; i}$ is defined as positive when net heat is provided to the system, and $W = W_{in, \; i} - W_{out, \; i}$ is positive when work is added to the system. Thus, the resulting energy equation can be formulated as:
\begin{equation}
    \frac{d(m_{i}u_{i})}{dt} = m\frac{d u_{i}}{dt} + u_{i}\frac{dm_{i}}{dt} =\Dot{m}_{in, \; i}u_{in, \; i} - \Dot{m}_{out, \; i}u_{out, \; i} + \Dot{Q} + \Dot{W}.
\end{equation}
Work $W\; [J]$ is organized transfer of energy. $W$ can be divided into several types of work \cite{skogestad2008chemical}:
\begin{equation}
    W = W_{flow} + W_{\Delta V} + W_s + W_{el} + W_{other}.
\end{equation}
$W_{flow}$ is the work associated with the volume displacements of streams that enter and exit the system, $ W_{\Delta V}$ is the work associated with changes of the volume of the system, $W_s$ is the mechanical work supplied using movable machinery, $W_{el}$ is the electrochemical work supplied when the system is connected to an external electric circuit. $W_{other}$ is the sum of other types of work, for example if surface areas changes or electromagnetic work. For an aluminum electrolysis, $W_{\Delta V} = W_s = W_{other} \approx 0 $. $W_{flow}$ is given by:
\begin{equation}
    W_{flow} = pV,
\end{equation}
where $p$ is pressure and $V$ is volume. Enthalpy $H \; [J]$ is given by:
\begin{equation}
    H = U + pV.
\end{equation}
where $U = m\cdot u \;[J]$. Furthermore, $H = h\cdot m$. Thus:
\begin{equation}
     m\frac{d u_{i}}{dt} + u_{i}\frac{dm_{i}}{dt} = \Dot{m}_{in, \; i}h_{in, \; i} -\Dot{m}_{out, \; i}h_{out, \; i} + \Dot{Q} + \Dot{W}_{el}.
\end{equation}
Assuming that the flow work is neglectable compared to the other quantities in the energy equation for aluminum electrolysis gives that $H \approx U$. Recall that $c_p = \frac{d h}{d T}$. Hence:
\begin{equation}
     \frac{d u_{i}}{dt} \approx \frac{d h_{i}}{dt} = \frac{\partial h_{i}}{\partial T_{i}}\frac{d T_{i}}{dt} = c_{p_{i}} \frac{d T_{i}}{dt},
\end{equation}
where $\frac{dT}{dt} = \Dot{T_{i}} \; [^\circ C]$ is the temperature derivative with respect to time. This yields:
\begin{equation}
    \frac{dT_{i}}{dt} = \frac{1}{m_{i} c_{p_{i}}}\left(\Dot{m}_{in, \; i}h_{in, \; i} - \Dot{m}_{out, \; i}h_{out, \; i} + \Dot{Q} + \Dot{W_{el}} - u_{i}\frac{dm_{i}}{dt}\right) 
\end{equation}
The mass rate equation can be formulated as:
\begin{equation}
    \frac{d m_i}{dt} = \Dot{m}_{in} - \Dot{m}_{out} + \sum_{j=1}^{n_j}r_{i,j},
    \label{eq:mass_rate_eq}
\end{equation}
where $r_{i,j}$ is the reaction rate of species $i$ being produced or consumed in reaction $j$. Assuming that the contents of the control volume to be perfectly mixed and assuming that the flow work is neglectable gives:
\begin{equation}
    u_i \approx h_i = h_{out, \; i}.
\end{equation}
Hence, the resulting temperature specific energy equation for component $i$ in a control volume is given by:
\begin{equation}
    \frac{d T_{i}}{dt} = \frac{1}{m_i c_{p_i}}\left(\Dot{m}_{in,i}(h_{in, i} - h_{out,i}) + \Dot{Q} + \Dot{W}_{el} - h_{out, \; i} \sum_{j=1}^{n_j}r_{i, \; j}.    \right)
    \label{eq:comp_temp_derivatives}
\end{equation}
The latter equation states that the time derivative of the temperature in the control volume is dependent on the composition of species in the control volume. It is assumed that the temperature is equal for all components in the control volume. Furthermore, it is assumed that there is a common heat loss $\Dot{Q}$ from a control volume to other control volumes, and that electrical power $\Dot{W}_{el}$ is performed on the whole control volume instead of on different components in the control volume. When different components are mixed in a control volume, the enthalpy of this mix is more complex than adding the enthalpy of individual species. However, the complexity of mixed enthalpy is left out of this simulation model. The heat capacity of a mix of components in a control volume $c_{p_{cv}}$ is simplified to be constant despite of that $c_{p_{cv}}$ varies with composition and temperature in the control volume. The values for different species and control volumes are taken from \cite{TAYLOR1996913} and \cite{drengstig1997process}.
Thus, the simplified simulation equation for the temperature derivative in a control volume is given by:  
\begin{align}
    \frac{d T_{cv}}{dt} &=\, \frac{1}{m_{cv} c_{p_{cv}}}\left ( \left[\sum_{i=1}^{n_i}\Dot{m}_{in,i}(h_{in, i} - h_{out,i})\right] + \Dot{Q} + \Dot{W}_{el} \right. \nonumber\\ 
    &\, \left. - \sum_{i=1}^{n_i}\left[h_{out, \; i} \sum_{j=1}^{n_j}r_{i, \; j}\right]\right)
    \label{eq:general_temp_derivatives}
\end{align}
where $m_{cp}$ is the sum of masses in the control volume.

\subsection{Heat transfer}
Heat transfer $\Dot{Q} [W]$ will from this point be referred to as $Q$, meaning that the dot is omitted. the In the process of aluminum electrolysis, the two most important principles for heat transfer are convection and conduction. \textbf{Conduction} is heat transfer through molecular motion within a solid material. The expression for conduction is given by
\begin{equation}
    Q = -k\cdot A \cdot \frac{\partial T}{\partial x}.
\label{eq:conduction_Fourier}
\end{equation}
 $Q \; [W]$ is heat transferred,  $A \; [m^2]$ is the area the heat is transferred through $\frac{\partial T}{\partial x} \; [^\circ C/m]$ is the temperature gradient in the direction $x$ that the heat is transferred, and $k \; [W/(m ^\circ C)]$ is the thermal conductivity, a material dependent proportionality constant. For a fixed cross-section area, the one dimensional \textbf{steady state heat flow} through a wall of thickness $x\; [m]$ from $x=0$ with temperature $T_1$ to $x=1$ with temperature $T_1$ integrates to:
 \begin{equation}
     Q = k \cdot A \cdot \frac{T_1 - T_2}{x},
 \end{equation}
 where $T_1>T_2$. Thermal conductive resistance for a plane wall can be extracted from the latter expression:
 \begin{equation}
     R_{cond} = \frac{x}{k\cdot A},
 \end{equation}
 where $R_{cond} [^\circ C/ W]$ is the thermal resistance,  $x[m]$ is the thickness of the solid material in the direction heat is transferred, and $k$ and $A$ are as mentioned above. Thermal conductive analysis is analogous to an electrical circuit, where the temperature difference is analogous to the potential difference $V$, the heat flow is analogous to the electrical current $I$ and thermal resistance is analogous to electrical resistance $R_{el}$. 
\textbf{Convection} is the heat transfer through the mass motion of a fluid. Heat transfer between a surface at temperature $T_s$ and a fluid at a bulk temperature $T_f$ is due to convection.  Convection can be formulated as:
 \begin{equation}
     Q = h \cdot A \cdot(T_s - T_f),
 \end{equation}
 where $A [m^2]$ is the contact surface between a solid surface and the liquid, the heat transfer coefficient $h \; [W/(m^2) ^\circ C]$ is the proportionality constant between the heat flux and the thermodynamic driving force for the flow of heat, i.e. the temperature difference $(T_s - T_f)$. 
 
Thermal resistance can be defined for a fluid $R_{conv}$, and is given by:
 \begin{equation}
     R_{conv} = \frac{1}{h \cdot A}.
 \end{equation}
 
 As for electrical circuits, thermal resistances can be coupled in series, and the reciprocal of the total resistance equals the sum of reciprocals of individual resistances:
 \begin{equation}
     \frac{1}{R_{tot}} =\sum_{i=1}^N \frac{1}{R_i}
     \label{eq:term_resist_series}
 \end{equation}
Eq. \ref{eq:term_resist_series} together with the assumption of stationary heat transfer makes it possible to calculate the heat transfer from one edge to the other through several resistors in series as the temperature difference between the edges divided by the sum of reciprocals of individual resistors:
 \begin{equation}
     Q = \frac{T_{1} - T_{N+1}}{\sum_{i=1}^N R_i}, 
     \label{eq:resist_series}
 \end{equation}
 where $T_1>T_2> ...>T_{N+1}$ is temperature and $R_i$ are the resistors. 
In the simulation model, the heat transfer is assumed to be piecewise stationary, meaning that the heat transfer is assumed to be constant from the middle of one control volume to the middle of the adjacent control volume. However, the heat transfer is not assumed to be stationary through several control volumes. Thus, there are separate energy balances for each control volume. Heat transfer is only considered through the side walls of the electrolysis cell. 

  \begin{figure}[H]
    \centering
    \includegraphics[width=\linewidth]{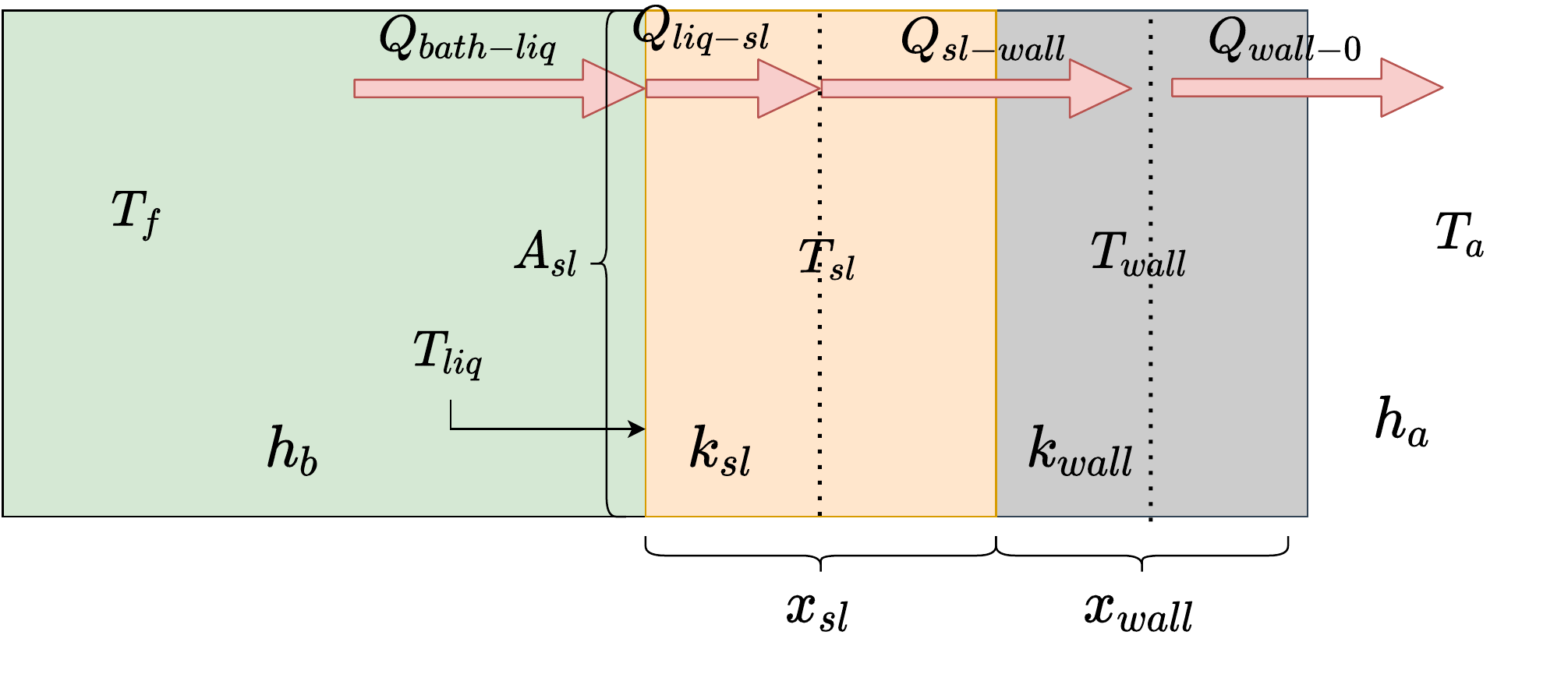}
    \caption{Convection and conduction trough several materials}
    \label{fig:conv_cond_heat_trans}
\end{figure}
Convective heat transfer $Q_{bath-liq}$ from the bath to the surface of the side ledge
\begin{equation}
    Q_{bath-liq} = h_{bath-sl}A_{sl}(T_{bath} - T_{liq}).
    \label{eq:Q_b_l}
\end{equation}
Conductive transfer $Q_{liq-sl}$ from surface of side ledge to the center of the side ledge
\begin{equation}
    Q_{liq-sl} = \frac{2k_{sl}A_{sl}(T_{liq} - T_{sl})}{x_{sl}}.
    \label{eq:Q_l_sl}
\end{equation}
Conductive heat transfer $Q_{sl-wall}$ from center of side ledge to the center of the side wall 
\begin{equation}
    Q_{sl-wall} = \frac{A_{sl}(T_{sl} - T_{wall}
    )}{(x_{wall}/2k_{wall}) + (x_{sl}/2k_{sl})}
\end{equation}
The heat transfer from the middle of the wall to the ambient $Q_{wall-0}$ consists of conductive heat transfer from the middle of the wall to the surface of the wall, and the convection from the surface of the wall to the ambient air
\begin{equation}
    Q_{wall-0} = \frac{A_{sl}(T_{wall} - T_0)}{(1/h_{wall-0}) + (x_{wall}/2k_{wall})}.
\end{equation}

\subsection{Electrochemical power}
Electrochemical power $\Dot{W}_{el} \; [W]$, from now referred to as $P_{el}$ is the amount of energy transferred to a system from a electrical circuit and is defined as:
\begin{equation}
    P_{el} = U_{cell} \cdot I_{line},
\end{equation}
where $U_{cell} \; [V]$ is the applied cell voltage and $I_{line} \; [A]$ is the line current sent through the electrolyte. The cell voltage is composed of three different types of voltage contributions. these are the \textbf{decomposition voltage}, which is the theoretical minimum potential for the decomposition of alumina, \textbf{overvoltage}, meaning the excess voltage  due to electrode polarization and \textbf{ohmic voltage drops}, due to resistance of various sections in the cell \cite{jessen2008mathematical}. These contributions can be divided into smaller contributions caused by different effects in different parts of the cells. To make the mathematical expression in the resulting nonlinear simulation model less comprehensive, only \textbf{ohmic voltage drop} contributions are included. These are:
\begin{itemize}
    \item Electrolyte voltage drop $U_{el} \; [V]$
    \item Bubble voltage drop $U_{bub}\; [V]$
\end{itemize}
The \textbf{voltage drop over the electrolyte} is due to the electrical resistivity of the electrolyte. Assuming uniform current density, the resistance of the elecrolyte is given by:
\begin{equation}
    R_{el} = \frac{1}{\kappa}\frac{d}{A}.
\end{equation}
$R_{el} \; [\Omega]$ is the electrical resistance, $\kappa \; [1/(\Omega m)]$ is electrical conductivity, $d \; [m]$ is the interpolar distance and $A \; [m^2]$ is the total surface of the anodes. The expression for electrical conductivity is given by \cite{grjotheim1993introduction}:
\begin{align}
    \kappa &=\, exp\biggl(2.0156 - \frac{2068.4}{T_{bath} + 273} + 0.4349\cdot BR - 2.07C_{al_2O_3} - 0.5C_{CaF_2} \nonumber \\
    &\,- 1.66 C_{MgF_2} + 1.78C_{LiF} + 0.77C_{Li_3AlF_6}   \biggr).
\end{align}
$T_{bath} \; [^\circ C]$ is the temperature of the electrolyte, $BR \; [-]$ is the bath ratio, while $C_{x} [-]$ is the concentration of substance $x$. $BR$ is assumed to be constant at $1.2$, $C_{MgF_2} = 0.01$, $C_{CaF_2} = 0.05$, $C_{LiF} = 0$ and $C_{Li_3 AlF_6} = 0$. Thus, $\kappa$ can be simplified to: 
\begin{equation}
    \kappa = exp\biggl( 2.496 - \frac{2068.4}{T_{bath} + 273} - 2.07C_{Al_2O_3}\biggr).
    \label{eq:el_cond}
\end{equation}
The voltage drop due to resistance in the electrolyte is given by:
\begin{equation}
    U_{EL} = R_{EL}\cdot I_{line}
\end{equation}
Gas accumulation beneath the anode surface which reduces the cross-sectional area of the electrolyte in that zone. Thus, the effective resistivity increases and causes the so called\textbf{ bubble voltage drop} $U_{bub}$ \cite{hyde1997gas}:
\begin{equation}
    U_{bub} = \frac{d_{bub} \cdot j_A}{\kappa}\frac{\phi}{1-\phi}.
    \label{eq:U_bub}
\end{equation}
$j_A \; [A/cm^2]$ is the anode current density $d_{bub} \; [cm] $ in the bubble layer thickness and $\phi \; [-] $ is the bubble coverage as a fraction of the anode:
\begin{equation}
    d_{bub} = \frac{0.5517 + j_A}{1 + 2.167j_A},
    \label{eq:d_bub}
\end{equation}
and
\begin{align}
    \phi &=\, 0.509 + 0.1823j_A - 0.1723j_A^2 + 0.05504j_A^3 \nonumber\\
    &\,+ \frac{0.4322 - 0.3781BR}{1-1.637BR} + \frac{0.431 - 0.1437(x_{Al_2O3} - x_{Al_2O3}^{AE})}{1 + 7.353(x_{Al_2O3} - x_{Al_2O3}^{AE})}
    \label{eq:bub_cov}
\end{align}
$x_{Al_2O_3} \; [-]$ is the weight percent of alumina in the bath and $x_{Al_2O_3}^{AE} \; [-]$ is the weight percent of alumina at where the anode effect occurs, in this case it is assumed that $x_{Al_2O_3}^{AE} = 2.0$. Since the simulation model is simplified to only include contributions from $U_{bub}$ and $U_{EL}$, the total applied cell voltage in the simulation model is given by:
\begin{equation}
    U_{cell} = U_{EL} + U_{bub}.
\end{equation}

\subsection{Mass rates}
The substances considered in the simulation model are alumina ($Al_2O_3$), aluminum fluoride ($AlF_3$) and cryolite ($Na_3AlF_6$) in the bath and liquid aluminum ($Al$) in the metal pad below the bath. Aluminum is extracted from alumina, which is dissolved in the electrolytic bath. In addition to alumina, carbon anodes are consumed in the net reaction, producing molten aluminum and carbon dioxide gas ($CO_2$):
\begin{equation}
    2Al_2O_3 + 3C \rightarrow 3CO_2 + 4Al.
    \label{eq:primary_reac}
\end{equation}
This reaction occurs at a rate according to Farraday's law for aluminum electrolysis \cite{drengstig1997process}:
\begin{equation}
    r_{al} = \frac{CE \cdot I_{line}}{z\cdot F},
\end{equation}
where $r_{al} \; [mol/s]$ is the reaction rate of the primary reaction of the Hall-H\'eroult process presented in Eq. \ref{eq:primary_reac}, $z=12$ is the number of electrones in the reaction, $F = 96486.7 [(A\cdot s)/mol]$ is the Faraday constant, and $CE [-]$ is the current efficiency, assumed constant at $CE = 0.95$. The fed alumina contains several impurities \cite{drengstig1997process}. In the simulation model derived used in this work, only sodium oxide ($Na_2O$) is considered as additions in the feeded alumina. The content of $Na_2O$ in alumina react as:
\begin{equation}
    3Na_2O + 4AlF_3 \rightarrow 2Na_3AlF_6 + Al_2O_3.
    \label{eq:reac_bath_prod}
\end{equation}
$3$ Mol $Na_2O$ reacts with $4$ Mol of $AlF_3$ and produces $2$ Mol of $Na_3AlF_6$ and $1$ Mol of $Al_2O_3$. The reaction rate of the latter reaction $r_{bath} \; [kmol/s]$ can be formulated as:
\begin{equation}
    r_{bath} = \frac{C_{Na_2O}}{3 M_{Na_2O}}u_{Al_2O_3},
\end{equation}
where  $C_{Na_2O} \; [-]$ is the weight percent of $Na_2O$ in the alumina feed, $M_{Na_2O} \; [g/mol]$ is the molar mass of $Na_2O$ and $u_{Al_2O_3} \; kg/s$ is the rate of alumina feed. The reaction in Eq. \ref{eq:reac_bath_prod} affects the mass balance of both $AlF_3$, $Na_3AlF_6$ and $Al_2O_3$. Therefore, $r_{bath}$ is included in the mass balance equations of all these species. The general mass rate Eq. \ref{eq:mass_rate_eq} is used in the derivation the mass rate of side ledge, cryolite, alumina, aluminum fluoride and metal. The control volumes in which there are nonzero mass rates are the bath/electrolyte, the metal pad and the side ledge. The \textbf{mass rates in the electrolyte} are:
\begin{equation}
    \Dot{m}_{Al_2O_3} = (1 - C_{Na_2O})u_{Al_2O_3} - \frac{2}{1000}r_{al}M_{Al_2O_3} + r_{bath}M_{Al_2O_3}.
    \label{eq:m_dot_al2o3}
\end{equation}
$\Dot{m}_{Al_2O_3} \; [kg/s]$ is the mass rate of alumina and $M_{Al_2O_3} \; [g/mol]$ is the molar mass of alumina. $\frac{2}{1000}r_{al}M_{Al_2O_3} \; [kg/s]$ is the reaction rate of alumina produced due to the reaction in Eq. \ref{eq:primary_reac}  and $r_{bath}M_{Al_2O_3} \; [kg/s]$ is the reaction rate of alumina due to the reaction in Eq. \ref{eq:reac_bath_prod}. The mass rate of $AlF_3$ is given by:
\begin{equation}
    \Dot{m}_{AlF_3} = u_{AlF_3} - 4r_{bath}M_{AlF_3},
    \label{eq:m_dot_alf3}
\end{equation}
where $\Dot{m}_{AlF_3} \; [kg/s]$ is the mass rate of aluminum fluoride, $u_{AlF_3} \; [kg/s]$ is the input rate of aluminum flouride and $4r_{bath}M_{AlF_3} \; [kg/s]$ is the reaction rate of produced aluminum fluoride from the reaction in equation (\ref{eq:reac_bath_prod}). The mass rate of cryolite in the bath is given by:
\begin{equation}
    \Dot{m}_{cry} = w_{fus} + 2r_{bath}M_{cry}.
    \label{eq:m_dot_cry}
\end{equation}
$\Dot{m}_{cry} \; [kg/s]$ is the mass rate of cryolite in the electrolyte, $ 2r_{bath}M_{cry} \; [kg/s]$ is the reaction rate of produced cryolite due to the reaction in Eq. \ref{eq:reac_bath_prod} and $w_{fus} \; [kg/s]$ is the mass rate of cryolite transferred between the side ledge and the bath. $w_{fus}$ is given by:
\begin{equation}
    w_{fus} = \frac{Q_{bath-liq} - Q_{liq-sl}}{\Delta_{fus}H_{cry}}.
\end{equation}
$Q_{bath-liq}$ and $Q_{liq-sl}$ is given in Eqs. \ref{eq:Q_b_l} and (\ref{eq:Q_l_sl}) respectively, and $\Delta_{fus}H_{cry}$ is the heat of fusion for cryolite, i.e. the amount of energy required to melt one $kg$ of cryolite. The heat of fusion for cryolite at $1000^\circ C$ is $\Delta_{fus}H_{cry} = 119495 \; [J/kg]$ \cite{drengstig1997process}, and is assumed to be constant in the simulation model. The \textbf{side ledge} is necessary to withstand the highly corrosive molten cryolite in the oven. The side ledge consists of \textit{frozen} cryolite \cite{drengstig1997process}. The mass rate of side ledge is therefore the transfer of cryolite between the electrolyte and side ledge due to melting and freezing:
\begin{equation}
    \Dot{m}_{sl} = - w_{fus}.
    \label{eq:m_dot_sl}
\end{equation}
$\Dot{m}_{sl} \; [kg/s]$ is the mass rate of side ledge, and $w_{fus} \; [kg/s]$ is given above. The \textbf{mass rate of aluminum} is given by:
\begin{equation}
\Dot{m}_{Al} = \frac{2}{1000}r_{al}M_{Al} - u_{tap}.
\label{eq:m_dot_al}
\end{equation}
$\Dot{m}_{Al} \; [kg/s]$ is the mass rate of aluminum, $\frac{2}{1000}r_{al}M_{Al} \; [kg/s]$ is the reaction rate of produced aluminum due to the reaction in equation (\ref{eq:primary_reac}), and $u_{tap} \; [kg/s]$ is the control input of tapping metal from the oven. 

\subsection{Temperature derivatives}
Eq. (\ref{eq:general_temp_derivatives}) is used to calculate the temperature derivatives in the electrolyte, side ledge and side wall. As mentioned above, $\Dot{Q}  = Q$ and $\Dot{W}_{el}=P_{el}$. In the \textbf{electrolyte}, the energy is transferred in and out of the control volume in many different ways. Heat $Q_{bath-sl}$ is transferred through convection from the bath to the side ledge surface ($Q_{bath-liq}$) and from the side ledge surface to the center of the side ledge with conductive heat $Q_{liq-sl}$. The resulting heat transfer can be formulated as:
\begin{equation}
    Q_{bath-sl} = \frac{T_{bath} - T_{sl}}{(x_{sl}/2k_{sl}A_{sl}) + 1/(h_{bath-sl}A_{sl})}.
\end{equation}
Energy is transferred through mass transfer in several ways. Energy needed to heat and melt substances fed as control input $u$ is given by:
\begin{align}
    E_{u} &=\ \Delta_{fus}H_{Al_2O_3}u_{Al_2O_3} + \Delta_{fus}H_{AlF_3}u_{AlF_3} \nonumber \\
    &\ + (T_{bath} - T_{in})(\Bar{c}_{p_{Al_2O_3}}u_{Al_2O_3} + \Bar{c}_{p_{AlF_3}}u_{AlF_3}). 
\end{align}
$\Delta_{fus}H_{i} \; [J/kg]$ is the specific heat of fusion for substance $i$, and $\Bar{c}_{p_i} \; [J/(^\circ C kg)]$ is the average heat capacity from $T_{in}$ to $T_{bath}$.
Energy is also transferred through mass transfer between the electrolyte and the side ledge. When side ledge (frozen cryolite) melts into the bath, energy is required both to heat and melt the frozen cryolite. The energy required to heat the frozen cryolite is:
\begin{equation}
    E_{tc, \; liq} = w_{fus}c_{p,cry, liq}(T_{bath} - T_{liq}).
\end{equation}
$w_{fus} \; [kg/s]$ is as mentioned above the mass rate of cryolite between bath and side ledge, and is positive when cryolite melts and is transferred to the bath. $c_{p,cry, liq} \; [J/(^\circ C kg)]$ is the heat capacity of molten cryolite $T_{bath} \; [^\circ C]$ is the bath temperature and $T_{liq} \; [^\circ C]$ is the liquidus temperature at which cryolite melts and freezes. $E_{tc, \; liq} \; [W]$ is the energy required to heat molten cryolite from liquidus temperature to bath temperature. The subscript $tc$ stands for temperature change and the subscript $liq$ indicates that the substance is liquid. When $E_{tc, liq}$ is positive, it is becauce $W_{fus}$ is positive, indicating that cryolite is melting. Thus, when $E_{tc, liq}$ is negative, cryolite is freezing. The energy needed to melt frozen cryolite is given by:
\begin{equation}
    E_{sc} = w_{fus}\Delta_{fus}H_{cry}. 
\end{equation}
$E_{sc} \; [W]$ is the energy required to melt the mass of frozen electrolyte. The subscript $sc$ stands for state change, meaning that it transitions between solid and liquid. $\Delta_{fus}H_{cry} \; [J/kg]$ is the heat of fusion for cryolite as mentioned above. When $E_{sc}$ is positive, cryolite is melted and energy is required whereas when $E_{sc}$ is negative, cryolite freezes, and energy is released. Energy is required for the primary reaction in Eq. \ref{eq:primary_reac} since it is endothermic:
\begin{equation}
    E_{r_{al}} = r_{al}\Delta H_{r_{al}}.
\end{equation}
$E_{r_{al}} \; [W]$ is the amount of energy required for the reaction to take place, $r_{al} [mol/s]$ is as mentioned above the reaction rate of the primary reaction in the process, and $\Delta H_{r_{al}} =2197582\; [J/mol]$ is the enthalpy of reaction for (\ref{eq:primary_reac}). Since the reaction in 
equation (\ref{eq:reac_bath_prod}) is exothermic, this reaction releases energy to its surroundings. This is given by:
\begin{equation}
    E_{r_{bath}} = 1000r_{bath}\Delta H_{r_{bath}}.
\end{equation}
$E_{r_{bath}} \; [W]$ is the energy released due to the reaction in (\ref{eq:reac_bath_prod}) $r_{bath} \; [kmol/s] = 1000[mol/s]$ is the reaction rate of (\ref{eq:reac_bath_prod}) and $\Delta H_{r_{bath}} = -993283 \; [J/mol]$ is the entalpy of reation for (\ref{eq:reac_bath_prod}).
The time derivative for the bath temperature is given by:
\begin{equation}
    \Dot{T}_{bath} = \frac{P_{el} - Q_{bath-sl} - E_u - E_{tc, \; liq} - E_{sc} - E_{r_{al}} - E_{r_{bath}}}{m_{bath} c_{p_{bath, \; liq}}}.
    \label{eq:T_dot_bath}
\end{equation}
$m_{bath} \; [kg]$ is the mass of the liquid bath and $c_{p_{bath, \; liq}} \; [J/(^\circ C \cdot kg)]$ is the enthalpy of the liquid bath.  For the \textbf{side ledge} control volume, the energy transfer is through melting and freezing of cryolite and heat transfer. The energy transfer related to melting and freezing of cryolite is given by:
\begin{equation}
    E_{tc \; sol} = w_{fus}c_{p_{cry, \; sol}}(T_{liq} - T_{sl}).
\end{equation}
$E_{tc \; sol} \; [W]$ is the energy required into the side ledge when frozen cryolite heats from side ledge temperature $T_{sl}$ to liquidus temperature $T_{liq}$. $w_{fus} \; [kg/s]$ is given above and $c_{p_{cry, \; sol}} \; [J/(^\circ C \cdot kg)]$ is the heat capacity of solid cryolite. The time derivative of the side ledge temperature is given by:
\begin{equation}
    \Dot{T}_{sl} = \frac{Q_{liq-sl} - Q_{sl-wall} - E_{tc \; sol}}{m_{sl}c_{p_{cry,\; sol}}}
    \label{eq:T_dot_sl}
\end{equation}
$\Dot{T}_{sl} \; [^\circ C/s]$ is the temperature change in the side ledge, $Q_{liq-sl} \;[W]$ and $Q_{sl-wall} \; [W]$ is the heat in and out of the side ledge respectively. $m_{sl} \; [kg]$ is the mass of the side ledge, and $c_{p_{cry,\; sol}} \; [J/(^\circ C \cdot kg)]$ is the heat capacity of solid cryolite. There is no mass transfer through the \textbf{wall}. Therefore, the only energy transfer through this control volume is through heat transfer. The time derivative of the wall temperature $\Dot{T}_{wall} \; [~\circ C/s]$ is given by:
\begin{equation}
    \Dot{T}_{wall} = \frac{Q_{sl-wall} - Q_{wall - 0}}{m_{wall} c_{p_{wall}}}.
    \label{eq:T_dot_wall}
\end{equation}
$Q_{sl-wall} \; [W]$ is the heat from the side ledge to the wall, $Q_{wall - 0} \; [W]$ is the heat from the wall to the ambient, $m_{wall} \; [kg]$ is the mass og the wall and $c_{p_{wall}} \; [J/(^\circ C \cdot kg)]$ is the heat capacity of the wall.
\subsection{Liquidus temperature}
In \cite{solheim2013liquidus}, the liquidus temperature $T_{liq}$ was determined for primary crystalization of cryolite $(Na_3AlF_6)$ in a system consisting of the bath components $Na_3AlF_6-AlF_3-LiF-CaF_2-MgF_2-KF$. The liquidus temperature was determined by thermal analysis in a vertical alumina tube furnace under argon atmosphere. An empirical cure was fitted, which is valid from temperatures $1011^\circ C$ to approximately $800^\circ C$. The curve is given by:
{\footnotesize
\begin{align}
    T_{liq} &=\ 1011 + 0.50[AlF_3] - 0.13 [AlF_3]^{2.2} \nonumber \\
    &\ -\frac{3.45[CaF_2]}{1 + 0.0173[CaF_2]} \nonumber \\
    &\ +0.124[CaF_2][AlF_3] - 0.00542\left([CaF_2][AlF_3]\right)^{1.5} \nonumber \\
    &\ -\frac{7.93[Al_2O_3]}{1 + 0.0936[Al_2O_3] - 0.0017[Al_2O_3]^2 - 0.0023[AlF_3][Al_2O_3]} \nonumber \\
    &\ -\frac{8.90[LiF]}{1 + 0.0047[LiF] + 0.0010[AlF_3]^2} \nonumber \\
    &\ -3.95[MgF_2] - 3.95[KF].\label{eq:T_liq}
\end{align}
}%
$[x]$ denote the weight-$\%$ of component $x$. In the simulator, it is assumed that the following components are constant at values $[MgF_2] = 1\%$, $[CaF_2]=5\%$, $[KF] = [LiF] = 0\%$. This yields:
{\footnotesize	
\begin{align}
    T_{liq} &=\ 991.2 + 1.12[AlF_3] - 0.13[AlF_3]^{2.2} + 0.061[AlF_3]^{1.5} \nonumber \\
    &\ - \frac{7.93[Al_2O_3]}{1 + 0.0936[AlF_3] - 0.0017[AlF_3]^{2} - 0.0023[AlF_3][Al_2O_3]}.
\end{align}
}%
\subsection{Further simplifications in the simulation model}
In addition to assumptions and simplifications accounted for in the article, some additional simplifications are made in the simulation model. The reason for this is to simplify the analytical expression in the ODE's used to simulate the dynamics of the simulation model. The ODE's will still describe a complex nonlinear system, but comparing predictive models with the simulation models, and thus analysing the performance of the novel predictive models will be clearer. From the expression for $\Dot{T}_{bath}$ in (\ref{eq:T_dot_bath}), the terms $E_u, \; E_{sc}, \; E_{r_{Al}}$ and $E_{r_{bath}}$ are omitted. Thus, in the simulation model, the expression for $\Dot{T}_{bath}$ is given by:
\begin{equation} 
    \Dot{T}_{bath, \; sim} = \frac{P_{el} - Q_{bath-sl} - E_{tc, \; liq}}{m_{bath}c_{p_{bath,\; liq}}}.
    \label{eq:T_dot_bath_sim}
\end{equation}
This neglects some essential physical effects in the process. This is justified by the argument that the main purpose of this work is to evaluate data driven models on a complex nonlinear system, rather than simulating the dynamics of an aluminum electrolysis cell as good as possible.
\end{document}